\documentclass{emulateapj}
\usepackage{apjfonts}
\journalinfo{The Astrophysical Journal, 2012, in press}

\shorttitle{SOLAR WIND PROTON, ELECTRON, AND ION HEATING}
\shortauthors{CRANMER AND VAN BALLEGOOIJEN}

\begin{document}

\title{Proton, Electron, and Ion Heating in the Fast Solar Wind from
Nonlinear Coupling Between Alfv\'{e}nic and Fast-Mode Turbulence}

\author{Steven R. Cranmer and Adriaan A. van Ballegooijen}
\affil{Harvard-Smithsonian Center for Astrophysics,
60 Garden Street, Cambridge, MA 02138, USA} 

\begin{abstract}
In the parts of the solar corona and solar wind that experience the
fewest Coulomb collisions, the component proton, electron, and
heavy ion populations are not in thermal equilibrium with one another.
Observed differences in temperatures, outflow speeds, and velocity
distribution anisotropies are useful constraints on proposed
explanations for how the plasma is heated and accelerated.
This paper presents new predictions of the rates of collisionless
heating for each particle species, in which the energy input is
assumed to come from magnetohydrodynamic (MHD) turbulence.
We first created an empirical description of the radial evolution
of Alfv\'{e}n, fast-mode, and slow-mode MHD waves.
This model provides the total wave power in each mode as a function
of distance along an expanding flux tube in the high-speed solar wind.
Next we solved a set of cascade advection-diffusion equations that
give the time-steady wavenumber spectra at each distance.
An approximate term for nonlinear coupling between the Alfv\'{e}n
and fast-mode fluctuations is included.
For reasonable choices of the parameters, our model contains enough
energy transfer from the fast mode to the Alfv\'{e}n mode to excite
the high-frequency ion cyclotron resonance.
This resonance is efficient at heating protons and other ions in
the direction perpendicular to the background magnetic field,
and our model predicts heating rates for these species that agree
well with both spectroscopic and in~situ measurements.
Nonetheless, the high-frequency waves comprise only a small part
of the total Alfv\'{e}nic fluctuation spectrum, which remains
highly two-dimensional as is observed in interplanetary space.
\end{abstract}


\section{Introduction}
\label{sec:intro}

The energy that heats the solar corona and accelerates the solar
wind originates in convective motions beneath the Sun's surface.
However, even after many years of investigation, the physical
processes that transport a fraction of this energy to the corona and
convert it into thermal, magnetic, and kinetic energy are still not
understood.
In order to construct and test theoretical models, a wide range of
measurements of relevant plasma parameters must be available.
In the low-density, open-field regions that reach into
interplanetary space, the number of plasma parameters that need
to be measured is larger because the plasma becomes
collisionless and individual particle species (e.g., protons,
electrons, and heavy ions) can exhibit divergent properties.
Such differences in particle velocity distributions are valuable
probes of kinetic processes of heating and acceleration.

The spectroscopic instruments aboard the
{\em Solar and Heliospheric Observatory} ({\em{SOHO}})---e.g.,
the Ultraviolet Coronagraph Spectrometer (UVCS) and
Solar Ultraviolet Measurements of Emitted Radiation (SUMER)---have
measured several key collisionless plasma properties for a variety
of solar wind source regions \citep{Ko95,Ko97,Ko06,Wk95,Wk97}.
These observations augment decades of in~situ plasma and
field measurements that show similar departures from thermal
equilibrium in the collisionless solar wind
\citep[e.g.,][]{N82,Ma99,Ma06,Ka08}.
In the high-speed solar wind, both coronal and heliospheric
measurements point to the existence of preferential ion heating
and acceleration, as well as protons being hotter than electrons.
There are also marked departures from Maxwellian velocity distributions
for protons and other ions, with the temperature measured in
directions perpendicular to the background magnetic field often
exceeding the temperature parallel to the field (i.e.,
$T_{\perp} > T_{\parallel}$).

A large number of different processes have been suggested to
explain the measured proton and ion properties.
Many of these processes are related to the dissipation of
magnetohydrodynamic (MHD) waves,
and many involve multiple steps of energy conversion between
waves, reconnection structures, and other nonlinear plasma features.
It was noticed several decades ago that the damping of ion
cyclotron resonant Alfv\'{e}n waves could naturally give rise to
many of the observed plasma properties
\citep[see reviews by][]{HI02,Ho08}.
The problem in the solar corona, though, is how these extremely
high-frequency ($10^2$--$10^4$ Hz) waves could be generated from
pre-existing MHD fluctuations that appear to have much lower
frequencies ($< 0.01$ Hz).

One likely source of high-frequency waves and kinetic dissipation
is an MHD turbulent cascade.
There is ample evidence that turbulence provides substantial heat
input to the plasma in interplanetary space
\citep[see][]{Co68,Ge95,TM95,Mt03}.
Furthermore, self-consistent models of turbulence-driven coronal
heating and solar wind acceleration have begun to succeed in reproducing
a wide range of observations without the need for ad~hoc free
parameters \citep[e.g.,][]{SI06,CvB07,Rp08,Br08,VV10,BP11,vB11,Ch11}.
The general scenario is that convection jostles open magnetic flux
tubes that are rooted in the photosphere and produces Alfv\'{e}n waves
that propagate into the corona.
These waves undergo partial reflection, and the resulting
``colliding wave packets'' drive a turbulent cascade which heats
the plasma when the eddies reach small enough spatial scales.

It has been known for many years that Alfv\'{e}nic turbulence
in a strong magnetic field produces a cascade to small scales mainly
in the two-dimensional plane perpendicular to the field
\citep{MT81,Sb83}, and thus is {\em not} likely to produce
high-frequency ion cyclotron waves.
In other words, MHD turbulence leads to eddies with large
perpendicular wavenumbers $k_{\perp}$ and not large parallel
wavenumbers $k_{\parallel}$.
Under typical plasma conditions in the corona and inner heliosphere,
the linear dissipation of high-$k_{\perp}$ Alfv\'{e}n waves
would lead to the preferential parallel heating of electrons
\citep{Lm99,CvB03,GB08}.
This apparently disagrees with the observational evidence for
perpendicular heating of positive ions.

There have been several proposed solutions to the apparent
incompatibility between the predictions of MHD turbulence and
existing measurements \citep[see also][]{Cr09a}.
For example, turbulent fluctuations may be susceptible to various
instabilities that cause ion cyclotron waves to grow \citep{Mk06,VP08}
or they may induce stochastic perpendicular motions in ions
if they reach nonlinear magnitudes \citep{VG04,WY07,Ch10}.
Nonetheless, heliospheric measurements have provided several pieces
of evidence for the existence of ion cyclotron resonance that gives
rise to perpendicular ion heating in the solar wind
\citep[e.g.,][]{MT01b,Bo10,He11,Sm12}.
The most direct solution to the problem still appears to be for
turbulence to transport some fraction of the fluctuation energy
to high-$k_{\parallel}$ cyclotron resonant waves.

The goal of this paper is to investigate the idea proposed by
\citet{Ch05} for the turbulent generation of ion cyclotron waves.
In this scenario, nonlinear couplings between Alfv\'{e}n waves and
other modes such as fast magnetosonic waves produce an enhancement
in the high-$k_{\parallel}$ power-law tail of the Alfv\'{e}nic
fluctuation spectrum.
This is made possible by the ability of fast-mode waves to cascade
nearly isotropically in wavenumber space.
Thus, the gradual nonlinear generation of ion cyclotron waves may
provide enough heat to protons and other ions in the corona and
inner solar wind \citep[see also][]{LM06,Ch08a,YF08}.

We note that it is not currently possible to produce a rigorous model
that contains a fully self-consistent description of MHD wave transport
(from the corona to 1 AU), turbulent cascade, mode coupling, and
dissipation.
In order to make some progress in trying to understand this complex
system, we have created models that include a range of simplifying
assumptions.
One key approximation is that we divide the modeling into two
separate components: (1) a large-scale model of the radial dependence
of fluctuation energy densities, and (2) a small-scale description
of how the ``local'' fluctuations at each radius evolve in wavenumber
space and heat the plasma.
Feedbacks from the latter to the former are not included, and we
discuss their potential importance in Section \ref{sec:conc}.

We model the plasma conditions in a representative magnetic flux
tube that is rooted in a polar coronal hole and that exhibits a
steady-state fast solar wind outflow.
In Section \ref{sec:global} we describe a model of background
plasma conditions and large-scale wave transport in this flux tube.
We take an empirical approach to the solar generation of Alfv\'{e}n,
fast, and slow mode MHD waves by specifying their amplitudes as
free parameters at a lower coronal boundary height of 0.01 solar
radii ($R_{\odot}$) above the photosphere.
Section \ref{sec:cascade} gives a summary of how we model the
small-scale transport of cascading wave energy in wavenumber space,
and Section \ref{sec:coupling} describes our treatment of the
nonlinear coupling between high-frequency Alfv\'{e}n and fast-mode
waves.
In Section \ref{sec:disp} we apply quasilinear kinetic theory to
predict the net rates of particle heating from the cascading waves.
Section \ref{sec:heating} presents a selection of results for the
collisionless rates of proton, electron, and heavy ion heating.
Finally, Section \ref{sec:conc} concludes this paper with a brief
summary of our major results, a discussion of some of the wider
implications of this work, and suggestions for future improvements.

\section{Large-Scale Model of Coronal Hole Conditions}
\label{sec:global}

We wish to better understand the global energy budget of MHD waves and
turbulence from the lower solar corona out to the interplanetary medium.
The work of this section builds on many earlier models of the radial
evolution of Alfv\'{e}n waves in the fast solar wind
\citep[e.g.,][]{Ho86,TM95,CvB05,CH09} and extends it to describe
the likely behavior of fast and slow magnetosonic waves as well.
Below, we describe an empirical model of how the time-steady
plasma properties vary with heliocentric distance
(Section \ref{sec:global:steady}) as well as a large-scale view
of the dispersion, propagation, and dissipation of linear waves in
such a system
(Sections \ref{sec:global:linear}--\ref{sec:global:results}).

\subsection{Background Time-Steady Plasma}
\label{sec:global:steady}

We model the plasma properties along an open magnetic flux tube
rooted in a polar coronal hole.
At solar minimum, large unipolar coronal holes are associated with
superradially expanding magnetic fields and the acceleration of the
high-speed solar wind.
Because we only consider a field line along the polar axis of symmetry,
we do not need to include the rotational generation of azimuthal magnetic
fields \citep[e.g., the Parker spiral effect; see][]{WD67,PP74} or
other geometrical effects of streamer-like flux tube curvature
\citep{Li11}.
We do not distinguish between dense polar plumes and the more
tenuous interplume regions between them.
The radial dependence of plasma parameters is described as a function
of either the heliocentric distance ($r$) or the height above the
solar photosphere ($z = r - R_{\odot}$).

To specify the radial variation of the time-steady magnetic field
strength $B_0$, mass density $\rho_0$, and solar wind outflow speed
$u_0$, we used the empirical description of \citet{CvB05}.
This model combined a broad range of observational constraints
with a two-dimensional magnetostatic model of the expansion of thin
photospheric flux tubes into a supergranular network canopy.
At $r = 1$ AU in this model, the solar wind outflow speed $u_{0}$
is 781 km s$^{-1}$ and the proton density $n_{p}$ is 2.56 cm$^{-3}$.
This model also specifies the Alfv\'{e}n speed
$V_{\rm A} = B_{0} / (4\pi\rho_{0})^{1/2}$, which decreases from a
maximum value of 2890 km s$^{-1}$ at $r = 1.53 \, R_{\odot}$ down to
31 km s$^{-1}$ at 1 AU.
There is a local minimum in $V_{\rm A}$ at $r \approx 1.02 \, R_{\odot}$
that is the result of the assumed shape of network ``funnels'' that
expand superradially into the corona.

We also need to know the plasma temperature $T$ in order to determine
the relative importance of gas pressure versus magnetic pressure.
Despite observational evidence for different particle species having
different temperatures (and departures from Maxwellian velocity
distributions), we generally assume that the majority proton-electron
magnetofluid is close enough to thermal equilibrium that strong plasma
microinstabilities are not excited \citep[e.g.,][]{Ga91,Ma06}.
Thus, we specify a one-fluid temperature $T$ that is assumed to be equal
to both the proton temperature $T_p$ and the electron temperature $T_e$,
and we assume temperature isotropy ($T_{\parallel} \approx T_{\perp}$)
for both species.

We used the polar coronal hole model of \citet{CvB07} as a starting
point to describe $T(r)$, but this model was modified in two ways.
First, we moved the sharp transition region (TR) down from a height
$z$ of 0.01 to 0.003 solar radii ($R_{\odot}$) to better match the
conditions of semi-empirical models \citep[e.g.,][]{FAL90,CvB05,AL08}.
Thus, in the adopted model, at $z = 0.01 \, R_{\odot}$ the temperature
has risen to 0.48 MK, and it continues to rise to a maximum value of
1.36 MK at $z = 0.89 \, R_{\odot}$.
We also increased the temperature slightly at distances greater
than $\sim$0.2 AU in order to better agree with the mean of the in~situ
$T_p$ and $T_e$ measurements of \citet{Ce09}.
At $r = 1$ AU, $T = 0.17$ MK and it declines as $T \propto r^{-0.6}$.
The one-fluid sound speed $c_s$ is defined as
$c_{s}^{2} = \gamma k_{\rm B} T / m_{\rm H}$, where
$\gamma = 5/3$ is the monatomic ratio of specific heats,
$k_{\rm B}$ is Boltzmann's constant, and $m_{\rm H}$ is the hydrogen
atomic mass.

\begin{figure}
\epsscale{1.13}
\plotone{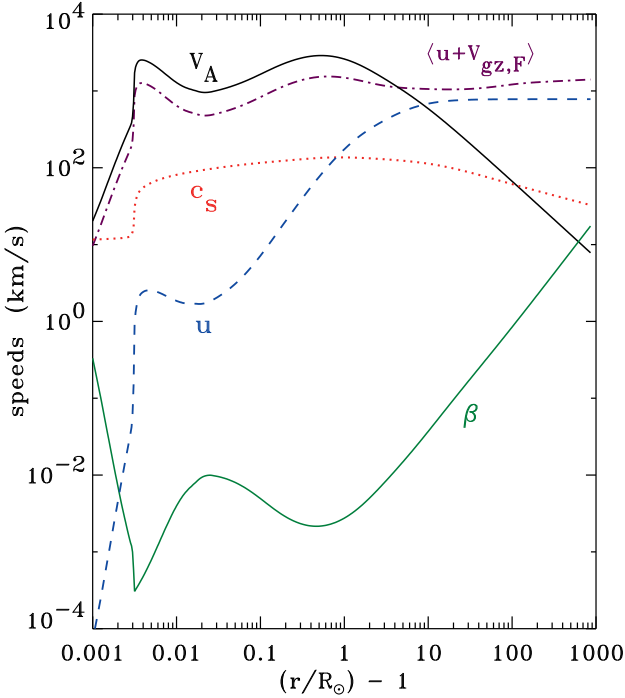}
\caption{Radial dependence of the Alfv\'{e}n speed (black solid
curve), solar wind outflow speed (blue dashed curve),
one-fluid sound speed (red dotted curve), and the
angle-averaged, intertial-frame group velocity of fast-mode waves
(violet dot-dashed curve), all in units of km s$^{-1}$.
Also shown is the dimensionless plasma $\beta$ parameter
(green solid curve).
\label{fig01}}
\end{figure}
Figure \ref{fig01} shows the radial dependence of a selection of
the background plasma properties defined above.
It also shows the dimensionless plasma beta parameter, which is usually
defined as the ratio of gas pressure to magnetic pressure, with
\begin{equation}
  \beta_{0} \, = \, \frac{P_{\rm gas}}{P_{\rm mag}} \, = \,
  \frac{2}{\gamma} \left( \frac{c_s}{V_{\rm A}} \right)^{2}  \,\, .
\end{equation}
However, we will often use a simpler dimensionless parameter $\beta$
given by
\begin{equation}
  \beta \, = \, \left( \frac{c_s}{V_{\rm A}} \right)^{2}
  \, = \, \frac{\gamma \beta_{0}}{2} \,\, ,
\end{equation}
where $\beta$ and $\beta_{0}$ differ only by a factor of 1.2
when $\gamma = 5/3$.
The range of heights shown in Figure \ref{fig01} extends down into
the solar chromosphere, but the wave models discussed below start at
a lower boundary condition in the low corona; i.e., they specify
the wave and turbulence properties only for $z \geq 0.01 \, R_{\odot}$.

\subsection{Linear Properties of MHD Waves}
\label{sec:global:linear}

In this section we briefly summarize the dispersion properties of
linear MHD waves (i.e., phase and group speeds for the Alfv\'{e}n
mode and the fast and slow magnetosonic modes) and the partitioning
between fluctuations in kinetic, magnetic, and thermal energy.
In Sections \ref{sec:global:action}--\ref{sec:global:results} we assume
that all three types of MHD waves are present, and we vary their
relative strengths arbitrarily in order to match the observations.

The phase speed $V_{\rm ph} = \omega / k$ is defined in terms of the
frequency $\omega$ and the magnitude of the wavenumber $k$.
In general, $V_{\rm ph}$ is a function of the Alfv\'{e}n speed, the
sound speed, and the angle $\theta$ between the background field
direction and the wavevector ${\bf k}$.
We follow the standard convention of defining a Cartesian coordinate
system with the background magnetic field along the $z$ axis
and the ${\bf k}$ vector having components only in the $x$-$z$ plane.
Also, for now we express $\omega$ and ${\bf k}$ in the frame
comoving with the solar wind.
For Alfv\'{e}n waves,
\begin{equation}
  V_{\rm ph}^{2} \, = \, V_{\rm A}^{2} \cos^{2} \theta
\end{equation}
and for the magnetosonic modes,
\begin{equation}
  V_{\rm ph}^{2} \, = \, \frac{V_{\rm A}^{2} + c_{s}^{2}}{2}
  \left( 1 \pm \Sigma \right)
  \label{eq:Vphfs}
\end{equation}
applies with the upper sign corresponding to the fast mode and the
lower sign corresponding to the slow mode, and with
\begin{equation}
  \Sigma \, = \, \sqrt{1 - \sigma \cos^{2} \theta}
  \,\, , \,\,\,\,
  \sigma \, = \, \frac{4\beta}{(1+\beta)^{2}}
\end{equation}
\citep[see, e.g.,][]{Wh97,GP04}.
In Section \ref{sec:global:action} we also need to know the component
of an MHD wave's {\em group velocity} in the direction parallel to
the background magnetic field.
We call this quantity $V_{\rm gz}$, and for the Alfv\'{e}n mode it is
identically equal to $V_{\rm A}$ no matter the value of $\theta$.
For the fast and slow modes,
\begin{equation}
  V_{\rm gz} \, = \, V_{\rm ph} \cos\theta \left( 1 \mp
  \frac{\sigma \sin^{2}\theta}{2 \Sigma \, (1 \pm \Sigma)}
  \right) \,\, .
\end{equation}

MHD waves excite oscillations in the plasma parameters.
We denote the root-mean-square (rms) fluctuation amplitudes in velocity
as $v_{x}, v_{y}, v_{z}$, in magnetic field as $B_{x}, B_{y}, B_{z}$,
and in density as $\delta \rho$.
We ignore fluctuations in the electric field because their
contribution to the total energy density tends to be negligible when
$V_{\rm A} \ll c$.
The kinetic, magnetic, and thermal energy densities associated
with each type of fluctuation are given as
\begin{equation}
  K_{i} = \frac{\rho_{0} v_{i}^2}{2}
  \,\, , \,\,\,\,
  M_{i} = \frac{B_{i}^2}{8\pi}
  \,\, , \,\,\,\,
  \Theta = \beta \frac{B_{0}^{2}}{8\pi}
  \left( \frac{\delta\rho}{\rho_0} \right)^{2}
\end{equation}
respectively, with $i = x,y,z$.
For linear Alfv\'{e}n waves, the total energy density $U_{\rm A}$ is
divided equally between transverse kinetic and magnetic fluctuations
along the $y$ axis, with $U_{\rm A} = K_{y} + M_{y}$ and
\begin{equation}
  \frac{K_y}{U_{\rm A}} \, = \,
  \frac{M_y}{U_{\rm A}} \, = \, \frac{1}{2} \,\, .
\end{equation}
For fast and slow mode waves,
\begin{equation}
  U_{\rm F,S} \, = \, K_{x} + K_{z} + M_{x} + M_{z} + \Theta
  \label{eq:fastpol0}
\end{equation}
and we follow \citet{Wh97} in expressing the partition fractions
as follows,
\begin{equation}
  \frac{K_x}{U_{\rm F,S}}  = 
  f_{n} \sin^{2}\theta + f_{t} \cos^{2}\theta
  \, , \,\,
  \frac{K_z}{U_{\rm F,S}}  = 
  f_{n} \cos^{2}\theta + f_{t} \sin^{2}\theta
  \label{eq:fastpol1}
\end{equation}
\begin{equation}
  f_{n} \, = \,
  \frac{V_{\rm ph}^{2} (V_{\rm ph}^{2} - V_{\rm A}^{2})}
  {c_{s}^{2} \Delta}
  \,\, , \,\,\,\,
  f_{t} \, = \,
  \frac{V_{\rm A}^{2} (V_{\rm ph}^{2} - c_{s}^{2}) \cos^{2}\theta}
  {V_{\rm ph}^{2} \Delta}
  \label{eq:fastpol2}
\end{equation}
\begin{equation}
  \frac{M_x}{U_{\rm F,S}} \, = \,
  \frac{(V_{\rm ph}^{2} - c_{s}^{2}) \cos^{2}\theta}{\Delta}
  \,\, , \,\,\,\,
  \frac{M_z}{U_{\rm F,S}} \, = \,
  \frac{(V_{\rm ph}^{2} - c_{s}^{2}) \sin^{2}\theta}{\Delta}
  \label{eq:fastpol3}
\end{equation}
\begin{equation}
  \frac{\Theta}{U_{\rm F,S}} \, = \,
  \frac{V_{\rm ph}^{2} - V_{\rm A}^{2}}{\Delta}
  \label{eq:fastpol4}
\end{equation}
where $\Delta = 4V_{\rm ph}^{2} - 2V_{\rm A}^{2} - 2c_{s}^{2}$.
The fast and slow velocity fluctuations ($K_{x} + K_{z}$) always occupy
exactly half of the total energy density, and the combination
of magnetic and thermal fluctuations ($M_{x} + M_{z} + \Theta$)
take up the other half.

\begin{figure}
\epsscale{1.10}
\plotone{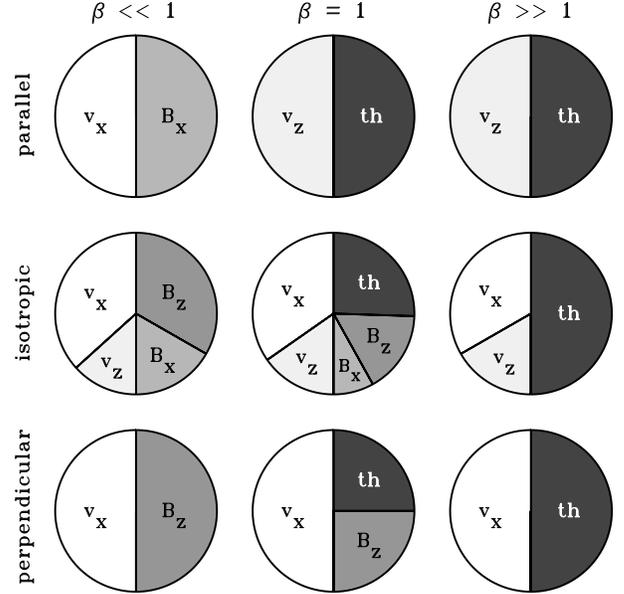}
\caption{Illustration of how fast-mode MHD waves divide their
total fluctuation energy into kinetic, magnetic, and thermal energy
in various regimes:
wavevectors parallel to ${\bf B}_0$ (top row),
an isotropic distribution of wavevectors (middle row),
wavevectors perpendicular to ${\bf B}_0$ (bottom row);
$\beta \ll 1$  (left column), $\beta = 1$  (middle column), and
$\beta \gg 1$  (right column).
Plotted areas are proportional to the
partition fractions given in Equations
(\ref{eq:fastpol1})--(\ref{eq:fastpol4}).
Kinetic energy fractions are denoted by $v_x$ and $v_z$,
magnetic energy fractions are denoted by $B_x$ and $B_z$, and
the thermal energy fraction is denoted by `th'.
\label{fig02}}
\end{figure}

The energy partition fractions given above are familiar components
of plasma physics and MHD textbooks \citep[e.g.,][]{Sx92,GP04}.
However, it is difficult to see intuitively how these fractions
vary throughout the heliosphere from Equations
(\ref{eq:fastpol1})--(\ref{eq:fastpol4}) alone.
Thus, in Figure \ref{fig02} we provide a schematic illustration of
the energy partitioning for fast-mode waves.
The three columns indicate the variation from low ($\beta \ll 1$)
to medium ($\beta=1$) and high ($\beta \gg 1$) beta plasmas.
The three rows show the results for purely parallel propagation
($\theta = 0$), an isotropic distribution of wavenumber vectors
(see below), and purely perpendicular propagation ($\theta = \pi/2$).
In general, all five terms on the right-hand side of of
Equation (\ref{eq:fastpol0}) are nonzero, but fractions less than
$\sim$1\% are not shown in Figure \ref{fig02}.
This diagram can be transformed to show the properties of slow-mode
waves by replacing $\beta$ with $1/\beta$ and interchanging the
$x$ and $z$ subscripts with one another.

\subsection{Radial Transport Equations}
\label{sec:global:action}

In order to determine how the total energy density of a
given wave mode evolves with heliocentric distance, we solve
equations of wave action conservation that contain multiple
sources of wave damping.
There have been many discussions of energy conservation for
both pure acoustic waves and incompressible Alfv\'{e}n waves
\citep[e.g.,][]{Dw70,IH82,Ve93,TM95,VV07,Sv09},
but general derivations that can also be applied to fast and slow
mode waves (for arbitrary $\theta$) are less frequently seen.
We utilize the results of \citet{J77} to write the damped
wave action conservation equation as
\begin{equation}
  \frac{\partial}{\partial t} \left( \frac{U_m}{\Omega} \right)
  + \frac{1}{A_0} \frac{\partial}{\partial r}
  \left( \frac{\langle u_{0} + V_{{\rm gz,} m} \rangle A_{0} U_m}
  {\Omega} \right) \, = \, -\frac{Q_m}{\Omega}
  \label{eq:action}
\end{equation}
where the subscript $m$ can be replaced by A, F, or S for the
relevant mode, $A_0$ is the cross sectional area of the
flux tube (i.e., $A_{0} \propto 1/B_{0}$), and $Q_m$ is the
total dissipation rate for the mode in question.
The dimensionless factor that takes account of the ``stretching''
effect of wavelengths in an accelerating reference frame is
\begin{equation}
  \Omega \, = \, \left\langle
  \frac{V_{{\rm ph,} m}}{u_{0} \cos\theta + V_{{\rm ph,} m}}
  \right\rangle
  \label{eq:Omega}
\end{equation}
and the angle brackets denote a weighted average over all angles,
\begin{equation}
  \langle f \rangle \, = \,
  \frac{\int d\theta \, \sin\theta \, f(\theta)}
  {\int d\theta \, \sin\theta}  \,\, ,
  \label{eq:anglebrack}
\end{equation}
where here we consider outward propagating waves with
$0 < \theta < \pi / 2$.
The factor of $\cos\theta$ in Equation (\ref{eq:Omega}) comes from
the difference between the wave frequency in the Sun's reference
frame ($\omega_0$) and the comoving-frame frequency
($\omega = \omega_{0} - {\bf k} \cdot {\bf u}_0$) that appears in
the definition of the wave action; see Section III of
\citet{J77}.\footnote{%
We note that the adopted form of
Equations (\ref{eq:action})--(\ref{eq:Omega})
is only one out of several possible ways of placing and grouping
the angle brackets.
For the fast and slow modes, there is also potential ambiguity
about whether one should use Lagrangian or Eulerian averages for
$u_0$ in the transport equation.
In future work we will explore the consequences of different
methods of averaging.}
Equation (\ref{eq:action}) implicitly assumes that $\omega_0$
remains constant, but it does not require the specification of
any given value of $\omega_0$.

Our use of weighted averages over $\theta$ is derived from the
assumption that wave power is distributed isotropically in
three-dimensional ${\bf k}$ space.
In Appendix \ref{appen:overview} we discuss the motivations for
assuming such an isotropic distribution of wavenumber vectors
(specifically for the fast-mode waves).
For Alfv\'{e}n waves, this assumption has no impact on
solving Equation (\ref{eq:action}), since the arguments of both
angle-bracketed quantities given above are independent of $\theta$
\citep[see also][]{Ho74}.
Thus, one obtains the same result for Alfv\'{e}n waves whether one
assumes a single value of $\theta$ or the isotropic distribution.
For fast and slow mode MHD waves, some quantities depend strongly
on $\theta$ and others do not.
For example, slow-mode waves in low-beta plasmas have values of the
angle-dependent quantity $u_{0} + V_{{\rm gz,} m}$ that are
always nearly equal to $u_{0} + c_{s}$.
Figure \ref{fig01} shows the radial dependence of
$\langle u_{0} + V_{\rm gz, F} \rangle$ for the isotropic
distribution of fast-mode waves.

We solve Equation (\ref{eq:action}) for the energy densities of
the three MHD modes ($U_{\rm A}$, $U_{\rm F}$, $U_{\rm S}$), and
we compute the dispersion and energy partition properties of all
three wave types as given in Section \ref{sec:global:linear}.
At this stage, we neglect couplings between multiple modes and
other nonlinear effects.
This is an approximation that is likely to break down wherever
the wave amplitudes become large
\citep[e.g.,][]{CW72,Wz74,Gm78,LM80,Po98,VH99,DZ01,Gg07}.
In Section \ref{sec:coupling} we discuss the likelihood of rapid
coupling between the high-wavenumber tails of the Alfv\'{e}n and
fast-mode power spectra.
However, we continue to assume that the {\em total} energy densities
are given by the solution of the individual transport equations.

To specify the dissipation rates $Q_m$, we include both linear
collisional effects (e.g., viscosity, thermal conductivity,
and electrical resistivity) for all three modes and nonlinear
turbulent damping for the Alfv\'{e}n and fast mode.
Thus, we use
\begin{equation}
  Q_{\rm A} = \tilde{Q}_{\rm A} + 2 \gamma_{\rm A} U_{\rm A}
  \,\, , \,\,\,\,
  Q_{\rm F} = \tilde{Q}_{\rm F} + 2 \gamma_{\rm F} U_{\rm F}
  \,\, , \,\,\,\,
  Q_{\rm S} = 2 \gamma_{\rm S} U_{\rm S}  \,\, .
  \label{eq:Qafs}
\end{equation}
We give the amplitude damping rates $\gamma_m$, which include
an approximation for the transition from strongly collisional to
collisionless regimes, in Appendix \ref{appen:rates}.
The turbulent damping rates $\tilde{Q}_{\rm A}$ and
$\tilde{Q}_{\rm F}$ are described in more detail below.
In general, these rates depend on the parallel and perpendicular
components of the wavenumber $(k_{\parallel}, k_{\perp})$.
For the purposes of evaluating these rates in the global wave
transport equations, we assumed that
$k_{\perp} = 1 / \lambda_{\perp}$ for all three modes, where
$\lambda_{\perp}$ is the turbulent correlation length
described below.
For the fast and slow modes, our assumption of an isotropic
distribution of wavenumbers is consistent with also assuming
$k_{\parallel} = k_{\perp}$.
For the Alfv\'{e}n mode, we found that $\gamma_{\rm A}$ never
depended on the assumed value of $k_{\parallel}$ at all, but for
completeness we used the critical balance condition (introduced
in Appendix \ref{appen:overview}) to specify $k_{\parallel}$.

We adopt phenomenological forms for the turbulent dissipation rates
that are equivalent to the total energy fluxes that cascade from
large to small scales.
Thus, $\tilde{Q}_{\rm A}$ and $\tilde{Q}_{\rm F}$ are
constrained only by the properties of fluctuations at the
largest scales, and they do not specify the exact kinetic means
of dissipation once the energy reaches the smallest scales
(but see, however, Section \ref{sec:disp}).
Dimensionally, these are similar to the rate of cascading energy flux
derived by \citet{vK38} for isotropic hydrodynamic turbulence.
For the nonlinear dissipation of Alfv\'{e}nic fluctuations, we use
\begin{equation}
  \tilde{Q}_{\rm A} \, = \, \rho_{0} \, \tilde{\alpha}_{\rm A}
  {\cal E}_{\rm turb} \,
  \frac{Z_{-}^{2} Z_{+} + Z_{+}^{2} Z_{-}}{4 \lambda_{\perp}}
  \label{eq:QturbA}
\end{equation}
\citep[see also][]{Hs95,ZM90a,Mt99,Dm01,Dm02,Br08}.
For the fast-mode waves, we use
\begin{equation}
  \tilde{Q}_{\rm F} \, = \, \rho_{0} \, \tilde{\alpha}_{\rm F} \,
  \frac{(v_{x}^{2} + v_{z}^{2})^2}{V_{\rm A} \lambda_{\perp}}
  \label{eq:QturbF}
\end{equation}
where the quantity $(v_{x}^{2} + v_{z}^{2})$ collects together
the total kinetic energy in fast-mode velocity fluctuations
\citep{Ch05,Sz07}.
Many of the terms introduced in Equations
(\ref{eq:QturbA})--(\ref{eq:QturbF}) are defined throughout the
remainder of this subsection.

Equation (\ref{eq:QturbA}) depends on the magnitudes of the
\citet{E50} variables,
$Z_{\pm} = v_{y} \pm B_{y} / (4\pi\rho_{0})^{1/2}$,
which specify the power in outward ($Z_{-}$) and inward ($Z_{+}$)
propagating Alfv\'{e}nic fluctuations.
Alfv\'{e}nic turbulent heating occurs only when there is energy
in both modes.
In practice we compute an effective {\em reflection coefficient}
${\cal R} = |Z_{+}| / |Z_{-}|$ whose magnitude is always less than
unity, and thus we express the Elsasser variables in terms of
the Alfv\'{e}nic energy density as
\begin{equation}
  Z_{-} \, = \, \sqrt{\frac{4 U_{\rm A}}{\rho_{0} (1 + {\cal R}^{2})}}
  \,\, , \,\,\,\,
  Z_{+} \, = \, {\cal R} Z_{-} \,\, .
  \label{eq:Zpm}
\end{equation}
An accurate solution for $Z_{\pm}$ requires the integration of
non-WKB equations of Alfv\'{e}n wave reflection
\citep[e.g.,][]{HO80,VV07}.
However, our assumption that the total power $U_{\rm A}$ varies
in accord with straightforward wave action conservation has been
shown to be reasonable, even in environments where ${\cal R}$
is not small such as the chromosphere \citep{vB11} and
interplanetary space \citep{Za96,CvB05}.

We estimate the reflection coefficient ${\cal R}$ using a
modification of the low-frequency approximation of \citet{CH09}.
Specifically, we examine the magnitudes of terms in the
transport equation for the inward Elsasser variable,
\begin{equation}
  \frac{\partial Z_{+}}{\partial t}
  + (u_{0} - V_{\rm A}) \frac{\partial Z_{+}}{\partial r}
   =  (u_{0} + V_{\rm A}) \left(
  \frac{Z_{+}}{4 H_{\rm D}} +
  \frac{Z_{-}}{2 H_{\rm A}} \right) -
  \frac{Z_{+} Z_{-}}{2 \lambda_{\perp}}
  \label{eq:dZplus}
\end{equation}
where
\begin{equation}
  H_{\rm A} = \frac{V_{\rm A}}{\partial V_{\rm A} / \partial r}
  \,\, , \,\,\,\,
  H_{\rm D} = \frac{\rho_0}{\partial \rho_{0} / \partial r}
  \,\, .
\end{equation}
\citet{CH09} neglected both terms on the left-hand side of
Equation (\ref{eq:dZplus}) as well as the term containing $H_{\rm D}$,
and thus were able to solve for $Z_{+}$ straightforwardly.
However, in cases of strong reflection, the term containing
$H_{\rm D}$ may have a magnitude comparable to the other dominant
terms.
Thus, we keep all three terms on the right-hand side and solve for
\begin{equation}
  {\cal R} \, \approx \,
  \frac{2 h / | H_{\rm A} |}{1 + (h/ | H_{\rm D} | )}
  \label{eq:Refl}
\end{equation}
where
\begin{equation}
  h \, = \, \frac{\lambda_{\perp} (u_{0} + V_{\rm A})}{2 Z_{-}} \,\, .
  \label{eq:hdef}
\end{equation}
Equation (22) of \citet{CH09} is recovered in the limit of
$h \ll |H_{\rm D}|$, with ${\cal R} \approx 2h/ |H_{\rm A}|$.
In the case of purely linear reflection, \citet{Cr10} found that
the most accurate local estimates for ${\cal R}$ were obtained
when $H_{\rm A}$ was replaced with the positive-definite quantity
\begin{equation}
  \tilde{H}_{\rm A} \, = \, V_{\rm A} t_{\rm ref}
  \, = \,
  (r+R_{\odot}) \left( 1 - \frac{R_{\odot}}{r} \right) \,\, .
  \label{eq:tref}
\end{equation}
We used $\tilde{H}_{\rm A}$ instead of $H_{\rm A}$ in
Equations (\ref{eq:Refl})--(\ref{eq:hdef}) to compute ${\cal R}$.

The definitions of the turbulent dissipation rates contain the
perpendicular length scale $\lambda_{\perp}$, which is an effective
transverse correlation length of the turbulence for the largest
``outer scale'' eddies.
For simplicity we use the same correlation length for both the
Alfv\'{e}nic and fast-mode fluctuations, but this may not be
universally valid \citep[e.g.,][]{Sz07}.
In previous papers we assumed that $\lambda_{\perp}$ scales with
the transverse width of the magnetic flux tube; i.e., that
$\lambda_{\perp} \propto B_{0}^{-1/2}$ \citep{Ho86}.
Here we describe the evolution of the transverse correlation length
$\lambda_{\perp}$ with the following transport equation,
\begin{equation}
  \frac{\partial \lambda_{\perp}}{\partial r} \, = \,
  \frac{\lambda_{\perp}}{2A_0} \frac{\partial A_0}{\partial r}
  + \frac{\tilde{\beta}_{\rm A}}{u_{0} + V_{\rm A}} \left(
  \frac{Z_{-}^{2} Z_{+} + Z_{+}^{2} Z_{-}}{Z_{-}^{2} + Z_{+}^{2}}
  \right) \,\, ,
  \label{eq:dLdr}
\end{equation}
where $\tilde{\beta}_{\rm A}$ is a dimensionless constant that is
often assumed to be equal to $\tilde{\alpha}_{\rm A}/2$
\citep[e.g.,][]{Hs95}.
The first term on the right-hand side of Equation (\ref{eq:dLdr})
drives the correlation length to expand linearly with the
perpendicular flux-tube cross section \citep{Ho86}.
The second term takes account of the nonlinear coupling between
the fluctuations and the background plasma properties.
It is given in a form suggested initially by \citet{Mt94} and
later generalized to nonzero cross-helicity turbulence by
\citet{Br08} and others.
Our transport equation attempts to bridge together the effects of
the two terms.
In the lower solar atmosphere (between the photosphere and the
chosen lower boundary of 0.01 $R_{\odot}$ for the wave transport
models) we assumed that the first term in Equation (\ref{eq:dLdr})
is dominant, and thus $\lambda_{\perp} \propto A_{0}^{1/2}$.

The turbulent dissipation rates also depend on dimensionless
Kolmogorov-type constants $\tilde{\alpha}_{\rm A}$ and
$\tilde{\alpha}_{\rm F}$ that are often assumed to have values
of order unity.
For example, \citet{Hs95} and \citet{Br09} found that
$\tilde{\alpha}_{\rm A} \approx 0.5$ gives rise to dissipation rates
that agree well with both numerical simulations and heliospheric
observations.
In our case, we used this value as a starting point, but we also
varied $\tilde{\alpha}_{\rm A}$ as a free parameter in order to
produce the best match to the well-constrained Alfv\'{e}nic
fluctuations.
On the other hand, the properties of heliospheric fast-mode turbulence
are not known nearly as well as the Alfv\'{e}n-wave turbulence.
We thus relied on the independent wave-kinetic simulations of
\citet{Pg12} to fix $\tilde{\alpha}_{\rm F}$ at a value of 2.3.

The Alfv\'{e}nic cascade rate contains an efficiency factor
${\cal E}_{\rm turb}$ that attempts to account for regions where
the turbulent cascade may not have time to develop before the
fluctuations are carried away by the wind.
\citet{CvB07} estimated this efficiency factor to scale as
\begin{equation}
  {\cal E}_{\rm turb} \, = \, \frac{1}{1 +
  ( t_{\rm eddy} / t_{\rm ref} )}  \,\, ,
  \label{eq:Eturb}
\end{equation}
where the two timescales above are $t_{\rm eddy}$, a nonlinear
eddy cascade time, and $t_{\rm ref}$, a timescale for large-scale
Alfv\'{e}n wave reflection \citep[see also][]{Dm03,OD06}.
The reflection time is often defined as
$t_{\rm ref} = 1/ |\nabla \cdot {\bf V}_{A}|$,
but we solved Equation (\ref{eq:tref}) for $t_{\rm ref}$ in order
to remain consistent with the adopted model for ${\cal R}$.
The eddy cascade time is given by
\begin{equation}
  t_{\rm eddy} \, = \, \frac{\lambda_{\perp} \sqrt{3\pi}}
  {(1 + M_{\rm A}) \, v_{y}}  \,\, ,
\end{equation}
where the Alfv\'{e}n Mach number $M_{\rm A} = u_{0}/V_{\rm A}$ and
the numerical factor of $3\pi$ comes from the normalization
of an assumed shape of the turbulence spectrum
\citep[see Appendix C of][]{CvB05}.
The two limiting cases of ${\cal E}_{\rm turb} \ll 1$ and
${\cal E}_{\rm turb} \approx 1$ are roughly equivalent to the
``weak'' and ``strong'' cascade phenomenologies discussed in
Section \ref{sec:cascade}, but they are not precisely the same.

\subsection{Representative Solutions}
\label{sec:global:results}

We solved the transport equations given in
Section \ref{sec:global:action} by numerically integrating upwards
from a specified set of lower boundary conditions at
$z = 0.01 \, R_{\odot}$ and assuming time-steady conditions
(i.e., $\partial U_{m} / \partial t = 0$).
We used a logarithmic grid of 500 radial zones in $z$ that expands
out to a maximum distance of $860 \, R_{\odot} \approx 4$~AU.
The transport equations were solved with straightforward first-order
Euler steps.
The values of the Elsasser variables $Z_{\pm}$ in each zone
were determined by iteration, since Equations
(\ref{eq:Zpm}) and (\ref{eq:Refl}) do not give a simple
closed-form solution for $Z_{+}$ and $Z_{-}$ by themselves.

There are a number of free parameters in this model whose values
were not easily obtained from either theoretical calculations or
observations.
In addition to the lower boundary conditions on the wave energy
densities $U_{\rm A}$, $U_{\rm F}$, and $U_{\rm S}$, there is also
the lower boundary condition on the correlation length $\lambda_{\perp}$
and the values of the two von K\'{a}rm\'{a}n constants
$\tilde{\alpha}_{\rm A}$ and $\tilde{\beta}_{\rm A}$.
Initially, we varied these six parameters randomly in order to
build up a large Monte Carlo ensemble of trial solutions.
For each model, we synthesized the radial variation of observable
plasma fluctuations such as the root mean squared (rms) parallel
and perpendicular fluctuation speeds,
\begin{equation}
  v_{\parallel} \, = \, v_{z}  \,\, ,
  \,\,\,\,\,\,\,
  v_{\perp} \, = \, (v_{x}^{2} + v_{y}^{2})^{1/2} \,\, ,
  \label{eq:vpp}
\end{equation}
the Elsasser variables $Z_{\pm}$, and the rms fractional
density fluctuation amplitude $\delta \rho / \rho_0$.
The velocity amplitudes $v_{\parallel}$ and $v_{\perp}$ contain
contributions from all three MHD wave types.
In nearly all models produced here, $v_{\parallel}$ is dominated by
the fast mode and $v_{\perp}$ is dominated by the Alfv\'{e}n mode.
Observations of these quantities are discussed below.

\begin{deluxetable}{cc}
\tablecaption{Standard Model Parameters for Coronal Hole
MHD Wave Transport
\label{table01}}
\tablewidth{0pt}
\tablehead{
\colhead{Parameter} &
\colhead{Value}
}
\startdata
$\tilde{\alpha}_{\rm A}$ &
0.60 \\

$\tilde{\beta}_{\rm A}$ &
0.31 \\

$\tilde{\alpha}_{\rm F}$ &
2.3 \\

$(U_{\rm A}/\rho_{0})^{1/2}$ (at $z = 0.01 \, R_{\odot}$) &
29.0 km s$^{-1}$ \\

$(U_{\rm F}/\rho_{0})^{1/2}$ (at $z = 0.01 \, R_{\odot}$) &
24.3 km s$^{-1}$ \\

$(U_{\rm S}/\rho_{0})^{1/2}$ (at $z = 0.01 \, R_{\odot}$) &
9.17 km s$^{-1}$ \\

$\lambda_{\perp}$ (at photosphere) &
120 km
\enddata
\vspace*{0.05in}
\end{deluxetable}
There was no single set of parameter values that gave rise to
perfect agreement between all of the synthesized and observed
fluctuation quantities.
This is not surprising, since the models are certainly incomplete
and there are significant uncertainties in the observations
and their interpretation.
Also, even though we aimed to restrict ourselves to measurements
made in ``quiet'' high-latitude fast wind streams,
sometimes only low-latitude data were available.
Thus, in Table \ref{table01} we give a set of optimized parameters
that were chosen because they produce adequate agreement with
the full set of observed quantities.
There were other combinations of the six parameters that gave
better agreement on any single observation, but in most of these
cases the agreement became worse for other observations.
Although the ratio of the two von K\'{a}rm\'{a}n constants
$\tilde{\alpha}_{\rm A}/\tilde{\beta}_{\rm A}$ was allowed to vary
freely, the optimal value was nonetheless found to be close to the
commonly used value of 2 \citep{Hs95}.
The best photospheric value of $\lambda_{\perp} \approx 120$ km
is intermediate between the values of 75 km \citep{CvB07}
and 300 km \citep{CvB05} found from earlier models.

\begin{figure}
\epsscale{1.15}
\plotone{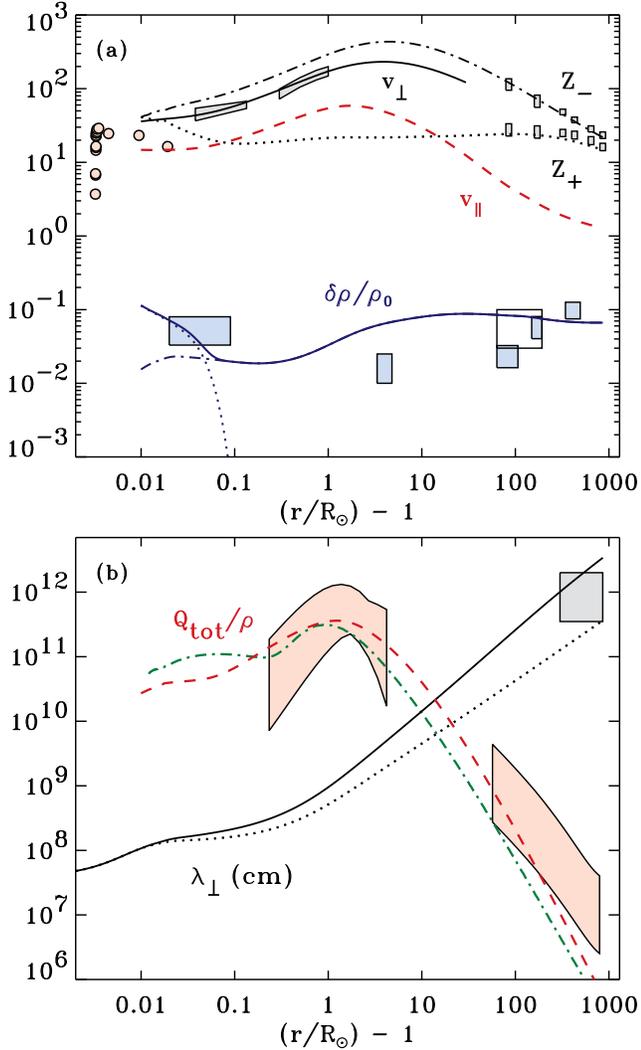}
\caption{(a) Model values for $v_{\perp}$ (black solid curve),
$Z_{-}$ (black dot-dashed curve), and $Z_{+}$ (black dotted curve)
compared with measurements (gray boxes).
Model values for $v_{\parallel}$ (red dashed curve) compared with
measurements (light red circles).
Velocities are plotted in units of km s$^{-1}$; see
Equation (\ref{eq:vpp}).
Total density amplitude $\delta\rho / \rho_0$ (blue solid curve) is
shown with its components from fast-mode (blue dot-dashed
curve) and slow-mode (blue dotted curve) waves, and compared
with observations (light blue regions and rectangles).
(b) Modeled total heating rate $Q_{\rm tot}/\rho_0$
(red dashed curve) compared with empirical constraints
(light red regions) and the total heating rate from \citet{CvB07}
(green dot-dashed curve), all in erg s$^{-1}$ g$^{-1}$.
Standard model value for $\lambda_{\perp}$ (solid black curve)
compared with earlier assumption
$\lambda_{\perp} \propto B_{0}^{-1/2}$ (black dotted curve)
and with in~situ estimates (gray region), shown in units of cm.
See text for data sources.
\label{fig03}}
\end{figure}
Figure \ref{fig03} shows the comparison between synthesized and
observed fluctuation quantities for the model parameters given
in Table \ref{table01}.
The observational constraints on $v_{\perp}$ at
$z \lesssim 0.1 \, R_{\odot}$ are a combination of the off-limb
nonthermal emission line widths given by \citet{Bj98} and \citet{LC09}.
The observations shown between 0.3 and 1 $R_{\odot}$ are from
\citet{Es99}.
At larger heights, $v_{\perp}$ becomes approximately equal to
$Z_{-}/2$, so we truncate the $v_{\perp}$ curve in favor of showing
the radial dependence of $Z_{+}$ and $Z_{-}$ more clearly.
The latter are compared directly with high-speed wind data from
{\em Helios} and {\em Ulysses} \citep{Bv00}.
Observations of longitudinal velocity fluctuations are more
difficult to find, and we show only the on-disk nonthermal line
width velocities of \citet{Ch98} as a way to compare with the
modeled values of $v_{\parallel}$.

Figure \ref{fig03}(a) shows how the modeled density fluctuation
amplitude $\delta \rho / \rho_0$ is dominated by slow-mode waves in
the low corona ($z \lesssim 0.1 \, R_{\odot}$) and by fast-mode waves
in the extended corona and solar wind ($z \gtrsim 1 \, R_{\odot}$).
The low-corona observations are drawn as an approximate boundary
region around the polar plume data given by \citet{Of99}.
The intermediate data point at $z = 4 \, R_{\odot}$ is an empirical
value of $\delta \rho / \rho_0$ estimated from radio sounding data
\citep{CH89,Sp02,HC05,Cq09}, but it is still unclear what fraction
of the measured density fluctuations are due to anything even close
to ideal MHD waves.
At larger distances, we show approximate ranges of density
fluctuations as reported by \citet{MT90}
(blue rectangles at $z < 200 \, R_{\odot}$),
\citet{TM94} (open rectangle), and
\citet{Is98} (blue rectangle at $z > 300 \, R_{\odot}$).

Figure \ref{fig03}(b) compares the result of solving
Equation (\ref{eq:dLdr}) for $\lambda_{\perp}$ with the simpler
approximation of $\lambda_{\perp} \propto B_{0}^{-1/2}$.
The plot also shows fast-wind estimates of $\lambda_{\perp}$ between
1.4 and 5 AU from {\em Ulysses} \citep{Br08}.
Figure \ref{fig03}(b) also compares the total heating
rate $Q_{\rm tot} = Q_{\rm A} + Q_{\rm F} + Q_{\rm S}$
with observational constraints and with the modeled coronal
heating rate from \citet{CvB07}.
The shaded area between 0.2 and 5 $R_{\odot}$ is an envelope
surrounding a collection of empirical and theoretical heating
curves from \citet{Wa94}, \citet{HL95}, and \citet{Al98}.
These rates illustrate what is needed to produce the observed
coronal heating and solar wind acceleration.
The area shown at larger distances ($z > 60 \, R_{\odot}$) is a
representation of the range of total (proton and electron) empirical
heating rates estimated by \citet{Ce09}.
Note that the turbulent heating rates $\tilde{Q}_{\rm A}$ and
$\tilde{Q}_{\rm F}$ dominate the total heating rate, with
approximately 70\% of the total coming from $\tilde{Q}_{\rm A}$
and 20\% from $\tilde{Q}_{\rm F}$.
Less than 10\% of $Q_{\rm tot}$ comes from the linear damping terms.

There are additional measurement techniques that may be used to
further constrain the model parameters, and in future work we will
incorporate as many of these as possible.
For example, \citet{Ho10} argued that radio measurements of Faraday
rotation fluctuations may put unique empirical constraints on the
value of $\lambda_{\perp}$ in the corona.
Also, \citet{Sr10} used multi-spacecraft data to tease out new
details of the wavenumber anisotropy of MHD fluctuations, which
may lead to better limits on, e.g., $v_{\parallel}/v_{\perp}$
in the heliosphere.
Unfortunately, the vast majority of these measurements have been
made for the slow solar wind and not the much less structured fast
wind associated with polar coronal holes.
Nearer to the Sun, \citet{Ki10} used the dispersive and energy
partition properties of thin-tube MHD waves to diagnose the presence
and strengths of various modes in active regions.
These techniques may be useful in open-field regions as well.

\begin{figure}
\epsscale{1.10}
\plotone{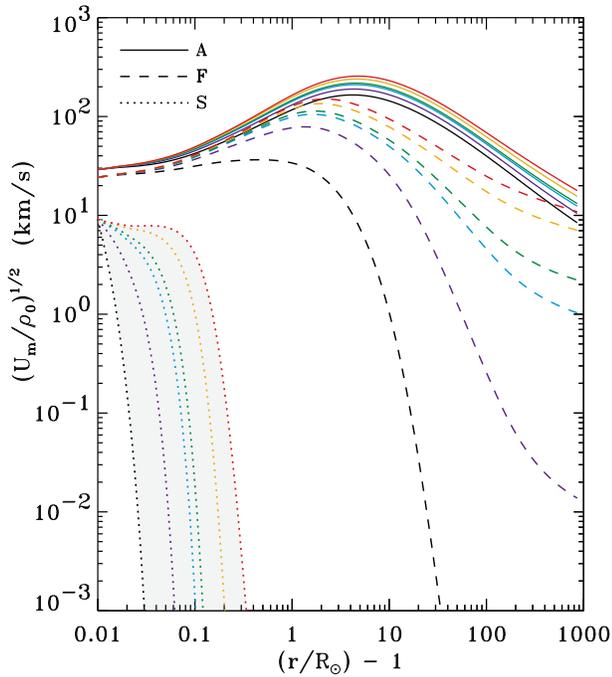}
\caption{Radial dependence of MHD wave energy densities per unit
mass, for a range of photospheric boundary conditions on
on $\lambda_{\perp}$.
From bottom to top in each set of curves, the values are:
30 km (black), 60 km (dark blue), 100 km (cyan), 120 km (green),
200 km (orange), and 300 km (red).
Different line styles denote Alfv\'{e}n waves (solid curves),
fast-mode waves (dashed curves), and slow-mode waves (dotted curves
surrounded by gray background).
\label{fig04}}
\end{figure}
Although we did not include any explicit multi-mode coupling in
the transport equations of Section \ref{sec:global:action}, there
is some feedback between the modes.
For example, the correlation length $\lambda_{\perp}$ is used in
both the Alfv\'{e}nic and fast-mode turbulent heating expressions,
and it is also used to set the wavenumbers $k_{\parallel}$ and
$k_{\perp}$ in the linear dissipation rates $\gamma_m$.
Thus, the choice of the lower boundary condition on $\lambda_{\perp}$
can have a significant impact on the radial evolution of all three
wave types.
Figure \ref{fig04} illustrates this by varying the photospheric
value of $\lambda_{\perp}$ between 30 and 300 km and using the
other standard parameters from Table \ref{table01}.
The integrated energy densities are plotted in velocity units as
$(U_{m} / \rho_{0})^{1/2}$.
The power in the Alfv\'{e}n waves changes by only a small amount
because the damping is never a strong contributor to the $U_{\rm A}$
transport equation.
However, damping is a major effect for the fast and slow modes, and
thus small changes in the damping rate's normalization can have
large relative impacts on the resulting energy densities.

Figure \ref{fig04} shows that, no matter the choice of normalization
for $\lambda_{\perp}$, it seems unlikely for the slow-mode waves
to have significantly large amplitudes anywhere but in the lowest
few tenths of a solar radius.
This appears to be consistent with models of slow-mode shock formation
and dissipation in polar plumes \citep{CS01}.
Therefore, in the remainder of this paper, our models of turbulence
in the fast solar wind {\em ignore the slow-mode waves altogether.}
We also note that Figure \ref{fig04} suggests that the actual
fast-mode wave properties in the high-speed solar wind may be more
highly variable than the Alfv\'{e}n wave properties.
Our use of a ``standard'' model for the  fast-mode waves
(using the parameters given in Table \ref{table01}) is thus
presented as an example case and not a definitive prediction.

\section{MHD Turbulent Cascade}
\label{sec:cascade}

In this section we begin constructing a model of the wavenumber
distribution of Alfv\'{e}n and fast-mode fluctuation power at each
radial distance.
We make use of a general assumption of ``scale separation;''
i.e., we presume that the turbulence becomes fully developed on
timescales short compared to the bulk solar wind outflow and the
large-scale expansion of open flux tubes.
This allows us to model the turbulence as spatially homogeneous
in a small volume element with constant background plasma properties.
This seems to be the general assumption made by the majority of MHD
simulations of turbulence in the solar wind.\footnote{%
See, however, ``expanding box'' type simulations \citep{GV96,Lw01}
that attempt to include some aspects of the large-scale radial
evolution of the plasma parcel undergoing a turbulent cascade,
and collisionless kinetic models that include expansion effects
together with local diffusion in velocity space \citep{IV09,IV11}.}
Whether or not this approximation is valid, it is useful to begin
studying the wavenumber dependence of the cascade in this manner.

\subsection{Wavenumber Advection-Diffusion Equations}
\label{sec:cascade:eqns}

We model the MHD fluctuations as time-steady Fourier distributions
of wave power in three-dimensional wavenumber space.
Although additional information about the physics of turbulence 
can be found in more complex statistical measures of the system
(e.g., higher-order structure functions), we limit ourselves to
describing the power spectrum because that is the basic quantity
needed to compute the quasilinear particle heating rates.

Because of the simplified flux-tube geometry discussed in
Section \ref{sec:global:steady}, we assume the background magnetic
field is parallel to the bulk flow velocity, and thus the system has
only one preferred spatial direction \citep[see, however,][]{Nr10}.
The random turbulent motions create a statistical equivalence between
the $x$ and $y$ directions transverse to the background field, so that
we can describe the power spectra as two-dimensional functions
of $k_{\parallel}$ and $k_{\perp}$ only.
By convention, we define the full three-dimensional power spectrum
$E_{m}$ in effective velocity-squared units; i.e., when integrated
over the full volume of wavenumber space, the spectrum gives
the fluctuation energy density per unit mass, or
\begin{equation}
  \frac{U_m}{\rho_0} \, = \, \int d^{3} {\bf k} \,\, E_{m} ({\bf k})
  \,\, .
\end{equation}
In Appendix \ref{appen:overview} we review some of the basic
physical processes that determine the shape of the spectrum for
Alfv\'{e}nic ($m = \mbox{A}$) and fast-mode ($m = \mbox{F}$)
fluctuations.

We describe the driven turbulent cascade as a combination of
advection and diffusion in wavenumber space.
At first, it may appear that a smooth and continuous description of
the spectral ``spreading'' of a cascade ignores too much of the
inherently stochastic and nonlocal nature of turbulence.
However, \citet{C43} showed that such a model can be made to
capture the essential statistics of a large ensemble of
random-walk-like (i.e., Brownian) processes.
Specific models of turbulent wavenumber transport using diffusion
or advection equations include those of
\citet{Pa65}, \citet{Lh67}, \citet{TPW84}, \citet{Tu88}, \citet{ZM90b},
\citet{Mi96}, \citet{St01}, \citet{Ch08b}, \citet{Mt09}, \citet{Jn09},
and \citet{GB10}.
For the cascade of Alfv\'{e}nic fluctuations, we generally follow
the approach taken by \citet{CvB03}.
The general forms of these equations are given as
\begin{displaymath}
  \frac{\partial E_{\rm A}}{\partial t} \, = \,
  \frac{1}{k_{\perp}} \frac{\partial}{\partial k_{\perp}}
  \left\{ D_{{\rm A}\perp} \left[ \frac{\alpha_{\perp}}{k_{\perp}}
  \frac{\partial}{\partial k_{\perp}}
  \left( k_{\perp}^{2} E_{\rm A} \right) - \mu_{\perp} E_{\rm A}
  \right] \right\}
\end{displaymath}
\begin{equation}
  + \,\, \alpha_{\parallel} \frac{\partial}{\partial k_{\parallel}}
  \left( D_{{\rm A}\parallel}
  \frac{\partial E_{\rm A}}{\partial k_{\parallel}} \right)
  + S_{\rm A} - 2 \gamma_{\rm A} E_{\rm A} + C_{\rm AF}
  \label{eq:difeqA}
\end{equation}
\begin{equation}
  \frac{\partial E_{\rm F}}{\partial t} \, = \,
  \frac{\alpha_{\rm F}}{k^2} \frac{\partial}{\partial k}
  \left( k^{2} D_{\rm F}
  \frac{\partial E_{\rm F}}{\partial k} \right)
  + S_{\rm F} - 2 \gamma_{\rm F} E_{\rm F} - C_{\rm AF}
  \label{eq:difeqF}
\end{equation}
and the terms on the right-hand sides of Equations
(\ref{eq:difeqA})--(\ref{eq:difeqF}) are defined throughout the
remainder of this subsection.
The mode coupling term $C_{\rm AF}$ is described further in
Section \ref{sec:coupling}, and the dissipation rates
$\gamma_{\rm A}$ and $\gamma_{\rm F}$ are described in
Section \ref{sec:disp}.

The perpendicular Alfv\'{e}nic cascade is described by the first term
on the right-hand side of Equation (\ref{eq:difeqA}), and we assume
an arbitrary linear combination of advection and diffusion.
\citet{CvB03} found that many key properties of the turbulence do not
depend on whether the cascade is modeled as advection, diffusion,
or both, so we retain all terms for maximum generality.
For both the parallel Alfv\'{e}nic spectral transport and the isotropic
fast-mode transport, a more standard diffusion coefficient is assumed.
The dimensionless multipliers to the $E_{\rm A}$ diffusion coefficients
are denoted $\alpha_{\perp}$ and $\alpha_{\parallel}$, to correspond
roughly to $\tilde{\alpha}_{\rm A}$ in Equation (\ref{eq:QturbA}), and
the dimensionless multiplier for the wavenumber advection coefficient
is denoted $\mu_{\perp}$.

For the Alfv\'{e}nic cascade, the overall behavior of wavenumber
transport in the perpendicular and parallel directions is specified
by the diffusion-like coefficients
\begin{equation}
  D_{{\rm A}\perp} \, = \, \frac{k_{\perp}^2}{\tau_{\rm A}}
  \,\, , \,\,\,\,
  D_{{\rm A}\parallel} \, = \, \left(
  \frac{v_{\perp}}{V_{\rm A}} \right)^{2} D_{{\rm A}\perp}
\end{equation}
where $\tau_{\rm A}$ is the cascade timescale defined below, and
$v_{\perp}$ is the $k_{\perp}$-dependent velocity response of the waves.
Note that $D_{{\rm A}\parallel}$ is independent of $k_{\parallel}$,
so it can be pulled out of the derivative in Equation (\ref{eq:difeqA}).
\citet{CvB03} showed that the above form for the diffusion
coefficients tends to reproduce the \citet{GS95} critical balance,
and \citet{Mt09} derived similar functional forms for the coefficients.
When specifying the properties of the wavenumber cascade, we apply
the scalings for ``balanced'' turbulence (i.e., zero cross helicity,
or $Z_{+} = Z_{-}$), which is more straightforward to implement but
is formally inconsistent with the large-scale transport model
of Section \ref{sec:global}.

For ideal MHD Alfv\'{e}nic fluctuations, $v_{\perp}^2$ is equal to
$b_{\perp}^2$, the latter representing the transverse magnetic
variance spectrum divided by $4\pi\rho_{0}$ to convert it to units
of velocity squared.
Following the usual convention, the power spectrum $E_{\rm A}$
tracks the magnetic fluctuations, so the reduced spectra are
defined formally as
\begin{equation}
  b_{\perp}^{2} \, = \, k_{\perp}^{2} \int dk_{\parallel} \, E_{\rm A}
  \,\, , \,\,\,
  v_{\perp}^{2} \, = \, \phi b_{\perp}^{2}  \,\, .
\end{equation}
The dimensionless factor $\phi$ describes the departure from
ideal MHD energy equipartition.
For small values of $k_{\perp}$, we assume $\phi \approx 1$.
However, as $k_{\perp}$ increases into the regime of kinetic
Alfv\'{e}n waves (KAWs), $\phi$ can become much larger than 1.
\citet{Ho99} described how the main difference between $v_{\perp}$ and
$b_{\perp}$ in the KAW regime comes from an enhanced response of
the electron velocity distribution to the electric and magnetic
fluctuations.
For simplicity, we use an approximate analytic expression
\begin{equation}
  \phi \, = \, \frac{\omega^2}{k_{\parallel}^{2} V_{\rm A}^2}
  \, \approx \, \frac{1 + k_{\perp}^{2} \rho_{p}^2}
  {1 + k_{\perp}^{2} \rho_{p}^{2} m_{e} / (\beta m_{p})}
  \,\, , \label{eq:phidef}
\end{equation}
where $\rho_{p} = w_{p} / \Omega_{p}$ is the proton thermal
gyroradius, with the proton most-probable speed given by
$w_{p} = (2 k_{\rm B} T_{p} / m_{p})^{1/2}$ and the
proton cyclotron frequency by $\Omega_{p} = e B / m_{p} c$.
Our term $\phi$ is equivalent to $\alpha^2$ as defined by \citet{Hw08}.

Inspired by Equation (\ref{eq:tauAbridge}), we define the
Alfv\'{e}nic spectral transport timescale as
\begin{equation}
  \tau_{\rm A} \, = \, \frac{1 + \chi_0}{k_{\perp} v_{\perp}}
  \label{eq:tauAformal}
\end{equation}
where we chose to replace the general critical balance parameter
$\chi$ by its value at the outer-scale parallel wavenumber
$k_{0 \parallel}$.  Thus,
\begin{equation}
  \chi_{0} \, = \, \frac{\omega_0}{k_{\perp} v_{\perp}}
  \, \approx \,
  \frac{k_{0 \parallel} V_{\rm A}}{k_{\perp} b_{\perp}} \,\, ,
  \label{eq:chi0}
\end{equation}
and $\chi_0$ is the appropriate critical balance parameter to use
when solving for the properties of the dominant low-frequency cascade.
From Equation (\ref{eq:phidef}) we see that
KAW outer-scale frequency $\omega_{0} \approx
\phi^{1/2} k_{0 \parallel} V_{\rm A}$, so that a factor of 
$\phi^{1/2}$ cancels out of both the numerator and denominator to
give the final approximate expression above.
The wavenumber $k_{0 \parallel}$ specifies the spatial scale
along the field at which energy is injected in the source term
$S_{\rm A}$ (see below).
Because $k_{0 \parallel}$ is assumed to be constant (at a given
heliocentric distance $r$), the parameters $\chi_{0}$ and
$\tau_{\rm A}$ are both functions of $k_{\perp}$ and not
$k_{\parallel}$.
The above form for Equation (\ref{eq:tauAformal}) was motivated
by the analysis of \citet{ZM90b}, \citet{Ch08b}, and \citet{Hw12},
who described how the cascade and wavenumber anisotropy change when
the system transitions from weak ($\chi_{0} \gg 1$) to strong
($\chi_{0} \ll 1$) turbulence.

As mentioned above, our expressions for $\tau_{\rm A}$,
$D_{{\rm A}\perp}$, and $D_{{\rm A}\parallel}$ assume zero
cross helicity (i.e., ${\cal R} = 1$).
There is still no agreement about how to generalize these terms
when inefficient wave reflection gives rise to nonzero cross helicity.
\citet{Lw07} found that the cascade timescales for outward and
inward wave modes are different from one another when
${\cal R} \neq 1$, but their parallel spatial scales are the same.
However, \citet{BL08,BL09} found that $k_{\parallel}$ for the
outward mode should be larger than $k_{\parallel}$ for the inward
mode, and thus the \citet{GS95} critical balance must be modified
(see Equation (\ref{eq:k0paraR}) below).
\citet{Ch08b} outlined a method for setting up the
advection-diffusion equations in the case of ${\cal R} \neq 1$,
but we defer a full implementation of that approach to future work.

Putting aside the issue of imbalanced turbulence, the dominant
perpendicular nature of the Alfv\'{e}nic cascade allows us to define
a reduced transport equation that follows the evolution
of the spectrum as a function of $k_{\perp}$ only.
If we ignore the mode coupling term $C_{\rm AF}$ for now, we can
multiply Equation (\ref{eq:difeqA}) by $k_{\perp}^2$ and integrate
over $k_{\parallel}$ to obtain
\begin{equation}
  \frac{\partial b_{\perp}^2}{\partial t} =
  k_{\perp} \frac{\partial}{\partial k_{\perp}}
  \left[ \frac{1}{\tau_{\rm A}} \left( \alpha_{\perp} k_{\perp}
  \frac{\partial b_{\perp}^2}{\partial k_{\perp}} -
  \mu_{\perp} b_{\perp}^{2} \right) \right] + \tilde{S}_{\rm A}
  - 2 \tilde{\gamma}_{\rm A} b_{\perp}^{2}
  \, .
  \label{eq:difbperp}
\end{equation}
This is essentially the same as Equation (11) of \citet{CvB03}.
The reduced source term $\tilde{S}_{\rm A}$ and dissipation rate
$\tilde{\gamma}_{\rm A}$ are defined similarly to the corresponding
terms in Equation (\ref{eq:difeqA}), but they are weighted toward
the low-$k_{\parallel}$ regions of wavenumber space that are
``filled'' by the cascade.
In Appendices \ref{appen:ansol:egrow}--\ref{appen:ansol:gy}
we derive analytic solutions for the time-steady Alfv\'{e}n-wave
power spectrum in various limiting cases.

The cascade of fast-mode waves, described by
Equation (\ref{eq:difeqF}), appears to be conceptually simpler
than the strongly anisotropic Alfv\'{e}n-wave cascade.
The diffusion coefficient is given by
$D_{\rm F} = k^{2} / \tau_{\rm F}$, where $\tau_{\rm F}$ is related
to the IK-like cascade time given by Equation (\ref{eq:tauIK})
with $p=1$.
There is increasing evidence \citep[e.g.,][]{Mk10} that
a fast-mode cascade is more rapid in the directions perpendicular
to the field than along the field.
However, the cascade does appear to proceed outward ``radially''
in the direction of increasing $k$.
Thus, it makes the most sense to use an isotropic diffusion
formalism as in Equation (\ref{eq:difeqF}), but scale the magnitude
of the diffusion timescale with $\theta$.
Following the weak turbulence model of \citet{Ch05}, we adopt
\begin{equation}
  \tau_{\rm F} \, = \, \frac{V_{\rm A}}{k v_{k}^{2} \sin\theta}
  \,\, ,
  \label{eq:tauF}
\end{equation}
which implies that
\begin{equation}
  D_{\rm F} \, = \, \frac{k^{3} v_{k}^{2} \sin\theta}{V_{\rm A}}
  \, = \, \frac{4\pi k^{6} E_{\rm F} \sin\theta}{V_{\rm A}} \,\, .
\end{equation}
\citet{Ch05} showed that the $\sin\theta$ dependence in the
denominator of $\tau_{\rm F}$ is consistent with an isotropic energy
flux for the cascade, but it does not guarantee an isotropic
wavenumber spectrum $E_{\rm F}(k)$.
More information about how we chose to implement the fast-mode
cascade is given in Appendices \ref{appen:ansol:fast} and
\ref{appen:ansol:fdamp}.

In order to fully describe the cascade in the advection-diffusion
equations, four dimensionless spectral transport constants
($\alpha_{\perp}$, $\alpha_{\parallel}$, $\mu_{\perp}$,
$\alpha_{\rm F}$) need to be specified.
\citet{Mt09} summarized the results of many MHD turbulence models
and found that $\alpha_{\perp}$ often takes on values between
0.2 and 0.5, and $\alpha_{\parallel} \approx 0.43 \alpha_{\perp}$
seems to be a useful parameterization
\citep[see Equation 13 of][]{Mt09}.
\citet{ZM90b} and \citet{Mt09} made a case for a classical form of
the diffusion operator that implies $\mu_{\perp}= 2\alpha_{\perp}$.
Alternately, \citet{vB86} found that a cascade of random-walk-like
displacements of magnetic flux tubes is described well by
$\mu_{\perp} = \alpha_{\perp}$.
\citet{Hw08} and \citet{Ch08b} used a straightforward advection
equation to model an Alfv\'{e}nic cascade, which sets
$\alpha_{\perp}=0$ and assumes $\mu_{\perp} \neq 0$.
For this type of model, \citet{Hw08} derived $\mu_{\perp} \approx 0.2$.

In our models, we are constrained by the values of the cascade
constants $\tilde{\alpha}_{\rm A}$ and $\tilde{\alpha}_{\rm F}$
used in the global transport equations of Section \ref{sec:global}.
We related these constants to the ones defined above
by integrating the cascade advection-diffusion terms over wavenumber
to find $\partial U_{m} / \partial t$.
By demanding this quantity be equal to the heating rate
$\tilde{Q}_{m}$, we obtained
\begin{equation}
  \frac{2 \alpha_{\perp}}{3} + \mu_{\perp} \, = \,
  \rule[-0.11in]{0in}{0.19in}
  \frac{3 \sqrt{6\pi}}{4} \, \tilde{\alpha}_{\rm A} \,\, ,
  \label{eq:alcomp}
\end{equation}
which assumes that the perpendicular cascade is dominant and
that ${\cal R} \approx 1$, and
\begin{equation}
  \alpha_{\rm F} \, = \, \frac{32}{7\pi} \, \tilde{\alpha}_{\rm F}
  \,\, .
\end{equation}
We keep the ratio $s = \mu_{\perp} / \alpha_{\perp}$ as a free
parameter and we explore the ramifications of varying it below.
Note, however, that if we used $s=2$ \citep[as assumed by][]{ZM90b},
then Equation (\ref{eq:alcomp}) gives $\alpha_{\perp} \approx 0.73$
and $\mu_{\perp} \approx 1.47$.
These are roughly consistent with the constants given by
\citet{ZM90b} and \citet{Mt09}.
To complete the system of cascade constants, we adopt the
\citet{Mt09} choice for $\alpha_{\parallel} = 0.43 \alpha_{\perp}$,
but we compute this quantity using the \citet{Mt09} assumption of
$s=2$.

The source terms, $S_{\rm A}$ in Equation (\ref{eq:difeqA}) and
$S_{\rm F}$ in Equation (\ref{eq:difeqF}), describe the outer-scale
injection of fluctuation energy.
The global energy balance of the waves is already described by the
radial transport model of Section \ref{sec:global}.
Thus, we specify the magnitudes of $S_{\rm A}$ and $S_{\rm F}$ by
demanding that the time-steady total energy densities $U_{\rm A}$ and
$U_{\rm F}$ be maintained at their known values at a given distance $r$.
From a physical standpoint, however, it is unclear whether the passive
propagation of waves dominates the source terms, or whether there is
significant local ``stirring'' that converts large-scale dynamical
motions (e.g., velocity shears in evolved corotating streams)
into new fluctuations.

\begin{figure}
\epsscale{1.10}
\plotone{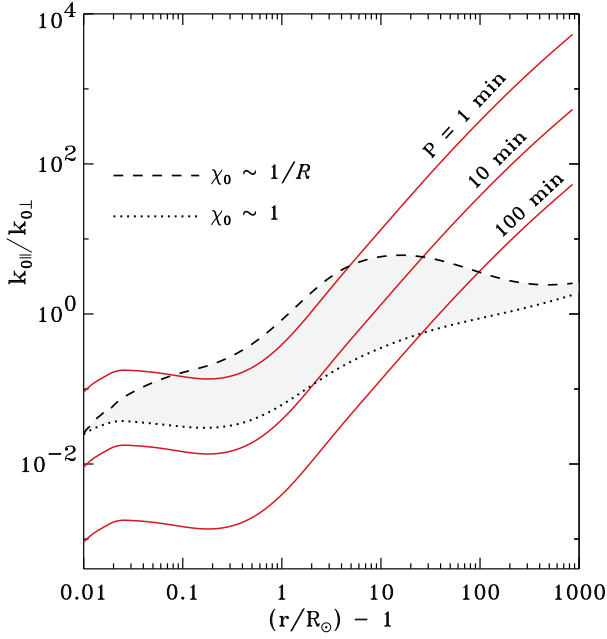}
\caption{Radial dependence of the modeled ratio of outer-scale
wavenumbers $k_{0 \parallel}/k_{0 \perp}$ computed under
various assumptions:
constant inertial-frame frequencies (red solid curves, labeled
by wave period), ideal \citet{GS95} critical balance (dotted
black curve), and modified \citet{BL08,BL09} critical balance
(dashed black curve).
\label{fig05}}
\end{figure}

We adopt specific functional forms for
$S_{\rm A} (k_{\parallel}, k_{\perp})$ and $S_{\rm F}(k)$
that are described in detail in Appendix \ref{appen:ansol}.
Generally, the source terms are nonzero only at the lowest
wavenumbers, at which the fluctuations are driven.
For the Alfv\'{e}n waves, we continue to use the assumption from
Section \ref{sec:global:action} that the perpendicular driving
scale is set by the turbulence correlation length; i.e.,
$k_{0 \perp} = 1 / \lambda_{\perp}$.
For the fast-mode fluctuations, we assume their outer-scale
wavenumber magnitude $k_{0 {\rm F}}$ is also equal to $k_{0 \perp}$,
since the largest-scale transverse stirring motions are likely
to be common to both Alfv\'{e}nic and fast-mode waves.
There are several ways that one could imagine defining the parallel
outer-scale Alfv\'{e}n wavenumber $k_{0 \parallel}$:
\begin{enumerate}
\item
Monochromatic Alfv\'{e}n waves that propagate up from the corona
retain a constant frequency $\omega_0$ in the Sun's inertial frame.
However, because the phase speed varies with distance, the
corresponding wavelength undergoes ``stretching'' commensurate
with the dispersion relation
\begin{equation}
  k_{0 \parallel} \, = \, \frac{\omega_0}{u_{0} + V_{\rm A}} \,\, .
  \label{eq:k0omega}
\end{equation}
\item
The fluctuations propagating up from the Sun may already be fully
turbulent \citep[see, e.g.,][]{vB11}.
Thus, the outer-scale parallel wavenumber may be coupled continuously
to the perpendicular wavenumber via critical balance \citep{GS95}, with
\begin{equation}
  k_{0 \parallel} \, \approx \, \frac{k_{0 \perp}}{V_{\rm A}}
  \sqrt{\frac{U_{\rm A}}{\rho_0}} \,\, .
  \label{eq:k0paraG}
\end{equation}
\item
In flux tubes with nonzero cross helicity (i.e., ${\cal R} < 1$),
\citet{BL08,BL09} found that the inward waves should obey the
\citet{GS95} critical balance, but the outward waves
(which are generally what we intend to model) obey a modified
version of critical balance, which we approximate as
\begin{equation}
  k_{0 \parallel} \, \approx \, \frac{k_{0 \perp}}{V_{\rm A}}
  \sqrt{\frac{U_{\rm A}}{\rho_0}} \frac{1}{\cal R} \,\, .
  \label{eq:k0paraR}
\end{equation}
\item
In some cases we assume that the dimensionless ratio
$k_{0 \parallel} / k_{0 \perp}$ remains fixed at a constant
specified value.
Many studies of MHD turbulence assume isotropic forcing at the
outer scale, which is consistent with the fixed ratio
$k_{0 \parallel} / k_{0 \perp} = 1$.
The lack of a physical justification for this approximation is
offset by its simplicity.
\end{enumerate}
Figure \ref{fig05} illustrates the ratio $k_{0 \parallel}/k_{0 \perp}$
for several of the above methods of setting the parallel outer scale.
For example, it shows the result of evaluating
Equation (\ref{eq:k0omega}) for a range of wave periods
$P = 2\pi / \omega_0$ between 1 and 100 minutes.
Constant assumed values of $k_{0 \parallel}/k_{0 \perp}$ would
correspond to horizontal lines in Figure \ref{fig05}.

\subsection{Solutions in the Absence of Coupling}
\label{sec:cascade:results}

Here we present some example results for the power spectra
$E_{\rm A} (k_{\parallel}, k_{\perp})$ and
$E_{\rm F} (k_{\parallel}, k_{\perp})$.
These spectra are computed from Equations (\ref{eq:difeqA}),
(\ref{eq:difeqF}), and (\ref{eq:difbperp}) in the limiting cases
of time independence and no mode coupling ($C_{\rm AF} = 0$).
The Alfv\'{e}nic spectrum was first computed in its reduced
form using the solutions for $b_{\perp}(k_{\perp})$ given in
Appendices \ref{appen:ansol:egrow} and \ref{appen:ansol:edecay},
and then it was expanded into full wavenumber space by using the
results of Appendix \ref{appen:ansol:gy}.
The shape of the fast-mode spectrum was determined from the analytic
solutions given in Appendices \ref{appen:ansol:fast} and
\ref{appen:ansol:fdamp}.
 
To illustrate the wavenumber dependence of the power spectra, we
chose a single coronal height $z = 10 \, R_{\odot}$ at which
$\beta \approx 0.04$.
We typically plot the wavenumbers in terms of dimensionless
quantities $k_{\parallel} V_{\rm A} / \Omega_p$ and
$k_{\perp} \rho_p$.
Dissipative wave-particle interactions tend to become important when
these quantities reach order-unity values, and ideal MHD conditions
apply when these quantities are small.
Typically, the driving scale for Alfv\'{e}nic turbulence occurs at
$k_{0 \perp} \rho_{p} \approx 10^{-6}$ to $10^{-4}$, with the larger
values generally occurring at larger heliocentric distances.

\begin{figure}
\epsscale{1.10}
\plotone{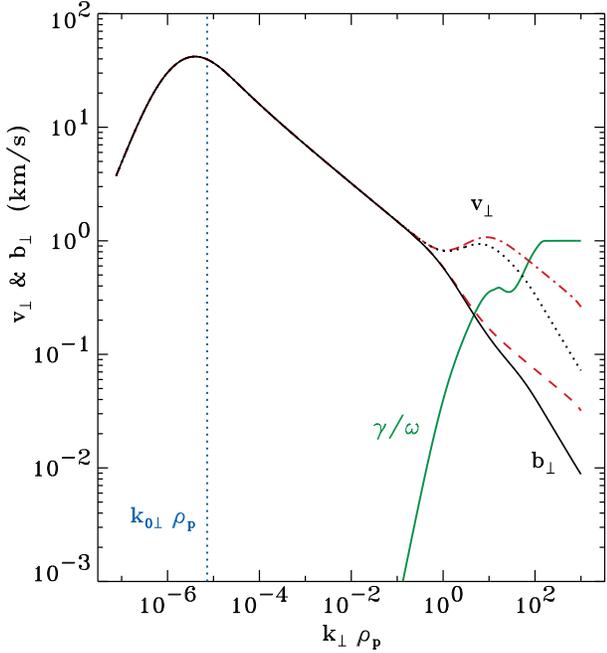}
\caption{Reduced Alfv\'{e}nic fluctuation spectra for magnetic
field and velocity fluctuations at $z = 10 \, R_{\odot}$, plotted
as a function of $k_{\perp} \rho_{p}$.
Undamped spectra for $b_{\perp}$ (red dashed curve) and $v_{\perp}$
(red dot-dashed curve) are compared with damped spectra for $b_{\perp}$
(black solid curve) and $v_{\perp}$ (black dotted curve).
The dimensionless KAW dissipation rate $\tilde{\gamma} / \omega$ used
to compute the damped spectra is also shown (green solid curve), as
is the location of the perpendicular outer scale
$k_{0 \perp} \rho_{p}$ (blue dotted line).
\label{fig06}}
\end{figure}
In Figure \ref{fig06} we show the time-steady $k_{\perp}$ dependence
for the Alfv\'{e}nic $b_{\perp}$ and $v_{\perp}$ fluctuations, both
with and without KAW dissipation.
To set the cascade properties, we utilized the values of the
constants given in Section \ref{sec:cascade:eqns}, and we also
assumed $s = \mu_{\perp} / \alpha_{\perp} = 2$ and
$k_{0 \parallel} / k_{0 \perp} = 0.1$.
The KAW damping ratio $\tilde{\gamma} / \omega$ appropriate for
the assumed value of $\beta$, which was used in
Equation (\ref{eq:epsgamma}), is also shown in green (see also
Section \ref{sec:disp}).
At the outer scale, the peak value of $v_{\perp}$ is 42 km s$^{-1}$.
We caution that this value should not be assumed to be equivalent
to the full rms velocity amplitude.
In this case, $(U_{\rm A} / \rho_{0})^{1/2} = 196$ km s$^{-1}$, which
is almost a factor of 5 larger than the maximum value of $v_{\perp}$
at this height.

The damped spectra shown in Figure \ref{fig06} have several features
that resemble those of measured KAWs in the solar wind.
Using the conventional form of the reduced energy spectrum
($e_{\rm A} \approx b_{\perp}^{2} / k_{\perp}$) we found that the
magnetic fluctuation power made a transition from a Kolmogorov-like
power law $k_{\perp}^{-5/3}$ to a steeper spectrum with
$k_{\perp}^{-2.5}$ at $k_{\perp} \rho_{p} \approx 1$.
The spectrum becomes shallower again around
$k_{\perp} \rho_{p} \approx 40$ because the wavenumber dependence
of $\phi$ flattens out at low values of $\beta$.
This behavior is reminiscent of that predicted by \citet{VD11}.
At larger radial distances where $\beta \gtrsim 1$, the KAW 
dispersion relation (Equation (\ref{eq:phidef})) gives rise to
a more sustained increase in $\phi$ with increasing $k_{\perp}$.
This in turn produces spectra that remain steep, with
$e_{\rm A} \propto k_{\perp}^{-2.5}$ persisting over several
orders of magnitude of $k_{\perp}$ in agreement with both
measurements \citep{Sm06,Sr10} and other models \citep{Hw08}.
We note that the predicted undamped KAW power-law decline of
$k_{\perp}^{-7/3}$ (see Appendix \ref{appen:ansol:egrow}) was
not seen for any sustained range of $k_{\perp}$.

\begin{figure}
\epsscale{1.10}
\plotone{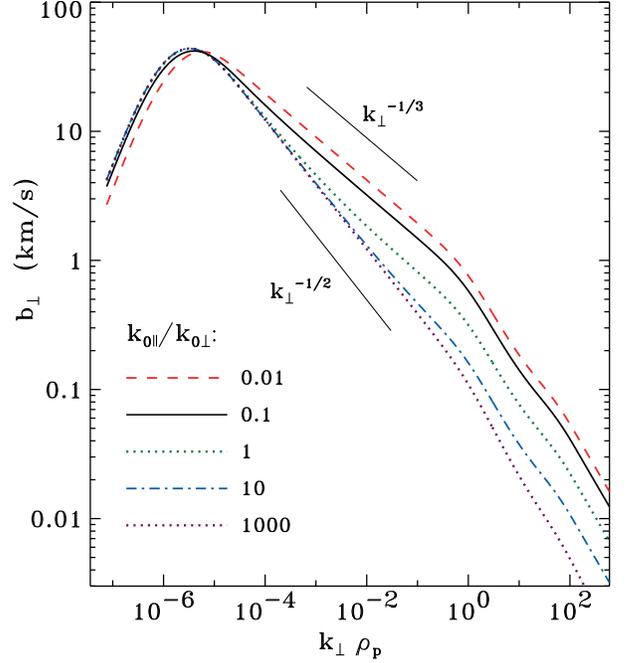}
\caption{Reduced Alfv\'{e}nic magnetic spectra at
$z = 10 \, R_{\odot}$, computed assuming different values of
$k_{0 \parallel} / k_{0 \perp} = 0.01$ (red dashed curve),
0.1 (black solid curve), 1 (green dotted curve), 10 (blue
dot-dashed curve), and 1000 (violet dotted curve).
\label{fig07}}
\end{figure}
Figure \ref{fig07} shows the result of varying the normalization 
of the parallel outer scale wavenumber $k_{0 \parallel}$ on the
shape of $b_{\perp} (k_{\perp})$.
We kept the same value of $s=2$ that was used in Figure \ref{fig06},
but we varied the constant ratio $k_{0 \parallel} / k_{0 \perp}$
over five orders of magnitude.
For the lowest values of $k_{0 \parallel}$ the outer-scale critical
balance ratio $\chi_{0}$ always remains much smaller than unity.
This means that the stirring or forcing takes place well within the
``filled'' region of wavenumber space, and thus strong turbulence occurs.
In this case, $b_{\perp} \propto k_{\perp}^{-1/3}$ and thus
$e_{\rm A} \propto k_{\perp}^{-5/3}$.
The opposite extreme case of large $k_{0 \parallel}$ corresponds to
$\chi_{0} \gg 1$ and weak turbulence with less anisotropic driving.
In that limit, the inertial range spectra are given by
$b_{\perp} \propto k_{\perp}^{-1/2}$ and
$e_{\rm A} \propto k_{\perp}^{-2}$.
Our model shows the gradual transition between these two extreme cases.

\begin{figure}
\epsscale{1.10}
\plotone{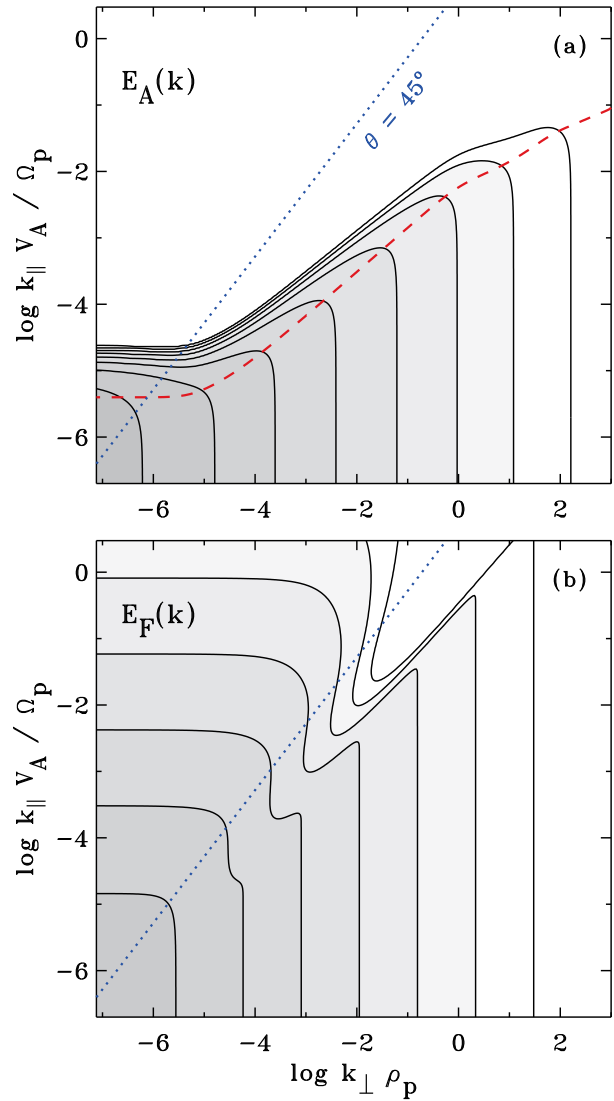}
\caption{Comparison of uncoupled power spectra at $z = 10 \, R_{\odot}$
for (a) Alfv\'{e}nic fluctuations, $E_{\rm A}(k_{\parallel}, k_{\perp})$,
and (b) fast-mode fluctuations, $E_{\rm F}(k_{\parallel}, k_{\perp})$.
Contours are plotted one per $10^4$ (i.e., one every four decades in
power) from $10^{-1}$ down to $10^{-29}$ times the maximum value of
$E_{\rm A}$.  Darker shading denotes higher power levels.
Also shown is a line denoting $\theta = 45^{\circ}$ (blue dotted curve)
and the critical balance locus of points that obey
$\chi_{\rm eff} = 1$ (red dashed curve).
\label{fig08}}
\end{figure}

In Figure \ref{fig08} we compare the Alfv\'{e}n and fast-mode
spectra with one another.
As above, we used the background conditions at a
coronal height of $z = 10 \, R_{\odot}$ and we assumed
$k_{0 \parallel} / k_{0 \perp} = 0.1$.
We illustrate the most extreme case of a {\em lack} of high-frequency
Alfv\'{e}nic power by showing the contours of
$E_{\rm A} (k_{\parallel}, k_{\perp})$ for the case
$s \rightarrow \infty$.
In this limit, Equation (\ref{eq:ggauss}) describes an exponential
decrease of power with increasing $\chi$.
Other comparable examples of this kind of spectrum can be found
in Figure 4b of \citet{CvB03} and Figures 1 and 2 of \citet{Jn09}.
We computed the Alfv\'{e}nic and fast-mode spectra with the kinetic
sources of damping that were described in Section \ref{sec:disp}.
Note that $E_{\rm F}$ experiences the strongest damping at
intermediate values of $\theta$.
For $\theta \lesssim 10^{\circ}$ or $\theta \gtrsim 85^{\circ}$,
the transit-time damping described by Equation (\ref{eq:gamttd})
is relatively weak.

\section{Coupling Between Alfv\'{e}n and Fast-Mode Waves}
\label{sec:coupling}

\subsection{Basic Physics and Phenomenological Rates}
\label{sec:coupling:rate}

There are several ways that the ideal linear MHD wave modes can
become coupled to one another in the corona and solar wind:
\begin{enumerate}
\item
Inhomogeneities in the background plasma can blur the definitions
of the individual modes.
For example, linear reflection due to radial variations in
$V_{\rm A}$ \citep{FP58,HO80} may produce not only incoming Alfv\'{e}n
waves (i.e., $0 < {\cal R} < 1$), but also fast and slow magnetosonic
waves \citep[e.g.,][]{St71,MH07}.
In addition, large-scale bends in the background magnetic field
${\bf B}$ \citep{Fr64,Wz74}, density variations between flux tubes
\citep{Vy74,Mk01,MM08}, or velocity shears \citep{Po98,Gg07}
can drive instabilities that partially convert Alfv\'{e}n waves
into other modes.
\item
Even in a homogeneous medium, the MHD waves begin to lose their ideal
linear character when their amplitudes become large.
Nonlinear Alfv\'{e}n waves naturally drive second order fluctuations
in $v_{\parallel}$ and $\delta \rho$ that mimic the properties of
both slow and fast magnetosonic waves \citep{Ho71,Sp89,VH99}.
Large-amplitude waves also excite a range of wave-wave interactions
that can often be characterized either as two modes giving birth to a
third, or one mode splitting into several others
\citep[e.g.,][]{CW72,Gm78,DZ01,SK10}.
Models of weak turbulence, in which the wave-wave interactions
describe the cascade process \citep{Ch05,Ch08a,LM06,YF08} also create
this kind of coupling.
\item
Although not strictly a multi-mode coupling, when
$k_{\perp} \rho_{p} \gtrsim 1$ the Alfv\'{e}n mode begins to exhibit
oscillations in density, parallel electron velocity, and the
parallel electric and magnetic fields \citep{HC76,Ho99}.
Observationally, it has proved difficult to separate such dispersive
KAW density fluctuations from those arising from independent sources
of fast or slow MHD waves \citep[e.g.,][]{HC05,Cq09}.
\end{enumerate}

In this paper we take account of one particular nonlinear effect
from the second entry in the above list.
Specifically, \citet{Ch05} suggested that weak turbulence couplings
between Alfv\'{e}n and fast-mode fluctuations may provide enough power
at high $k_{\parallel}$ to induce substantial ion cyclotron heating.
\citet{Sz07} argued that this effect may be relatively
unimportant because the fast-mode cascade timescale $\tau_{\rm F}$
is long in comparison to the Alfv\'{e}n cascade timescale
$\tau_{\rm A}$.
This may be the case in the low-frequency regime of wavenumber
space where $\chi \ll 1$, but at the cyclotron resonant frequencies
of interest ($k_{\parallel} \sim \Omega_{p} / V_{\rm A}$) the
Alfv\'{e}nic cascade is quenched because $\chi \gg 1$.
The fast-mode cascade may in fact even be {\em faster} than any
intrinsic Alfv\'{e}nic spectral transfer in this region of wavenumber
space.
Therefore, we proceed using the \citet{Ch05} results for
Alfv\'{e}n/fast-mode coupling.

We express the coupling term in
Equations (\ref{eq:difeqA})--(\ref{eq:difeqF}) as
\begin{equation}
  C_{\rm AF} \, = \, \frac{E_{\rm F} - E_{\rm A}}{\tau_{\rm AF}}
  \label{eq:caf}
\end{equation}
such that, in the absence of other processes, the power spectra at
a given wavenumber ${\bf k}$ are driven toward a common value over
a coupling timescale $\tau_{\rm AF} ({\bf k})$.
The weak turbulence model of \citet{Ch05} gave an approximate value
for this timescale of
\begin{equation}
  \tau_{\rm AF} \, \approx \, \frac{15}{23 \pi^2} \tau_{\rm F}
  \, \sin^{2} \theta
  \label{eq:tafben}
\end{equation}
which holds in the limiting cases of $E_{\rm F} > E_{\rm A}$
and nearly parallel propagation ($\theta \ll 1$).
In the opposite case of $E_{\rm A} \gg E_{\rm F}$, it's likely that
$\tau_{\rm AF}$ would no longer depend linearly on $\tau_{\rm F}$,
and may scale instead with $\tau_{\rm A}$.
However, the region of wavenumber space with which we are most
concerned is the high-$k_{\parallel}$, low-$\theta$ ion cyclotron
regime.
At those wavenumbers, we know that in the absence of coupling the
condition $E_{\rm F} \gg E_{\rm A}$ is likely to be satisfied,
and the coupling will be a transfer of energy from the dominant
fast-mode spectrum to the much less intense Alfv\'{e}n mode.

The wave-wave conditions of frequency and wavenumber matching 
\citep[e.g.,][]{SG69} confirm that the most rapid coupling should occur
when the dispersive properties of the Alfv\'{e}n and fast-mode waves
are the most similar to one another; i.e., at $\theta \rightarrow 0$.
Note that Equation (\ref{eq:tauF}) gave
$\tau_{\rm F} \propto 1 / \sin\theta$, so the combined dependence
for the coupling time is $\tau_{\rm AF} \propto \sin\theta$.
In practice, however, we found that using this ideal expression for
$\tau_{\rm AF}$ could lead to an unphysical singularity at $\theta = 0$.
We removed this singularity by replacing $\theta$ in
Equation (\ref{eq:tafben}) by $\theta + \delta\theta$.
We set $\delta\theta$ to a constant value of 0.01 to avoid having
an infinitely fast coupling rate at parallel propagation.\footnote{%
Also note that the magnetic field in MHD turbulence undergoes a
complex, multi-scale ``wandering,'' such that the direction
corresponding to $\theta = 0$ is continuously varying in time
and space \citep[see, e.g.,][]{Rg06,SK07}. Thus, the plasma may
seldom ``see'' exactly parallel wavenumber conditions.}
To retain the most generality, we chose to reparameterize the
coupling timescale as
\begin{equation}
  \tau_{\rm AF} \, = \, \frac{1}{\Phi} \,
  \tau_{\rm F} \sin^{2} (\theta + \delta\theta)
  \label{eq:tauaf}
\end{equation}
where we find it useful to vary the constant coupling strength
$\Phi$ up or down from the value of $23 \pi^{2} / 15 \approx 15.1$
derived by \citet{Ch05}.
The case $\Phi = 0$ corresponds to ignoring the coupling altogether.

Note that the above form for the coupling timescale implies that
$\tau_{\rm AF} \propto k_{\perp} / k^{3/2}$, so that the coupling
is rapid at wavenumbers corresponding to ion cyclotron resonance
(large $k_{\parallel}$, small $k_{\perp}$).
The coupling is much slower at KAW wavenumbers favored by the
pure Alfv\'{e}nic cascade (small $k_{\parallel}$, large $k_{\perp}$).
Thus, the bulk of the Alfv\'{e}nic spectrum at $\chi \ll 1$ is
likely to be more or less unaffected by the coupling.
This seems to be consistent with our assumption that the
integrated energy densities $U_{\rm A}$ and $U_{\rm F}$
also remain uncoupled from one another.
We realize that this may be a severe {\em underestimate} of the
degree of energy transfer between Alfv\'{e}n and magnetosonic
modes in the corona and solar wind.
However, one main purpose of this paper is to investigate how much
can be accomplished with only this small degree of coupling in
the high-$k_{\parallel}$ tails of the power spectra.

\subsection{Approximate Solutions for Coupled Spectra}
\label{sec:coupling:results}

The exact solutions to Equations (\ref{eq:difeqA}) and (\ref{eq:difeqF})
with coupling ($C_{\rm AF} \neq 0$) must be found numerically.
Here we present an approximate solution that is both
(1) likely to reflect the proper behavior of more rigorous numerical
solutions in many limiting regimes of parameter space, and
(2) efficient to implement on a large grid of model spectra spanning
a wide range of heliocentric distances.
We begin by approaching the problem iteratively; i.e.,
we solve Equation (\ref{eq:difeqA}) for $E_{\rm A}$ under
the assumption that $E_{\rm F}$ is known, and
we then solve Equation (\ref{eq:difeqF}) for $E_{\rm F}$ under
the assumption that $E_{\rm A}$ is known.
The analytic solutions derived below suggest a natural way
to terminate this iteration after only one round.

When solving the advection-diffusion equation for Alfv\'{e}nic
fluctuations, let us temporarily ignore the outer-scale source
term $S_{\rm A}$ and the dissipation-range damping term that
depends on $\gamma_{\rm A}$.
Since we are most concerned with the generation and transport of
wave power in the high-$k_{\parallel}$ regions that undergo
ion cyclotron resonance, we consider the weak turbulence regime
of $\chi \gg 1$, in which the transport of energy is mainly from
low to high $k_{\perp}$ and there is negligible parallel spreading
\citep[see also][]{OD06}.
Thus, we solve the advection-diffusion equation for discrete,
non-interacting ``strips'' of wavenumber space each having
constant $k_{\parallel}$.
The nonlinear coupling supplies wave energy locally, and the
Alfv\'{e}nic cascade takes it from low to high $k_{\perp}$.
If we simplify further by assuming pure advection (i.e.,
$\alpha_{\perp} = 0$), the time-steady version of
Equation (\ref{eq:difeqA}) becomes
\begin{equation}
  \frac{\mu_{\perp}}{k_{\perp}} \frac{\partial}{\partial k_{\perp}}
  \left( \frac{k_{\perp}^{2} E_{\rm A}}{\tau_{\rm A}} \right)
  \, = \, \frac{E_{\rm F} - E_{\rm A}}{\tau_{\rm AF}} \,\, ,
\end{equation}
where we use Equation (\ref{eq:tauAbridge}) to give the timescale
$\tau_{\rm A} \approx \chi / (k_{\perp} v_{\perp})$
in the weak turbulence regime, and we use Equation (\ref{eq:tauaf})
for $\tau_{\rm AF}$.

The above advection-coupling equation can be rewritten as
a first-order ordinary differential equation,
\begin{equation}
  \frac{\partial E_{\rm A}}{\partial k_{\perp}} + \left(
  \frac{10}{3k_{\perp}} + \frac{f_0}{k_{\perp}^{10/3}} \right)
  E_{\rm A} \, = \, \frac{f_{0} E_{\rm F}}{k_{\perp}^{10/3}}
  \label{eq:EA1st}
\end{equation}
where
\begin{equation}
  f_{0} \, = \, \frac{\Phi}{\mu_{\perp}} \left(
  \frac{v_{0 {\rm F}}}{v_{0 \perp}} \right)^{2}
  \frac{k_{0 {\rm F}}^{1/2} k_{\parallel}^{5/2}}{k_{0 \perp}^{2/3}}
  \,\, .
\end{equation}
To obtain Equation (\ref{eq:EA1st}), we made several power-law
assumptions for the timescales $\tau_{\rm A}$ and $\tau_{\rm AF}$,
which depend on the velocity spectra $v_{\perp}$ (for Alfv\'{e}n
waves) and $v_k$ (for fast-mode waves), respectively, with
\begin{equation}
  v_{\perp} \, = \, v_{0 \perp}
  \left( \frac{k_{\perp}}{k_{0\perp}} \right)^{-1/3}
  \,\, , \,\,\,\,
  v_{k} \, = \, v_{0 {\rm F}}
  \left( \frac{k}{k_{0 {\rm F}}} \right)^{-1/4} \,\, .
\end{equation}
We also assumed that we are solving for $E_{\rm A}$ mainly in
the small-$\theta$ region of wavenumber space in which
$k \approx k_{\parallel}$.

With the above assumptions taken into account,
Equation (\ref{eq:EA1st}) can be solved by means of an
integrating factor.
We first define the dimensionless independent variable
\begin{equation}
  y \, = \, \frac{3f_0}{7 k_{\perp}^{7/3}} \, = \,
  \left( \frac{k_c}{k_{\perp}} \right)^{7/3}
\end{equation}
which is a measure of the relative strength of the nonlinear coupling.
When $y \gg 1$ (or $k_{\perp} \ll k_{c}$) the coupling is strong and
we should expect $E_{\rm A} \approx E_{\rm F}$.
When $y \ll 1$ (or $k_{\perp} \gg k_{c}$) the coupling is weak in
comparison to the cascade and we expect $E_{\rm A} \ll E_{\rm F}$.
Note also that $y$ depends much more sensitively on $\theta$ than
on the magnitude $k$.
Working through the integrating factor method and choosing an
integration constant of zero (to avoid the solution diverging to
infinity when $y \gg 1$), we obtain
\begin{equation}
  E_{\rm A} \, = \, \frac{7y}{3} \left[
  1 - e^{y} y^{3/7} \, \Gamma \left( \frac{4}{7} , \, y \right)
  \right] E_{\rm F}
  \label{eq:EAigamma}
\end{equation}
where $\Gamma(a,y)$ is the incomplete gamma function.
This function behaves as expected in the limits of strong and weak
coupling as discussed above.

Next we solve the coupled fast-mode advection-diffusion equation
for $E_{\rm F}$ under the assumption that $E_{\rm A}$ is known.
Making use of many of the same simplifications that were used to
solve the $E_{\rm A}$ equation, we include only the cascade and
coupling terms, with
\begin{equation}
  \frac{\alpha_{\rm F}}{k^2} \frac{\partial}{\partial k}
  \left( k^{2} D_{\rm F}
  \frac{\partial E_{\rm F}}{\partial k} \right) \, = \,
  C_{\rm AF} \, = \, \left[ \frac{1}{\tau_{\rm AF}} \left(
  1 - \frac{E_{\rm A}}{E_{\rm F}} \right) \right] E_{\rm F}
  \,\, .
  \label{eq:EFgeff}
\end{equation}
Noticing that the quantity in square brackets above is an effective
damping rate $\gamma_{\rm eff}$, we use Equation (\ref{eq:EAigamma})
to write the ratio $E_{\rm A}/E_{\rm F}$ as a known function of
$k_{\parallel}$ and $k_{\perp}$.
After substituting in the wavenumber dependence for $\tau_{\rm AF}$,
we found that $\gamma_{\rm eff} \propto (k/k_{0{\rm F}})^{1/2}$.
The analytic solution of $E_{\rm F}(k)$ for this special case is
given in Equation (\ref{eq:EFzhalf}), and the constant $c_{\gamma}$
in that expression is specified here to be
\begin{equation}
  c_{\gamma} \, = \, \frac{8 \Phi}{49 \, \alpha_{\rm F}
  \sin^{2}\theta} \left( 1 - \frac{E_{\rm A}}{E_{\rm F}} \right)
  \,\, .
\end{equation}
The solution of Equation (\ref{eq:EFzhalf}) is applied only for
$k \geq k_{0 {\rm F}}$, and the uncoupled/undamped fast-mode power
spectrum $E_{0 {\rm F}}$ is used for $k < k_{0 {\rm F}}$.

Since our solution for the ratio $E_{\rm A}/E_{\rm F}$ depends only
on wavenumber and not on any prior solutions of $E_{\rm A}$ or
$E_{\rm F}$, we found that there is no need for further iteration.
We solve first for $E_{\rm F}$ as described above, using
Equation (\ref{eq:EAigamma}) for the ratio $E_{\rm A}/E_{\rm F}$,
and then we use this ratio to solve for $E_{\rm A}$.
Note that the complete solution for $E_{\rm F}$ must take account of
both coupling and transit-time damping (i.e., the damping rate given
by Equation (\ref{eq:gamttd})).
In practice, we apply both types of damping separately to the
uncoupled and undamped fast-mode power spectrum $E_{0 {\rm F}}$
and we use the solution that gives rise to stronger local damping
at any given wavenumber.
At high enough values of $k_{\parallel}$, the complete solution
for $E_{\rm A}$ must take into account the effects of ion cyclotron
damping.
We use the approximate prescription given by
Equation (\ref{eq:EAcycdamp}) to implement this damping.

If the original uncoupled spectra obey
$E_{0 {\rm A}} \leq E_{0 {\rm F}}$, then the coupled spectra follow
\begin{displaymath}
  E_{0 {\rm A}} \leq E_{\rm A} \leq E_{\rm F} \leq E_{0 {\rm F}}
\end{displaymath}
at wavenumbers in the high-$k_{\parallel}$ regime where the
coupling is applied.
Usually, the relative increase in $E_{\rm A}$ from its uncoupled
solution is greater than the relative decrease in $E_{\rm F}$
from its uncoupled solution.
In all cases, however, we found that the variations in the spectra
introduced by the coupling do not significantly affect the total
wavenumber-integrated power in either $E_{\rm A}$ or $E_{\rm F}$.

\begin{figure}
\epsscale{1.13}
\plotone{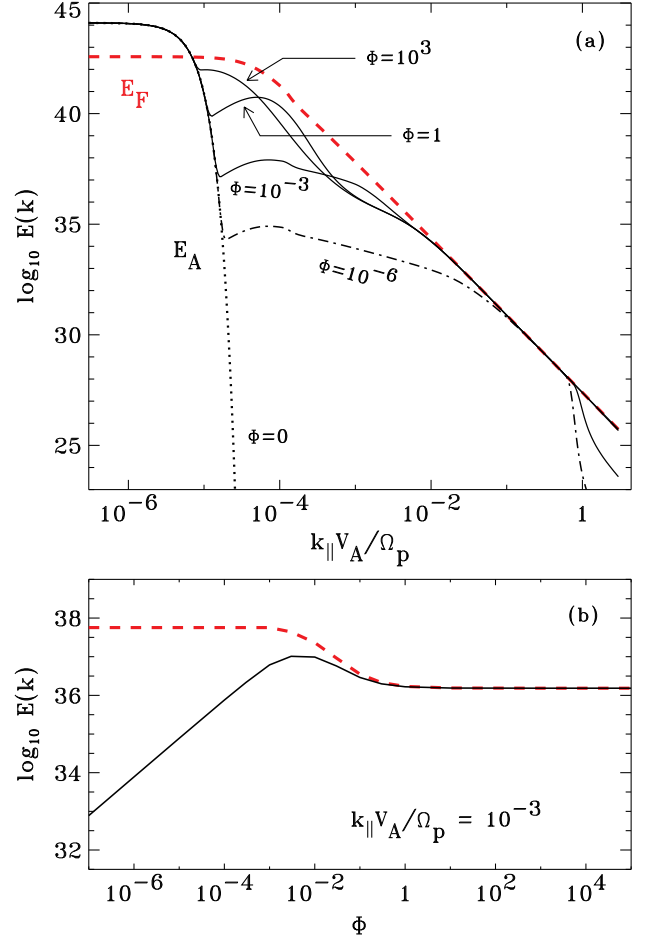}
\caption{(a) Slices of time-steady spectra at $z = 10 \, R_{\odot}$,
shown at constant $k_{\perp} = k_{0 \perp}$:
uncoupled spectra $E_{0 {\rm A}}$ (black dotted curve) and
$E_{0 {\rm F}}$ (red dashed curve), and coupled $E_{\rm A}$
spectra that were computed with a range of $\Phi$ values
(black solid curves).
(b) Variation of $E_{\rm A}$ (black solid curve) and
$E_{\rm F}$ (red dashed curve) with $\Phi$, shown at constant
wavenumber $k_{\perp} = k_{0 \perp}$ and
$k_{\parallel} V_{\rm A} / \Omega_{p} = 10^{-3}$.
\label{fig09}}
\end{figure}
Figure \ref{fig09} illustrates the effects of including coupling
on $E_{\rm A}$.
As in earlier plots of spectrum results, we used the representative
height $z = 10 \, R_{\odot}$ and we assumed
$k_{0 \parallel} = k_{0 \perp}/10$.
In order to show that the coupling can be efficient even when the
uncoupled Alfv\'{e}n wave power $E_{0 {\rm A}}$ is negligibly
small, we assumed the extreme limiting case of $s \rightarrow \infty$.
In Figure \ref{fig09}(a) we show the $k_{\parallel}$ dependence of
the spectra along a slice taken at a constant value of
$k_{\perp} = k_{0 \perp}$.
We varied the parameter $\Phi$ between $10^{-6}$ and $10^{+3}$.
Even if the coupling is several orders of magnitude weaker than
estimated by \citet{Ch05}, it is still likely to be efficient
at generating some Alfv\'{e}nic wave power at
$k_{\parallel} \approx \Omega_{p} / V_{\rm A}$.
However, if the coupling constant $\Phi$ is significantly smaller
than $\sim$10$^{-3}$, the ion cyclotron damping at
$k_{\parallel} \approx \Omega_{p} / V_{\rm A}$ is likely to
overwhelm the ``local supply'' of wave energy from the coupling
and give rise to a low level of resonant wave power.

Figure \ref{fig09}(b) shows how the power at a given wavenumber
($k_{\perp} = k_{0 \perp}$ and
$k_{\parallel} V_{\rm A} / \Omega_{p} = 10^{-3}$)
varies as a function of $\Phi$.
The fast-mode power decreases monotonically as $\Phi$ is
increased, which confirms our treatment of the coupling in
Equation (\ref{eq:EFgeff}) as an effective damping.
The Alfv\'{e}nic power generally increases (from its uncoupled
value far below the lower edge of the plot) with increasing
$\Phi$, but there is some nonmonotonicity around
$\Phi \approx 10^{-2}$.
This gives rise to a slightly counter-intuitive result that
there may be {\em more} $E_{\rm A}$ power at high-$k_{\parallel}$
(and thus more proton and ion heating) at some values of $\Phi$
than in the $\Phi \rightarrow \infty$ limit.

\begin{figure}
\epsscale{1.15}
\plotone{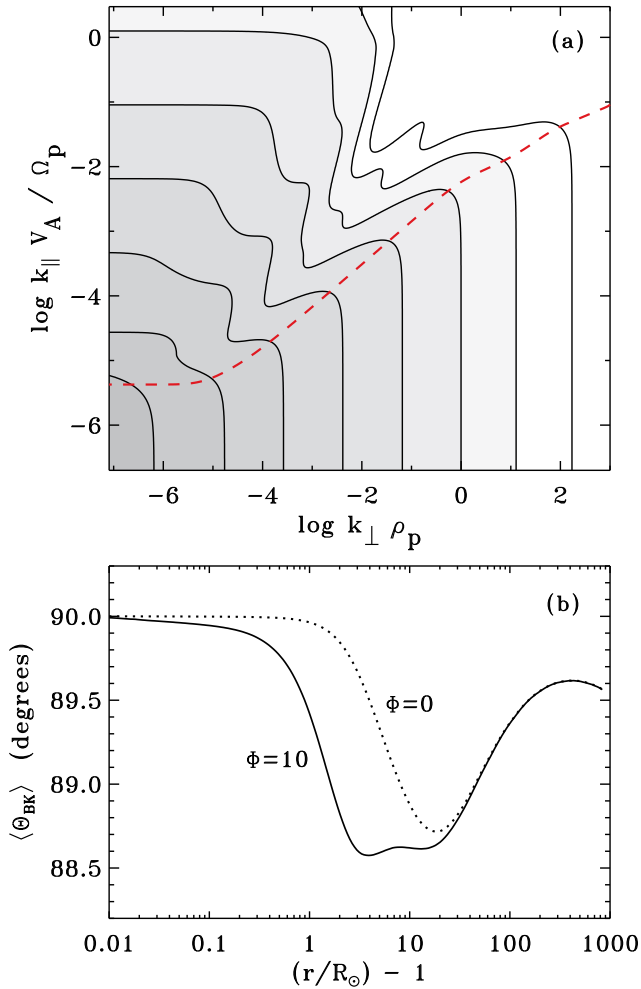}
\caption{(a) Contours of the $E_{\rm A}$ power spectrum, as
in Figure \ref{fig08} but computed with full fast-mode
coupling ($\Phi = 10$).
(b) Radial dependence of spectrum-averaged angle
$\Theta_{{\rm B}k}$ between the background field direction
and the wavenumber vector ${\bf k}$, computed for $\Phi=10$
(solid curve) and for $\Phi=0$ (dotted curve).
\label{fig10}}
\end{figure}
An example of the full wavenumber dependence of the coupled
$E_{\rm A}(k_{\parallel}, k_{\perp})$ spectrum is shown in
Figure \ref{fig10}(a) for a radial distance of
$r = 10 \, R_{\odot}$.
This model has the same parameters as the one shown in
Figure \ref{fig08}, except that we set $\Phi = 10$.
Despite the appearance of substantial wave power at large
values of $k_{\parallel}$, most of the power is still contained
within the critical balance locus of $\chi \lesssim 1$.
This is illustrated in another way by Figure \ref{fig10}(b),
in which we show the radial dependence of the spectrum-averaged
angle $\Theta_{{\rm B}k}$ between the background field direction
and the wavenumber vector ${\bf k}$.
We used a definition for the spectrum-averaged wavevector
anisotropy that is similar to that of \citet{Ga10},
\begin{equation}
  \tan^{2} \langle \Theta_{{\rm B}k} \rangle \, = \,
  \frac{\int d^{3}{\bf k} \,\, E_{\rm A}({\bf k}) \, k_{\perp}^2}
  {\int d^{3}{\bf k} \,\, E_{\rm A}({\bf k}) \, k_{\parallel}^2}
  \,\, .
\end{equation}
Note that the model result at $r = 1$~AU ({89.5\arcdeg}) is
reasonably close to the value of $\sim${88\arcdeg} measured by
\citet{Sr10} from the four {\em Cluster} satellites at 1 AU.
It is evident that a strongly perpendicular (``quasi-two-dimensional'')
sense of wavenumber anisotropy is not incompatible with the
existence of high-frequency ion cyclotron resonant wave power.

\section{Kinetic Dispersion and Dissipation}
\label{sec:disp}

When computing the dissipation rates $\gamma_{\rm A}$ and 
$\gamma_{\rm F}$, we are careful to distinguish between two 
conceptually different sources of damping.
First, there are the collisional and outer-scale cascade processes
that were included in Equation (\ref{eq:Qafs}).
These processes act at low wavenumber and drive the overall radial
evolution of the wave energy densities $U_{\rm A}$ and $U_{\rm F}$.
We do not include them in the damping terms in
Equations (\ref{eq:difeqA})--(\ref{eq:difeqF}) because their
net effects are already included in the source terms
$S_{\rm A}$ and $S_{\rm F}$.
Second, there are the largely collisionless kinetic processes
that become dominant at large wavenumbers.
These are the actual processes that dissipate the power and give
rise to heating, and we describe them in the remainder of this
section.

Once the power levels of Alfv\'{e}nic and fast-mode fluctuations
are specified as detailed functions of $k_{\parallel}$, $k_{\perp}$,
and radial distance, we compute their damping rates and
species-dependent heating rates from linear Vlasov theory.
Although it is known that strong MHD turbulence is far from
``wavelike'' (i.e., coherent wave packets do not survive for more
than about one period before being shredded by the cascade), there
is a long history of using damped linear wave theory to study the
small-scale dissipation of such a cascade
\citep[see, e.g.,][]{Ei79,Qt98,Lm99,QG99,MT01a,CvB03,GB04,HC05}.
A typical justification of this approach is that no matter the
strength of the fluctuations at the outer scale, once the cascade
reaches the high-$k$ dissipation range the magnitudes are much
smaller and quite linear; see also \citet{Sp91} and \citet{Lh09}.

For the Alfv\'{e}n waves, we utilize the Vlasov-Maxwell code
described by \citet{CvB03} and \citet{Ce09} to solve the ``warm''
linear dispersion relation for the real and imaginary parts of the
frequency in the solar wind frame ($\omega = \omega_{r} + i \gamma$)
assuming a known real wavevector ${\bf k}$.
The code uses the Newton-Raphson technique to isolate individual
solutions from a grid of starting guesses in
$\omega_r$, $\gamma$ space, and we select only the
left-hand-polarized (Alfv\'{e}nic) solutions.
We assumed homogeneous plasma conditions and isotropic Maxwellian
velocity distributions (with $T_{p}=T_{e}$), and we ran the
code for a range of assumed values of $\beta$ between $10^{-3}$
and $10^{2}$.
The code also provides the partition fractions of wave energy
in electric, magnetic, kinetic, and thermal perturbations for
each wave mode \citep[see also][]{KV94}.

\begin{figure}
\epsscale{1.10}
\plotone{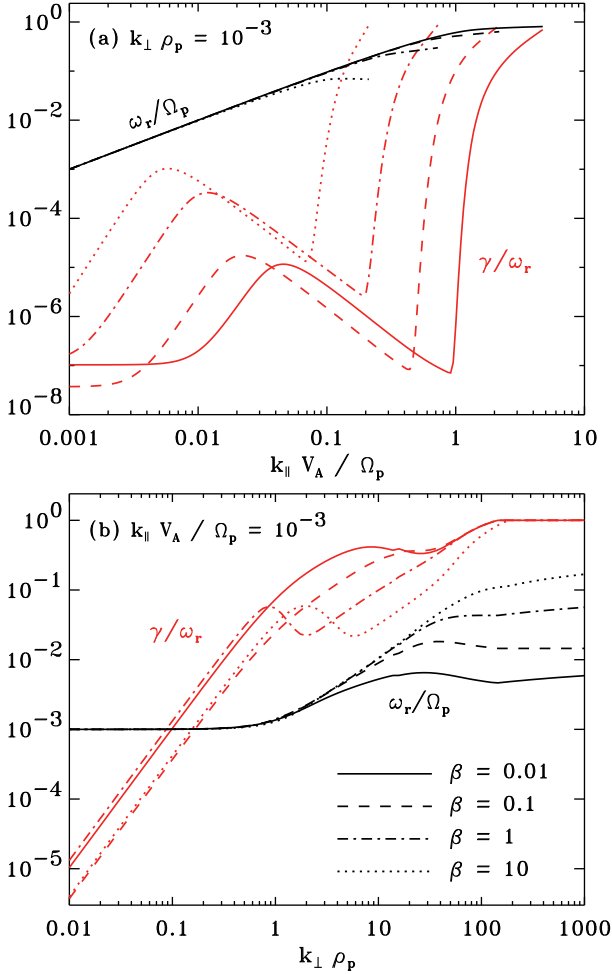}
\caption{Linear dispersion properties of Alfv\'{e}n waves computed
for a range of plasma $\beta$ values.
(a) Real frequencies $\omega_{r} / \Omega_p$ (black curves) and
damping rates $\gamma / \omega_r$ (red curves) plotted versus
$k_{\parallel}$ at constant $k_{\perp} \rho_{p} = 10^{-3}$, for
$\beta = 0.01$ (solid curves),
$\beta = 0.1$ (dashed curves),
$\beta = 1$ (dot-dashed curves),
$\beta = 10$ (dotted curves).
(b) Same quantities as in panel (a), but shown as a function
of $k_{\perp}$ at constant
$k_{\parallel} V_{\rm A} / \Omega_{p} = 10^{-3}$.
\label{fig11}}
\end{figure}
Figure \ref{fig11} shows several example solutions for the real
and imaginary parts of the frequency along one-dimensional cuts
through wavenumber space.
For simplicity, we present all damping rates $\gamma$ as their
absolute values, since strictly speaking the solutions from the
Vlasov-Maxwell code all have $\gamma < 0$.
Figure \ref{fig11}(a) illustrates the approach to the ion cyclotron
resonance regime by holding $k_{\perp}$ constant at a small value
and plotting $\omega_r$ and $\gamma$ versus $k_{\parallel}$.
Note the cessation of weakly damped solutions
at $\gamma \approx \omega_{r} \approx \Omega_{p}$, which takes
place at lower values of $k_{\parallel {\rm max}}$ for higher
values of $\beta$.
Equation (\ref{eq:kparamax}) is a parameterized fit to the
$\beta$-dependence of this cutoff wavenumber.

Figure \ref{fig11}(b) shows the approach to the high-$k_{\perp}$
KAW dissipation limit for a constant small value of $k_{\parallel}$.
When solving the dispersion relation along a succession of
increasing values of $k_{\perp}$, there are sometimes small
discontinuities in slope between neighboring solutions (especially
in strongly damped regions where $|\gamma / \omega_{r}| \gtrsim 0.5$).
Nonetheless, the dispersion properties of our solutions remain
sufficiently ``KAW-like'' to represent a continuous set of damping
rates from low to high $k_{\perp}$.
The behavior of $\omega_r$ versus $k_{\perp}$ agrees reasonably well
with the approximate expression given by Equation (\ref{eq:phidef}).
For values of $\beta \gtrsim 1$, there are secondary maxima in
$\gamma / \omega_{r}$ at $k_{\perp} \rho_{p} \approx 1$ that
come from proton Landau damping, whereas the larger rates at
$k_{\perp} \rho_{p} > 10$ are dominated by electron Landau damping.
The damping rates shown in Figure \ref{fig11}(b) were also used as
the effective KAW ratios $\tilde{\gamma}_{\rm A} / \omega_{r}$
described in Appendix \ref{appen:ansol:edecay}.
These rates were used to compute the high-$k_{\perp}$ dissipation
of $b_{\perp}$ and $v_{\perp}$ as shown in Figures \ref{fig06}
and \ref{fig07}.

For the fast-mode waves, we make use of a parameterized expression
for the rate of transit-time damping, which in several 
studies was found to be the dominant kinetic process to dissipate
this wave mode \citep[e.g.,][]{Ba66,Pk73,YL04}.
Thus, we assume
\begin{displaymath}
  \frac{\gamma_{\rm F}}{\omega_r} \, = \,
  \frac{\theta^{2} \sqrt{\pi \beta}}{4}
  \left( 1 + \frac{\theta^2}{\sqrt{\theta^{4} +
  4 \Omega_{p}^{2}/\omega_{r}^{2}}} \right)
\end{displaymath}
\begin{equation}
  \times \,\, \sqrt{\frac{m_e}{m_p}}
  \exp \left( - \frac{m_e}{m_{p} \beta \cos^{2}\theta}
  \right) \,\, ,
  \label{eq:gamttd}
\end{equation}
where $\omega_{r}$ is given by the ideal fast-mode dispersion
relation of Equation (\ref{eq:Vphfs}).
This expression was given by \citet{YL04} based on initial
calculations of \citet{St58}.
Equation (\ref{eq:gamttd}) is valid strictly for only $\theta \ll 1$,
but it does not diverge from the more exact solution at larger
$\theta$ by more than about a factor of two.

The remainder of this section describes how the dissipated
Alfv\'{e}n wave energy is partitioned between protons, electrons,
and heavy ions.
We ignore the particle heating that comes from fast-mode wave
dissipation because its overall magnitude was found to be small
in comparison to that from Alfv\'{e}n waves.
In a pure hydrogen plasma, we separate the damping rate $\gamma$
into components attributed to the kinetic effects of protons and
electrons.
To zeroth order, the contribution to $\gamma$ from other ions is
negligibly small and can be estimated separately (see below).
Thus, we define $\gamma = \gamma_{p} + \gamma_{e}$, where
\begin{equation}
  \gamma_{s} \, = \, \gamma \,
  \frac{\psi_s}{\psi_{p} + \psi_{e}} \,\, ,
\end{equation}
where $s = p,e$ denotes either the protons or electrons, and the
species-dependent resonance functions are given by
\begin{equation}
  \psi_{s} \, = \, \frac{\omega_{r} \omega_{ps}^2}
  {k_{\parallel} w_{\parallel s}}
  \sum_{\ell = -\infty}^{+\infty}
  \exp \left( - \xi_{\parallel}^{2} - \xi_{\perp}^{2} \right)
  \sum_{m = \ell - 1}^{\ell + 1}
  a_{m} I_{m} (\xi_{\perp}^{2}) \,\, ,
  \label{eq:psi}
\end{equation}
where $\omega_{ps}^{2} = 4 \pi e^{2} n_{s} / m_{s}$ is the squared
plasma frequency, $w_{\parallel s}$ and $w_{\perp s}$
are parallel and perpendicular thermal speeds of species $s$, and
$I_m$ is the $m$-order modified Bessel function of the first
kind.
The dimensionless coefficients $a_m$ depend on the electric-field
polarization vector that is output from the Vlasov-Maxwell dispersion
code of \citet{CvB03}, and they are given in full by
Equations (43)--(45) of \citet{MT01a}.
Equation (\ref{eq:psi}) is valid for an isotropic Maxwellian
distribution, for which $w_{\parallel s} = w_{\perp s}$ and there is
assumed to be zero differential bulk flow between the protons and
electrons.
The dominance of ion cyclotron or Landau damping depends on the
values of the dimensionless resonance factors,
\begin{equation}
  \xi_{\parallel} =
  \frac{\omega - \ell \Omega_s}{k_{\parallel} w_{\parallel s}}
  \,\, , \,\,\,\,
  \xi_{\perp} =
  \frac{k_{\perp} w_{\perp s}}{\Omega_{s} \sqrt{2}} \,\, .
\end{equation}
In practice, we truncate the infinite sum in Equation (\ref{eq:psi})
at $-10 \leq \ell \leq +10$.
Test runs made with a larger range of summation indices produced no
substantial differences from those using the default range.

\begin{figure}
\epsscale{1.15}
\plotone{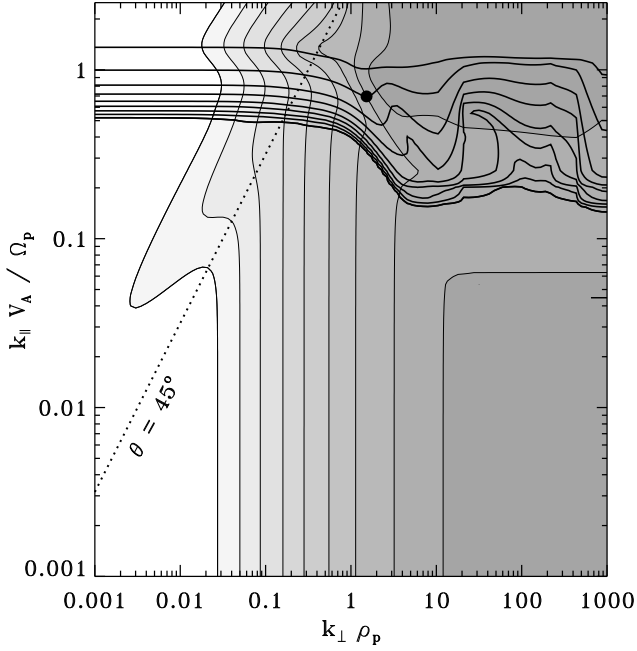}
\caption{Contours of $\gamma_{p} / \omega_r$ (thick curves) and
$\gamma_{e} / \omega_r$ (thin curves separated by varying gray
shading) plotted versus $k_{\parallel}$ and $k_{\perp}$.
Contours are plotted twice per decade from
$3 \times 10^{-5}$ to $3 \times 10^{-1}$ and generally go from
low to high values with increasing wavenumber.
A line denoting $\theta = 45^{\circ}$ (dotted curve) and a point
illustrating where $\theta_{ep}$ is defined (filled circle)
are also shown.
\label{fig12}}
\end{figure}
Figure \ref{fig12} shows separate sets of contours for
$\gamma_{p} / \omega_r$ and $\gamma_{e} / \omega_r$
in wavenumber space for an example value of $\beta = 0.1$.
These contours can be compared with Figure 4(a) of \citet{CvB03},
which was computed for $\beta \approx 0.01$.
The proton damping rate $\gamma_{p} / \omega_r$ increases rapidly
as $k_{\parallel} V_{\rm A} / \Omega_p$ approaches unity, and the
electron damping rate $\gamma_{e} / \omega_r$ increases more slowly
as $k_{\perp} \rho_p$ increases from 0.1 to 100.
The complex behavior of the contours in region of wavenumber space
with both high $k_{\parallel}$ and high $k_{\perp}$ is the result
of the dispersion relation being affected by the presence of strongly
damped ion Bernstein modes \citep[see, e.g.,][]{Sx92,Hw08}.

It is evident from Figure \ref{fig12} that, in the solar corona, the
region of nearly parallel Alfv\'{e}n wave propagation in wavenumber
space (i.e., $\theta \ll 1$) is dominated by proton damping and the
region of nearly perpendicular propagation ($\theta \rightarrow \pi/2$)
is dominated by electron damping.
The observational evidence for preferential proton and ion heating 
\citep{Ko06} thus presents a problem when confronted with the
dominant perpendicular anisotropy of Alfv\'{e}nic turbulence.

\begin{figure}
\epsscale{1.11}
\plotone{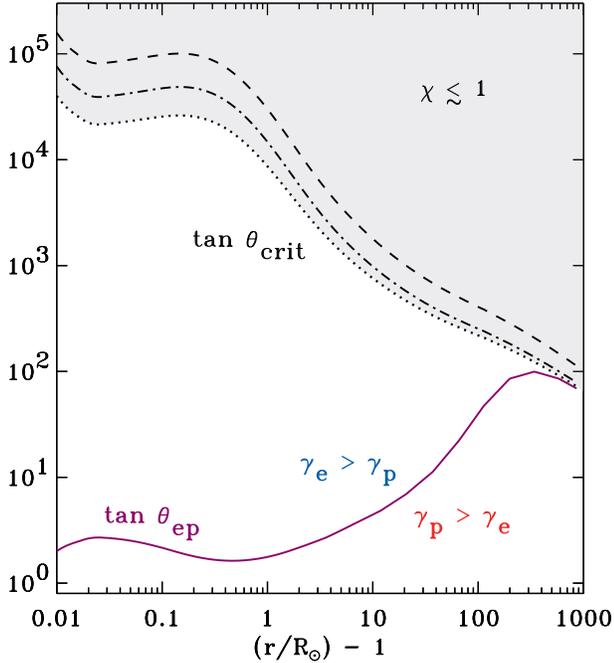}
\caption{Radial dependence of the tangents of $\theta_{ep}$
(solid purple curve) and $\theta_{\rm crit}$ (black curves), the
latter computed for
$k_{0 \parallel} / k_{0 \perp} = 0.01$ (dotted),
$k_{0 \parallel} / k_{0 \perp} = 0.1$ (dot-dashed), and
$k_{0 \parallel} / k_{0 \perp} = 1$ (dashed).
The gray region denotes the approximate region of parameter space
expected to be ``occupied'' by a purely Alfv\'{e}nic turbulent
cascade.
\label{fig13}}
\end{figure}
Figure \ref{fig13} illustrates the magnitude of this apparent
discrepancy by comparing the large-scale radial dependence of two
key angles.
The strongly anisotropic Alfv\'{e}nic cascade is illustrated by
$\theta_{\rm crit}$, which is the angle between ${\bf k}$ and
${\bf B}_0$ at which occurs both the \citet{GS95} critical
balance ($\chi = 1$) and the onset of KAW dispersion
($k_{\perp} \rho_{p} = 1$).
We find that $\tan \theta_{\rm crit} \approx V_{\rm A} / b_{\perp}$,
where $b_{\perp}$ is evaluated at $k_{\perp} \rho_{p} = 1$, and
we plot $\tan \theta_{\rm crit}$ for three example values of the
outer-scale wavenumber ratio $k_{0 \parallel} / k_{0 \perp}$.
Figure \ref{fig13} also shows the radial dependence of $\theta_{ep}$,
which is defined as the angle at which the contours for
$\gamma_{p}/\omega = 0.1$ intersect with those of
$\gamma_{e}/\omega = 0.1$ in wavenumber space.
(This point is shown in Figure \ref{fig12} with a filled circle.)
For $\theta < \theta_{ep}$ the damping is dominated by protons and ions;
for $\theta > \theta_{ep}$ the damping is dominated by electrons.
Note that $\theta_{ep} \ll \theta_{\rm crit}$ in the solar corona
and much of the inner heliosphere, so that it is difficult to see
how the cascade of linear Alfv\'{e}n waves alone can be responsible
for the observed proton and ion heating.

We computed the rates of proton and electron plasma heating from
the modeled values of $\gamma_{p}$ and $\gamma_{e}$ by using the
quasilinear framework outlined by \citet{MT01a} and \citet{CvB03}.
The volumetric heating rates $Q_{s}$ (e.g., expressed in units of
erg s$^{-1}$ cm$^{-3}$) are given
by integrals over vector wavenumber $\bf k$ of the form
\begin{equation}
  \frac{Q_s}{\rho} \, = \, \int d^{3} {\bf{k}} \,\,
  E_{\rm A}({\bf{k}}) \,\, 2 \gamma_{s}
  \label{eq:Qsdef}
\end{equation}
where $s = p,e$ denotes the particle type of interest.
For now, we ignore differences between parallel and perpendicular
heating and only compute the summed heating rate
$Q_{s} = Q_{s \parallel} + Q_{s \perp}$.
In order to perform the wavenumber integration in
Equation (\ref{eq:Qsdef}), we constructed two-dimensional numerical
grids of $\gamma_p$ and $\gamma_e$ for values of
$10^{-3} \Omega_{p} / V_{\rm A} \leq k_{\parallel} \leq
k_{\parallel {\rm max}}$ and
$10^{-3} \leq k_{\perp} \rho_{p} \leq 10^{3}$.
We used 200 points in $k_{\parallel}$ and 100 points in $k_{\perp}$,
and we constructed a total of 14 grids for values of $\beta$
ranging from $10^{-3}$ to 22 (with $\beta$ varying logarithmically
with three samples per decade).
Linear interpolation was used to evaluate the damping rates at
values of $k_{\parallel}$, $k_{\perp}$, and $\beta$ between the
discrete grid points.
We assumed that the ratios $\gamma_{p} / \omega_r$ and
$\gamma_{e} / \omega_r$ remain constant as one extrapolates into
the weakly-damped regions defined by $k_{\perp} \rho_{p} < 10^{-3}$
and $k_{\parallel} V_{\rm A} / \Omega_{p} < 10^{-3}$.

To estimate the heating rates experienced by heavy ions, we assume
that most low-abundance ions do not have a significant effect on
the overall wave dispersion relation.
This allows us to use an ``optically thin'' resonance condition for
the ion cyclotron wave-particle interaction \citep{Cr00}, which
results in a perpendicular heating rate
\begin{equation}
  \frac{Q_{\perp i}}{m_{i} n_{i}} \, \approx \,
  \frac{\pi \Omega_{i}^2}{V_{\rm A}}
  \left( 1 - \frac{Z_i}{A_i} \right)
  \int d^{3} {\bf{k}} \,\, E_{\rm A}({\bf{k}}) \,\,
  \delta ( k_{\parallel} - \Omega_{i}/V_{\rm A} ) \,\, ,
  \label{eq:Qion}
\end{equation}
where $Z_i$  and $A_i$ are the ion charge and mass in proton units
\citep[see also][]{Cr01,TM01,LC09}.
The Dirac delta function extracts a one-dimensional ``strip''
of the power spectrum that is in resonance with the ion Larmor
motions at $\omega_{r} \approx k_{\parallel} V_{\rm A} = \Omega_{i}$.
Thus, Equation (\ref{eq:Qion}) can be evaluated with just a single
integration along the $k_{\perp}$ direction.

\section{Results for Collisionless Particle Heating}
\label{sec:heating}

\begin{figure}
\epsscale{1.10}
\plotone{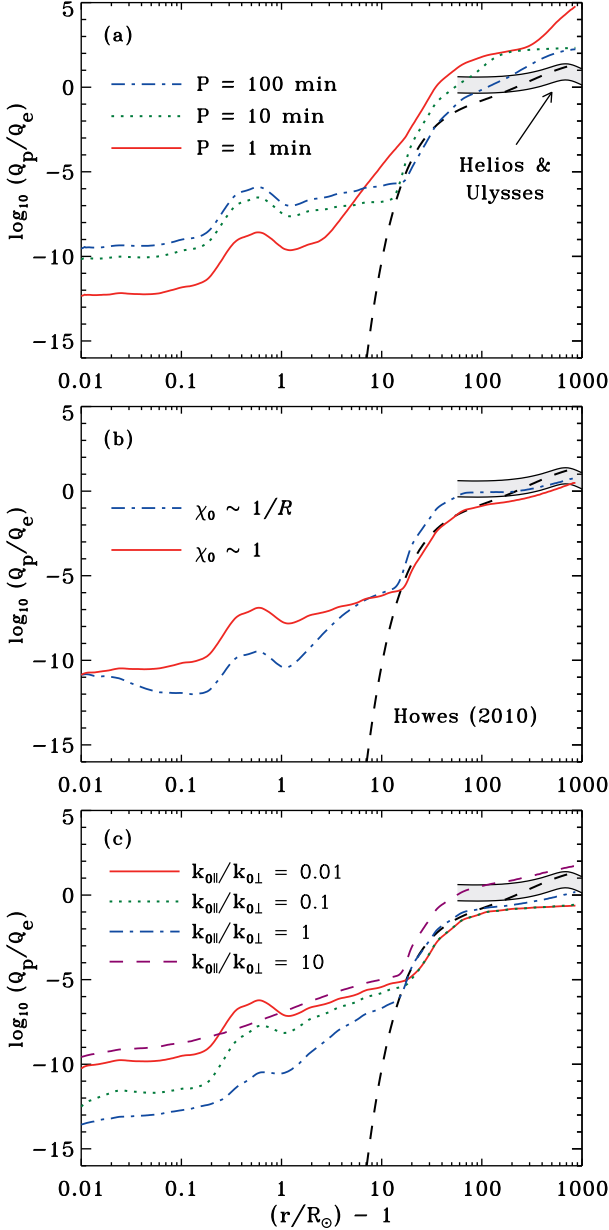}
\caption{Radial dependence of $\log Q_{p}/Q_{e}$
for: (a) constant Alfv\'{e}n wave periods $P = 1$ min (solid
red curve), $P = 10$ min (green dotted curve), and $P = 100$ min
(blue dot-dashed curve);
(b) outer-scale $k_{0\parallel}$ determined from ideal
\citet{GS95} critical balance (red solid curve) and from
modified \citet{BL08,BL09} critical balance (blue dot-dashed curve);
(c) constant ratios
$k_{0 \parallel} / k_{0 \perp} = 0.01$ (red solid curve), 
$k_{0 \parallel} / k_{0 \perp} = 0.1$ (green dotted curve),
$k_{0 \parallel} / k_{0 \perp} = 1$ (blue dot-dashed curve), and
$k_{0 \parallel} / k_{0 \perp} = 10$ (purple dashed curve).
Also shown in (a)--(c) are the \citet{Ce09} measurements
(gray region) and the \citet{Hw10} model prediction (black dashed
curve).
\label{fig14}}
\end{figure}

\begin{figure}
\epsscale{1.10}
\plotone{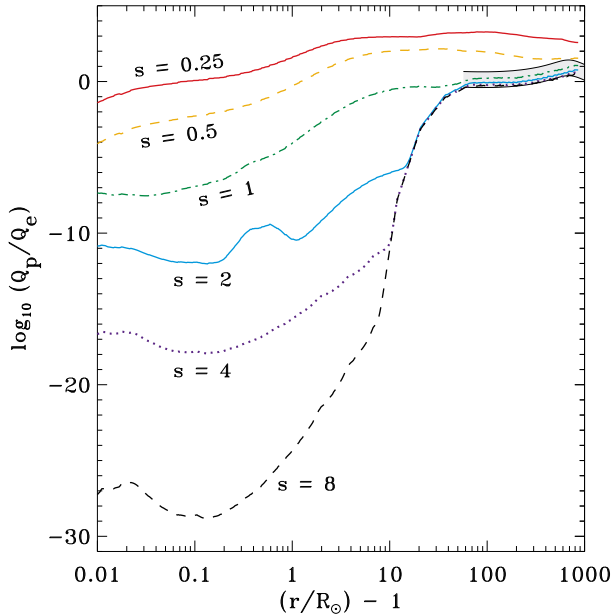}
\caption{Radial dependence of $\log Q_{p}/Q_{e}$
for $\chi_{0} \approx 1 / {\cal R}$ and
$s = 0.25$ (red solid curve),
$s = 0.5$ (orange dashed curve),
$s = 1$ (green dot-dashed curve),
$s = 2$ (cyan solid curve),
$s = 4$ (dark blue dotted curve), and
$s = 8$ (black dashed curve).
Also shown are the \citet{Ce09} measurements (gray region).
\label{fig15}}
\end{figure}

\begin{figure}
\epsscale{1.15}
\plotone{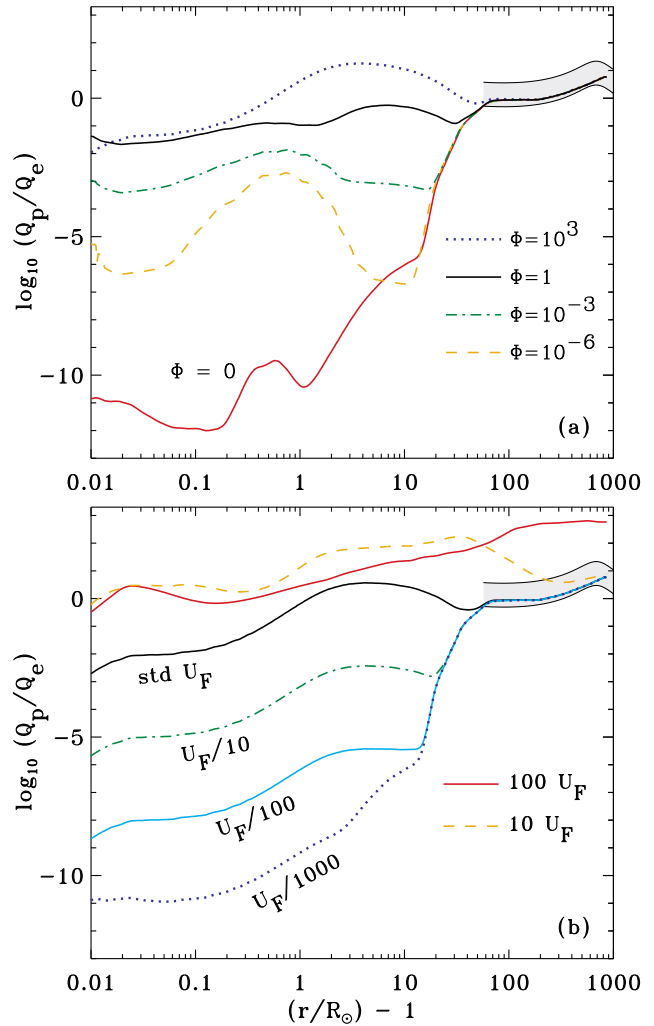}
\caption{Radial dependence of $\log Q_{p}/Q_{e}$
for varying properties of Alfv\'{e}n/fast mode coupling, with:
(a) standard model for $U_{\rm F}$ and a range of coupling
constants:
$\Phi = 0$ (red solid curve),
$\Phi = 10^{-6}$ (orange dashed curve),
$\Phi = 10^{-3}$ (green dot-dashed curve),
$\Phi = 1$ (black solid curve),
$\Phi = 10^{3}$ (dark blue dotted curve);
(b) constant value of $\Phi=10$ and a range of modified values
for fast-mode power:
$U_{\rm F}/10^{3}$ (dark blue dotted curve),
$U_{\rm F}/10^{2}$ (cyan solid curve),
$U_{\rm F}/10$ (green dot-dashed curve),
the standard model of $U_{\rm F}$ (black solid curve),
$10 U_{\rm F}$ (orange dashed curve),
$100 U_{\rm F}$ (solid red curve).
Also shown in both panels are the \citet{Ce09} measurements
(gray regions).
\label{fig16}}
\end{figure}

Here we present results for $Q_{p}/Q_{e}$, the ratio of proton to
electron heating rates, computed from Equation (\ref{eq:Qsdef})
with various assumptions for the shape of the turbulent
Alfv\'{e}n-wave spectrum $E_{\rm A} (k_{\parallel}, k_{\perp})$.
Figures \ref{fig14} and \ref{fig15} show how this ratio behaves for
pure ``uncoupled'' Alfv\'{e}n waves, and Figure \ref{fig16}
summarizes the outcome of coupling the Alfv\'{e}n and fast-mode
waves as discussed in Section \ref{sec:coupling}.
Table \ref{table02} summarizes the specific values of cascade
and coupling parameters that were assumed in each of these plots.
\begin{deluxetable}{ccccc}
\tablecaption{Choices for Cascade and Coupling Parameters
\label{table02}}
\tablewidth{0pt}
\tablehead{
\colhead{Figure} &
\colhead{$s$} &
\colhead{Prescription for $k_{0 \parallel}$} &
\colhead{$\Phi$} &
\colhead{Multiplier to $U_{\rm F}$}
}
\startdata

14    & 2      & varies & 0 & 1 \\

15    & varies & $\chi_{0} = 1/{\cal{R}}$ & 0 & 1 \\

16(a) & 2      & $\chi_{0} = 1/{\cal{R}}$ & varies & 1 \\

16(b) & 2      & $\chi_{0} = 1/{\cal{R}}$ & 10 & varies \\

17    & 2      & $\chi_{0} = 1/{\cal{R}}$ & 10 & varies
\enddata
\vspace*{0.05in}
\end{deluxetable}

In Figure \ref{fig14} we show the radial dependence of $Q_{p}/Q_{e}$
for various methods of computing the outer-scale parallel wavenumber
$k_{0 \parallel}$.
In all panels, the turbulence spectra were computed with constant
values of $s = 2$ and $\Phi = 0$, as well as the other default
parameter choices discussed in Section \ref{sec:cascade:eqns}.
Figure \ref{fig14}(a) assumes a range of radially constant wave
frequencies which determine $k_{0 \parallel}$ from
Equation (\ref{eq:k0omega}).
Figure \ref{fig14}(b) applies the \citet{GS95} conditions of
critical balance for both zero and nonzero cross helicity at the
outer scale; see Equations (\ref{eq:k0paraG})--(\ref{eq:k0paraR}).

Figure \ref{fig14}(c) shows the relative heating rates $Q_{p}/Q_{e}$
for a range of constant ratios $k_{0 \parallel} / k_{0 \perp}$.
At the coronal base ($z = 0.01 \, R_{\odot}$), note that
$Q_{p}/Q_{e}$ behaves non-monotonically as a function of this
wavenumber anisotropy ratio.
The minimum value of $Q_{p}/Q_{e}$ occurs at
$k_{0 \parallel} / k_{0 \perp} \approx 0.55$.
The non-monotonic behavior occurs because of two competing effects.
At large values of $k_{0 \parallel}$, the weak-turbulence critical
balance curve ($\chi_{0} = 1$) begins to approach the ion cyclotron
frequencies.
This has the result of increasing $Q_{p}$ while leaving $Q_{e}$
unchanged.
However, when $k_{0 \parallel}$ becomes very small, the wave power
becomes concentrated into narrower ``cones'' that provide more
energy to the KAWs.
This has the result of increasing both $Q_{p}$ and $Q_{e}$, but
the smaller rate $Q_{p}$ receives a larger fractional change.

Each panel of Figure \ref{fig14} also shows the empirically determined
range of $Q_{p}/Q_{e}$ ratios from the {\em Helios} and {\em Ulysses}
measurements described by \citet{Ce09}.
The plotted error range of $\pm 0.3$ in $\log (Q_{p}/Q_{e})$
accounts for both modeling and observational uncertainties.
Also, we show the theoretical prediction for $Q_{p}/Q_{e}$ from the
gyrokinetic model of \citet{Hw10} as a dashed black curve.
As discussed by \citet{Hw11}, this model agrees well with the
\citet{Ce09} measurements at $r \gtrsim 200 \, R_{\odot}$, but
underestimates the proton heating at $r \lesssim 100 \, R_{\odot}$.
The \citet{Hw10} gyrokinetic model includes the same sources of
high-$k_{\perp}$ KAW damping that we use, but not the
high-$k_{\parallel}$ sources of ion cyclotron damping.
In Figure \ref{fig14}, we find that the best agreement with the
\citet{Ce09} measured ratio comes from the model that assumes
critical balance with the \citet{BL08,BL09} modification for nonzero
cross helicity; i.e., $\chi_{0} \approx 1 / {\cal R}$.

In Figure \ref{fig15} we vary the ratio $s$ used in the Alfv\'{e}nic
parallel cascade function $g(\chi)$; see Equation (\ref{eq:gkappa}).
We retain the $\chi_{0} \approx 1 / {\cal R}$ approximation
for $k_{0 \parallel}$ that was found to be an optimal choice for
agreement with observations at $r \gtrsim 60 \, R_{\odot}$.
For lower heights in the low-$\beta$ corona, we find that large
values of $s$ give insufficient wave power at the ion cyclotron
resonant values of $k_{\parallel}$ to provide significant energy
to the protons.
One would need to specify $s \lesssim 0.5$ in order for there to be
enough high-$k_{\parallel}$ power to give protons a substantial
fraction of the dissipated energy.
\citet{CvB03} and \citet{LC09} came to this same essential
conclusion.
Although there are still no firm experimental or theoretical
bounds on the expected value of $s$ in MHD turbulence, it is
generally believed that values as low as $s \lesssim 0.5$ are
unrealistic.

Figure \ref{fig16} shows the results of mode coupling between
the Alfv\'{e}n and fast-mode fluctuations.
The curves in Figure \ref{fig16}(a) were computed for a range
of constant values of the coupling constant $\Phi$ from
$10^{-6}$ to $10^{+3}$.
At large distances ($r \gtrsim 0.3$ AU), it is clear that the
presence or absence of coupling has very little effect on the
$Q_{p}/Q_{e}$ ratio.
This insensitivity occurs because much of the proton heating
at intermediate and high values of $\beta$ comes from the Landau
and transit-time damping of KAWs.
The low-k$_{\parallel}$, high-$k_{\perp}$ part of the $E_{\rm A}$
spectrum is there no matter the value of $\Phi$, and it dominates
the proton and electron heating in this case.
The results are similar to those of \citet{Hw10,Hw11}
who did not include mode coupling.

In the low-$\beta$ corona, Figure \ref{fig16} indicates that
$\Phi$ needs to be at least of order unity to excite sufficient
power in high-$k_{\parallel}$ ion cyclotron waves to heat protons
on par with the electrons (i.e., $Q_{p}/Q_{e} \sim 1$).
For low values of $\Phi$, the plotted ratio undergoes several
increases and decreases as a function of radius that we cannot
trace to any one simple cause.
The local maximum that appears at $z \approx 0.5 \, R_{\odot}$
corresponds to the local minimum in plasma $\beta$ (see
Figure \ref{fig01}). 
In the low-$\beta$ regime, it is likely that the relative
``competition'' between mode coupling, transit-time damping
(for $E_{\rm F}$), and ion cyclotron damping (for $E_{\rm A}$)
undergoes numerous reversals as a function of radius.

In Figure \ref{fig16}(b) we fix the coupling constant at $\Phi = 10$,
which is of the same order of magnitude as predicted by \citet{Ch05},
and we vary the normalization of the fast-mode wave power.
It was evident from Figure \ref{fig04} that small changes in the
large-scale wave transport properties could give rise to large
changes in the fast-mode power in much of the corona and
solar wind.
Thus, we take the standard model for $U_{\rm F}(r)$ and multiply
it by constant factors ranging from $10^{-3}$ to $10^{+2}$.
We note, however, that we do not have excessive freedom to increase
the $U_{\rm F}$ normalization too far above the standard model.
A significantly higher coronal population of fast-mode waves would
contribute to a larger $v_{\perp}$ that may exceed the
observational constraints shown in Figure \ref{fig03}(a).
Nonetheless, Figure \ref{fig16}(b) shows that the standard model
ends up being a reasonable solution that matches the observed
in~situ heating ratio \citep{Ce09} and also gives appreciable
proton heating in the extended corona (as required qualitatively
from UVCS proton temperature measurements); see \citet{CvB03}.

\begin{figure}
\epsscale{1.15}
\plotone{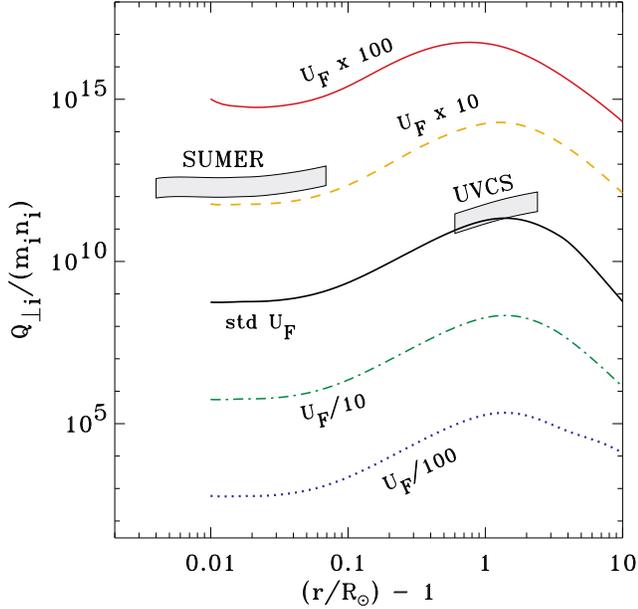}
\caption{Radial dependence of the perpendicular heating rate per unit
mass $Q_{\perp i}/(m_{i} n_{i})$, in units of erg s$^{-1}$ g$^{-1}$,
for O$^{+5}$ ions.
Model results shown for a range of modified values
for fast-mode power:
$U_{\rm F}/100$ (dark blue dotted curve),
$U_{\rm F}/10$ (green dot-dashed curve),
standard $U_{\rm F}$ (black solid curve),
$10 U_{\rm F}$ (orange dashed curve),
$100 U_{\rm F}$ (solid red curve).
Also shown are empirical constraints from SUMER
and UVCS emission line measurements (gray regions).
\label{fig17}}
\end{figure}

An example calculation of preferential heavy ion heating is shown
in Figure \ref{fig17}.
The ion used for the model was O$^{+5}$, whose properties have
been measured in the corona from emission in the
\ion{O}{6} 1032, 1037 {\AA} spectral line doublet \citep{Ko06}.
We used the parameters corresponding to the best agreement with
observational constraints on $Q_{p}/Q_{e}$ (see Table \ref{table02}).
We then adjusted the fast-mode wave power $U_{\rm F}(r)$ by changing
the multiplicative constant that was varied in Figure \ref{fig16}(b).
As in Figure \ref{fig16}(b), values of this multiplicative constant
between about 1 and 10 appear to bracket the observational constraints.

The plotted ranges for the observationally determined $Q_{\perp i}$
rates were derived by combining observations from both the UVCS
\citep{CFK99} and SUMER \citep{LC09} instruments on {\em SOHO}
with semi-empirical solutions of the perpendicular internal energy
conservation equations.
These heating rates were not given explicitly by either \citet{CFK99}
or \citet{LC09}, but they were computed and saved from the models
that produced agreement with the observed radial behavior of
$T_{\perp i}$.
The SUMER and UVCS data were obtained for off-limb measurements of
\ion {O}{6} emission, in which the line widths are primary
diagnostics of $T_{\perp i}$.
Note that the radial dependence of the two observationally
determined regions is similar to that in the plotted model curves.
However, the SUMER data corresponds to about a factor of 10 higher
fast-mode wave power normalization than the UVCS data.

If the postulated mode-coupling explanation for ion cyclotron
proton/ion heating is correct, then the results given in
Figures \ref{fig16} and \ref{fig17} constrain the required levels
of fast-mode wave power.
In the low corona ($z \lesssim 0.1 \, R_{\odot}$), there may need
to be up to a factor of 10 higher value of $U_{\rm F}$ than in
the standard model of Section \ref{sec:global}, but in the extended
corona and heliosphere the standard model may be close to correct.
Of course, it is only the high-$k_{\parallel}$ tail of the
fast-mode spectrum that matters to the calculation of available
Alfv\'{e}nic power at the ion cyclotron resonances, not its
outer-scale normalization.
Therefore it is possible that $U_{\rm F}$ may depart significantly
from the values predicted by the standard model of
Section \ref{sec:global}, but still produce agreement with the
various observations by having different values for the spectral
slope and angle-dependence of $E_{\rm F}({\bf k})$.

\section{Discussion and Conclusions}
\label{sec:conc}

The aim of this paper was to explore the consequences of
\citeauthor{Ch05}'s (\citeyear{Ch05})
conjecture that nonlinear couplings between Alfv\'{e}n and
fast-mode waves may produce sufficient ion cyclotron wave power
to heat protons and heavy ions in the corona.
To test this idea, we constructed a semi-empirical model of
the background plasma and MHD wave properties in a flux tube
connected to a polar coronal hole.
For the sake of practicality, we utilized several approximations
when solving the wave energy transport equations for the
energy densities of Alfv\'{e}n, fast, and slow modes:
\begin{enumerate}
\item
The equations themselves were adapted from standard WKB
``wave action conservation'' theory, which does not take
account of the effects of linear wave reflection in a fully
self-consistent manner.
We also assumed the associated WKB limiting case of equipartition
between the kinetic and magnetic energy densities for the dominant
Alfv\'{e}n waves (i.e., $K_{y} = M_{y}$).
Roughly speaking, these approximations are consistent with an
assumption that the wave frequencies are higher than
$\sim 10^{-3}$ Hz in the corona.
However, it has also been shown that the radial behavior of
Alfv\'{e}nic wave power in the solar wind is never far from the
predictions of WKB theory even in the heliosphere where
reflection is not negligible \citep{Za96,CvB05}.
\item
Because of other evidence that the dominant inertial-frame
frequencies in coronal MHD turbulence may be lower than
$\sim 10^{-4}$ Hz \citep[see, e.g.,][]{CH09,Cr10}, we
made use of a low-frequency approximation for the Alfv\'{e}n
wave reflection coefficient ${\cal R}$.
This also involved an analytic approximation for the radial
dependence of the Alfv\'{e}n speed scale height $H_{\rm A}$
(Equation (\ref{eq:tref})).
\item
For the fast and slow magnetosonic waves, we modeled the
radial transport of an isotropic ensemble of propagation directions
$\theta$ using a single wave action conservation equation.
We chose one reasonable method to perform the averages over
$\theta$, but other methods may yield different results.
We also used the Eulerian average for the outflow speed $u_0$
and neglected the second-order effects of ``Stokes drift'' that
would enter into the associated Lagrangian version of the mean
\citep[see][]{Cr09b}.
\end{enumerate}
Although the effects of removing these approximations should be
investigated further, we do not believe their use invalidates
the results of the wave transport models presented above.

With the above caveats taken into account, we produced a standard
model of the Alfv\'{e}n, fast, and slow mode energy densities
between 0.01 and 1000 $R_{\odot}$ above the solar photosphere.
In agreement with earlier results,
we found that slow-mode MHD waves of solar origin probably
cannot survive into the extended corona.
In addition, we found that the amplitudes of fast-mode waves at
large distances are more sensitive to the assumed model parameters
than are the amplitudes of Alfv\'{e}n waves.
For this reason the standard model of fast-mode wave energy density
was treated as a representative example and not a definitive
prediction.
Thus, other reasonable models of the available fast-mode power
can be obtained by multiplying or dividing the standard model's
energy density by factors of order 10--100 without sacrificing
too much realism.

At each radial distance, we simulated the time-steady wavenumber
power spectra of Alfv\'{e}nic and fast-mode turbulent fluctuations.
We included the effects of nonlinear coupling and collisionless
kinetic wave dissipation.
We also computed the time-steady heating rates for protons,
electrons, and a representative minor ion species (O$^{+5}$) for
comparison with observational constraints.
The resulting heating rates for the standard model of fast-mode
wave power was found to provide both substantial heating for
coronal protons as well as produce agreement with the preferential
O$^{+5}$ ion heating measured by UVCS/{\em{SOHO}}.
However, if the fast-mode wave power in the corona is significantly
{\em lower} than was assumed in the standard model, the proposed
idea of mode coupling is probably not a viable mechanism for
the ion heating.

In order to match some of the observations---such as the need
for $Q_{p}/Q_{e}$ to be of order unity at
$z \lesssim 0.1 \, R_{\odot}$ and for the O$^{+5}$ heating
rate to agree with that measured by SUMER/{\em{SOHO}} at similar
heights---we found that approximately 10 times the standard model's
assumed fast-mode wave energy density may need to be present.
This could be accounted for in several ways.
First, we neglected the effects of Alfv\'{e}n waves giving rise to
second order fluctuations that mimic the properties of both fast
and slow magnetosonic waves \citep{Ho71,VH99}.
It is possible that these secondary oscillations could behave
similarly enough to ideal fast-mode waves that they enable the
same kinds of cascade and coupling.
Second, we also neglected nonlinear couplings that involve slow-mode
MHD waves, which appear to dominate the density fluctuations in the
low corona.
It may be possible for these couplings \citep[see, e.g.,][]{YF08}
to also power the high-$k_{\parallel}$ part of the Alfv\'{e}nic
fluctuation spectrum.

To make further progress with the proposed set of ideas,
it will be important to better understand the origin of the fast,
slow, and Alfv\'{e}n waves in the solar photosphere and chromosphere.
\citet{Ho78a}, \citet{Su81}, and others studied the wavelike
oscillations induced by convective jostling in small-scale flux
tubes that extend up into the chromosphere.
However, once waves reach the sharp and ``corrugated'' TR boundary,
they can undergo reflection, refraction, and multiple types of mode
conversion \citep{Ho78b,Bg02,HvB08,Fe11,CH11}.
The types and strengths of MHD waves that survive the chaotic
lower atmosphere probably also depend on the nature of the region
underlying the solar wind flux tubes of interest (i.e., coronal
hole, active region, or quiet loops).

Future work must also involve more physical realism for the model
of turbulent cascade.
Replacing our hodge-podge collection of analytic solutions with a
fully self-consistent numerical simulation is an obvious priority.
A key part of this improvement will be to remove the assumption
of scale separation that prevents different radial zones from
interacting with one another in wavenumber space
\citep[see, e.g.,][]{VV09}.
In addition, we note that the advection-diffusion terms in
Equations (\ref{eq:difeqA})--(\ref{eq:difeqF}) contain the
limiting assumption that the spectral transfer is ``local''
in ${\bf k}$-space.
It has been shown that true MHD turbulence is not so local because
of intermittent high-order wave-wave interactions and nonlinear
steepening effects \citep[e.g.,][]{Mv00,Ch08b,Co10,Hw12}.
We also assumed energy equipartition between the
$v_{\perp}$ and $b_{\perp}$ spectra in the MHD inertial range,
but in~situ measurements show that not to be the case in actual
solar wind turbulence \citep{Gr83,Wy11}.

We also intend to improve upon the kinetic treatment of collisionless
particle heating described in Section \ref{sec:disp}.
We assumed isotropic Maxwellian velocity distributions when
solving for the linear damping rates, but \citet{Bs10} showed
how non-Maxwellian temperature anisotropies can significantly
affect the KAW dispersion relation.
The ultimate rate of electron heating from KAW Landau damping
can also be affected by nonlinearity and Coulomb collision effects
that we did not include \citep[e.g.,][]{BG11}.
The time evolution of proton and ion velocity distributions,
under the influence of cyclotron resonant heating, is also
decidedly non-Maxwellian \citep{GS00,Is01,Cr01,IV09,IV11}.

Finally, we emphasize that the proposed idea of nonlinear coupling
between Alfv\'{e}n and fast-mode waves is only one proposed solution
to the problem of preferential proton/ion heating.
Some of the other suggested explanations were listed briefly in
Section \ref{sec:intro}.
One recent example that has received significant attention is
the stochastic energization of protons and ions that occurs when
KAW amplitudes become sufficiently high \citep{JC01,Ch10,Ch11}.
To excite this proposed stochasticity, the dimensionless ratio
$v_{\perp}/c_{s}$ (evaluated at $k_{\perp}\rho_{p} = 1$)
should exceed values of order 0.1.
However, in this paper's standard model of Alfv\'{e}nic fluctuations
(either with or without nonlinear couplings), this ratio never
exceeds a value of 0.003.
The main factor responsible for this dramatic mismatch is our
assumption of the \citet{GS95} scaling in the limit of strong
turbulence (i.e., $v_{\perp} \propto k_{\perp}^{-1/3}$).
Alternate theories of the strong Alfv\'{e}nic cascade
\citep[e.g.,][]{Bo06,Pd11} give a shallower dependence of
$v_{\perp} \propto k_{\perp}^{-1/4}$.
This would allow larger values of $v_{\perp}$ to survive to the
onset of KAW dispersion at $k_{\perp}\rho_{p} \approx 1$.
We await improved theoretical descriptions of MHD turbulence and
conclusive empirical tests of such models.

\acknowledgments
The authors are indebted to Ben Chandran for indispensable
contributions to this work.
We also acknowledge Greg Howes, Phil Isenberg,
Peera Pongkitiwanichakul, Bill Matthaeus, Steve Spangler,
and the anonymous referee for many helpful comments and discussions.
This work was supported by the National Aeronautics and Space
Administration (NASA) under grants {NNX\-09\-AB27G},
{NNX\-10\-AC11G}, and {NNX\-10\-AQ27G}
to the Smithsonian Astrophysical Observatory.

\appendix

\section{A. Heuristic Overview of MHD Turbulence}
\label{appen:overview}

The cascade of energy from large to small eddies was first described
in the context of isotropic hydrodynamic turbulence
\citep{vK38,K41,Ob41,Ba53}.
The spectral transport timescale for energy to be transferred down
to the next order of magnitude of eddy size is estimated generally
as $\tau_{s} \approx (k v_{k})^{-1}$, where $k$ is the
magnitude of the local wavevector ${\bf k}$ and $v_k$ is the local
eddy velocity at this value of $k$.
For isotropic fluctuations that depend only on $k$ and not its
direction, we can define the reduced one-dimensional spectrum
$e_{m}(k) = v_{k}^{2} / k$.
Thus, since
\begin{equation}
  \frac{U_m}{\rho_0} \, = \, \int dk \,\, e_{m} (k)
  \,\, ,
\end{equation}
we relate the eddy velocity to the full three-dimensional spectrum
via $v_{k}^{2} = 4\pi k^{3} E_{m}$.
The cascade rate is estimated as $\varepsilon \sim v_{k}^{2} / \tau_s$.
Assuming that $\varepsilon$ is constant in the inertial range
leads to the time-steady Kolmogorov-Obukhov spectrum
$e_{m} \propto k^{-5/3}$, or $E_{m} \propto k^{-11/3}$.

When the background magnetic field becomes strong, other physical
processes become important.
\citet{Ir63} and \citet{Kr65} (hereafter IK) realized that the
``eddy'' description of hydrodynamic turbulence could be generalized
by referring to colliding MHD wave packets, and that the Alfv\'{e}n
speed $V_{\rm A}$ introduces a new absolute scale into the problem.
If one continues to treat the cascade isotropically in $k$-space,
a more generalized spectral transport time can be defined as
\begin{equation}
  \tau_{s} \, = \, \frac{1}{k v_{k}}
  \left( \frac{V_{\rm A}}{v_{k}} \right)^{p}
  \label{eq:tauIK}
\end{equation}
where $p=0$ gives the Kolmogorov-Obukhov limit and $p=1$ is the result
of the IK analysis.
Using the same assumption above that $\varepsilon$ is constant,
we obtain a more general one-dimensional power spectrum
$e_{m} \propto k^{-(p+5)/(p+3)}$.
For the IK value of $p=1$, the spectrum is $e_{m} \propto k^{-3/2}$
\citep[see also][]{Bo05}.

It has been known for several decades that a cascade of
Alfv\'{e}n-wave-like fluctuations does not lead to an isotropic
distribution of power in wavenumber space \citep{St76,MT81,Sb83,Hg84}.
The dominant energy cascade takes place mainly in the two-dimensional
plane perpendicular to the background field.
For the Alfv\'{e}nic fluctuations, we can define the local eddy
velocity as $v_{\perp}$ being mainly a function of $k_{\perp}$.
The one-dimensional spectrum in this case is given by
$e_{\rm A} = v_{\perp}^{2} / k_{\perp}$ and the integration over
wavenumber space is best done in cylindrical coordinates with
\begin{equation}
  \frac{U_{\rm A}}{2\pi \rho_0} \, = \,
  \int dk_{\parallel} \int dk_{\perp} \, k_{\perp} E_{\rm A}
  \, = \, \int dk_{\perp} \, e_{\rm A} \,\, .
\end{equation}
Taking into account the spectral anisotropy
($k_{\parallel} \neq k_{\perp}$) we can also write an even more
general perpendicular transport time for the Alfv\'{e}n waves as
\begin{equation}
  \tau_{\rm A} \, = \, \frac{1}{k_{\perp} v_{\perp}}
  \left( \frac{V_{\rm A}}{v_{\perp}} \right)^{p}
  \left( \frac{k_{\parallel}}{k_{\perp}} \right)^{q}  \,\, .
  \label{eq:tauA}
\end{equation}
A perpendicular generalization of the IK model is given by
$p=1$ and $q=0$, which gives $e_{\rm A} \propto k_{\perp}^{-3/2}$
\citep[see also][]{Nk99,Nk01,Bo06,Pd11}.
Weak three-wave couplings have been shown to give rise to the case
$p=q=1$, which yields $e_{\rm A} \propto k_{\perp}^{-2}$
\citep[e.g.,][]{Ga00,BN01,BP09}.
However, in that case nonlinear effects grow in magnitude as
$k_{\perp}$ gets larger, so it is generally believed that a weakly
turbulent inertial range must eventually become strongly turbulent
\citep[see also][]{GS97}.

\citet{GS95} described strong Alfv\'{e}nic turbulence with a
spectral transfer time given by $p=q=0$, and thus
$e_{\rm A} \propto k_{\perp}^{-5/3}$ reminiscent of the original
Kolmogorov-Obukhov model.
In this case of strong mixing between the turbulent motions
(perpendicular to the field) and the flow of Alfv\'{e}n wave packets
(parallel to the field) there is a so-called ``critical balance''
that couples $k_{\perp}$ and $k_{\parallel}$ to one another.
We define a critical balance parameter
\begin{equation}
  \chi \, \approx \,
  \frac{k_{\parallel} V_{\rm A}}{k_{\perp} v_{\perp}}
\end{equation}
such that the \citet{GS95} strong cascade is consistent with the
condition $\chi \approx 1$.
Combining this with the velocity scaling
$v_{\perp} \propto k_{\perp}^{-1/3}$ yields the wavenumber anisotropy
scaling $k_{\parallel} \propto k_{\perp}^{2/3}$.
Note that assuming $p=q$ in Equation (\ref{eq:tauA}) is equivalent
to $\tau_{\rm A}$ being given by $\chi^{p} / (k_{\perp} v_{\perp})$;
see also \citet{Ga05}.
An alternate way of describing the cascade was given by \citet{ZM90b},
who defined a triple correlation timescale equivalent to
\begin{equation}
  \tau_{\rm A} \, \approx \,
  \frac{1 + \chi}{k_{\perp} v_{\perp}} \,\, .
  \label{eq:tauAbridge}
\end{equation}
This expression naturally bridges the strong ($\chi \lesssim 1$) and
weak ($\chi \gg 1$) turbulence scaling limits, and we use a
similar relation in Section \ref{sec:cascade:eqns}.

The cascade of compressible fast-mode waves has received less attention
than that of the incompressible Alfv\'{e}n waves.
Because fast-mode waves propagate at a roughly constant phase speed
no matter the direction angle $\theta$, the fast-mode cascade has
been suspected to resemble isotropic hydrodynamic turbulence.
In fact, numerical simulations do tend to find that fast-mode waves
produce a more isotropic spectrum than do Alfv\'{e}n waves
\citep{Co03,Sd09}.\footnote{%
This is generally valid in the ideal MHD range, at which
$\omega \lesssim \Omega_p$.
We ignore the large literature of ``whistler turbulence,'' in which
dispersive effects may lead to wavenumber anisotropy
at higher frequencies $\omega \gg \Omega_p$.}
The rate of the cascade is generally assumed to follow the weak
IK-type scaling of Equation (\ref{eq:tauIK}) with $p=1$
\citep[see, e.g.,][]{Ch05,Sz07}.
Thus, because in most cases we expect $v_{k} \ll V_{\rm A}$,
the fast-mode cascade timescale $\tau_{\rm F}$ is likely to be
significantly longer than the Alfv\'{e}nic timescale $\tau_{\rm A}$.

It is important to emphasize that there is still no agreement
concerning the most realistic and universal way to describe MHD
turbulence.
There remains controversy about the applicability of the various
power law exponents (especially $5/3$ versus $3/2$) for the
Alfv\'{e}nic inertial range \citep{By11,Fm11,Ms11,Pd11}.
Simulations have not been able to accurately pin down the amount of
slow ``leakage'' of power to the high-$k_{\parallel}$ region of the
spectrum where $\chi \gg 1$.
Furthermore, the observed steepening of the spectrum at high values
of $k_{\perp}$ is still not well understood \citep{Lm98,St01,Hw08}.
In many models, the precise scalings depend on the degree of cross
helicity of the fluctuations (i.e., on the imbalance between $Z_{+}$
and $Z_{-}$) and on whether the turbulence is driven or decaying
\citep[e.g.,][]{Lw07,Ch08b,Cn11}.
In this paper, we attempt to identify the most controversial aspects
of the models and discuss how they can be modified once such issues
are resolved.

\section{B. Linear Damping Rates: Collisional and Collisionless}
\label{appen:rates}

\citet{Al47} and \citet{Os61} first proposed that MHD waves in the
solar atmosphere could be damped by collisional processes.
These processes include viscosity, thermal conductivity, electrical
resistivity (i.e., Joule or Ohmic dissipation), and ion-neutral friction.
In the fully collisional regime we make use of the basic expressions
derived by \citet{B65}.
Here we describe the total linear damping rate for MHD wave mode $m$
by a sum of three components,
\begin{equation}
  \gamma_{m} \, = \, \gamma_{{\rm vis,} m} +
  \gamma_{{\rm ohm,} m} + \gamma_{{\rm con,} m} 
\end{equation}
where $\gamma_{{\rm vis,} m}$ denotes damping due to kinematic viscosity,
$\gamma_{{\rm ohm,} m}$ denotes electrical resistivity, and
$\gamma_{{\rm con,} m}$ denotes thermal conductivity.
Since our main goal is to model the wave damping in the (almost completely
ionized) corona and solar wind, we ignore ion-neutral friction.
For Alfv\'{e}n waves,
\begin{equation}
  \gamma_{\rm vis, A} \, = \, \frac{1}{\rho_0} \left(
  \eta_{1} \, k_{\perp}^{2} +
  \eta_{2} \, k_{\parallel}^{2} \right)
  \label{eq:gamvisA}
\end{equation}
\begin{equation}
  \gamma_{\rm ohm, A} \, = \, 
  \left( \frac{c^2}{4\pi\sigma_{\parallel}} \right) k_{\perp}^{2} +
  \left( \frac{c^2}{4\pi\sigma_{\perp}} \right) k_{\parallel}^{2}
\end{equation}
\begin{equation}
  \gamma_{\rm con, A} \, = \, 0 \,\, .
\end{equation}
For fast and slow mode waves ($m = \mbox{F,S}$), we note that the
damping rates given explicitly by \citet{B65} are valid only in
the $\beta \ll 1$ limit.
The expressions given here are appropriate for arbitrary values
of $\beta$, but we made the assumption that
$k_{\parallel} \approx k_{\perp}$.
In other words, for the assumed isotropic distribution of fast and
slow wave vectors, the damping rates depend only on the magnitude
$k^{2} = k_{\parallel}^{2} + k_{\perp}^{2}$.
With that caveat, the damping rates are given by
\begin{equation}
  \gamma_{{\rm vis,} m} \, = \, \frac{2 k^{2}}{\rho_0} \left[
  \frac{\eta_0}{3} \left( 4f_{vz} - 4 f_{xz} + f_{vx} \right)
  + \eta_{1} f_{vx}
  + \eta_{2} \left( f_{vx} + 2 f_{xz} + f_{vz} \right) \right]
\end{equation}
\begin{equation}
  \gamma_{{\rm ohm,} m} \, = \, 
  \frac{c^{2} k^{2} f_{\rm B}}{4\pi\sigma_{\perp}}
\end{equation}
\begin{equation}
  \gamma_{{\rm con,} m} \, = \,
  \frac{4 (\gamma-1)^{2} k_{\rm B} T}{\rho_{0} c_{s}^2}
  \, f_{\rm th} k^{2} \left(
  \kappa_{\parallel} + \kappa_{\perp} \right)  \,\, ,
  \label{eq:gamcon}
\end{equation}
where the fractions $f$ specify the energy partition fractions
of Section \ref{sec:global:linear}.
Specifically, $f_{\rm th} = \Theta / U_{m}$,
$f_{\rm B} = (M_{x} + M_{z}) / U_{m}$,
$f_{vx} = K_{x} / U_{m}$, $f_{vz} = K_{z} / U_{m}$, and
$f_{xz} = (f_{vx} f_{vz})^{1/2}$.
We assume that protons dominate other ion species in the
viscosity and thermal conductivity terms, and that electrons
dominate the electrical resistivity terms
\citep[see also][]{Tu84,Wh97,Cp99}.

We use the \citet{B65} expressions for the transport coefficients
in the fully collisional limit.
These coefficients depend on the proton and electron Coulomb collision
timescales \citep[e.g.,][]{Sp62},
\begin{equation}
  \tau_{p} \, = \, \frac{3}{4} \sqrt{\frac{m_p}{\pi}}
  \frac{(k_{\rm B} T_{p})^{3/2}}{e^{4} n_{p} \ln\Lambda}
\end{equation}
\begin{equation}
  \tau_{e} \, = \, \frac{3}{4} \sqrt{\frac{m_e}{2\pi}}
  \frac{(k_{\rm B} T_{e})^{3/2}}{e^{4} n_{p} \ln\Lambda}
\end{equation}
where we approximate the Coulomb logarithm in the coronal regime
of temperatures and densities by
\begin{equation}
  \ln \Lambda \, = \, 23.2 + \frac{3}{2} \ln \left(
  \frac{T_e}{10^{6} \, \mbox{K}} \right) - \frac{1}{2} \ln \left(
  \frac{n_e}{10^{6} \, \mbox{cm}^{-3}} \right)  \,\, .
\end{equation}
We also specify the magnitudes of the proton and electron
gyrofrequencies,
\begin{equation}
  \Omega_{p} = \frac{e B_0}{m_{p} c}
  \,\, , \,\,\,\,
  \Omega_{e} = \frac{e B_0}{m_{e} c}
\end{equation}
and the dimensionless products
\begin{equation}
  x_{p} = \tau_{p} \Omega_{p}
  \,\, , \,\,\,\,
  x_{e} = \tau_{e} \Omega_{e} \,\, .
\end{equation}
Thus, the proton viscosity coefficients are given by
\begin{equation}
  \eta_{0} = 0.96 \, n_{p} k_{\rm B} T_{p} \tau_{p}
  \,\, , \,\,\,\,
  \eta_{1} = 2 x_{p} \eta_{2}
\end{equation}
\begin{equation}
  \eta_{2} = n_{p} k_{\rm B} T_{p} \tau_{p}
  \left( \frac{1.2 x_{p}^{2} + 2.23}
  {x_{p}^{4} + 4.03 x_{p}^{2} + 2.33} \right) \,\, .
\end{equation}
The electrical conductivities are given by
\begin{equation}
  \sigma_{\parallel} = \frac{1.95 \, e^{2} n_{e} \tau_{e}}{m_e}
\end{equation}
\begin{equation}
  \sigma_{\perp} = \frac{e^{2} n_{e} \tau_{e}}{m_e}
  \left( 1 - \frac{6.416 x_{e}^{2} + 1.837}
  {x_{e}^{4} + 14.79 x_{e}^{2} + 3.7703} \right)^{-1} \,\, .
\end{equation}
The thermal conductivities are given by
\begin{equation}
  \kappa_{\parallel} = \frac{3.906 \, n_{p} k_{\rm B} T_{p} \tau_{p}}{m_p}
\end{equation}
\begin{equation}
  \kappa_{\perp} = \frac{n_{p} k_{\rm B} T_{p} \tau_{p}}{m_p}
  \left( \frac{2 x_{p}^{2} + 2.645}
  {x_{p}^{4} + 2.70 x_{p}^{2} + 0.677} \right) \,\, .
\end{equation}

Finally, we need to take account of the transition from collisional
to collisionless wave damping.
In the low-density limit of the classical \citet{B65} expressions,
some of the transport coefficients (e.g., $\eta_0$, $\eta_1$,
$\kappa_{\parallel}$) become infinitely large as the mean time
between collisions becomes infinite.
This ``molasses limit'' has been recognized to be unphysical
\citep[see][]{Wi95,CvB05}.
Thus, we derived a simplified version of the general expressions of
\citet{CC92} and \citet{Ji09} to describe what happens when
collisions become infrequent.
We computed the $\gamma$ dissipation rates as above, but then
we multiplied them all by the following dimensionless factor $C$,
where
\begin{equation}
  C = \frac{\tau_{\rm esc}}{\tau_{\rm esc} + \tau_{p}}
\end{equation}
and the macroscopic expansion timescale for waves is estimated as
\begin{equation}
  \tau_{\rm esc} \approx
  \frac{\rho_0}{(u + V_{\rm A}) | \partial \rho_{0} / \partial r |}
  \,\, .  \label{eq:tauesc}
\end{equation}
For strong collisions, $C \approx 1$ and the \citet{B65} expressions
are valid.
For weak collisions, $C \approx \tau_{\rm esc} / \tau_{p} \ll 1$ and
the damping rates are quenched.

\begin{figure}
\epsscale{0.60}
\plotone{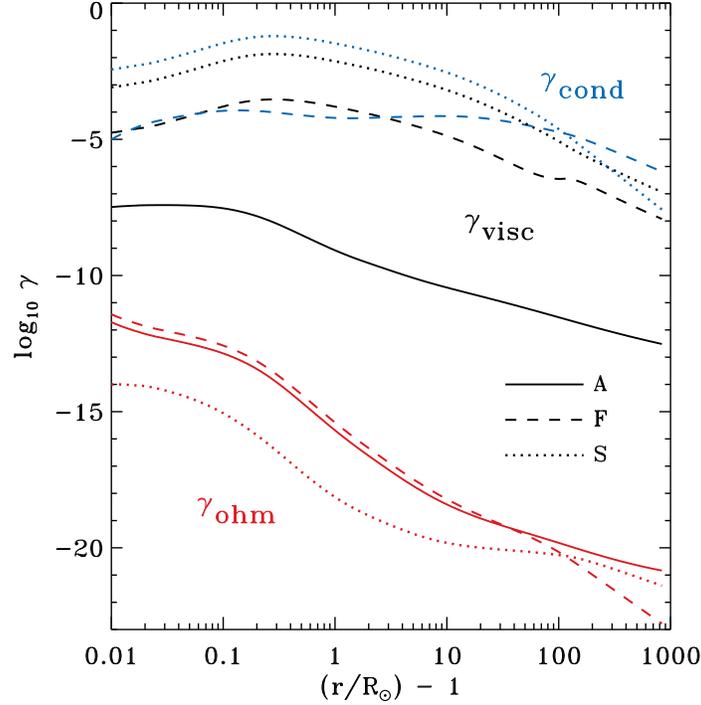}
\caption{Linear collisional damping rates for MHD waves.
Color denotes the physical dissipation process:
viscosity (black), electrical resistivity (red), and thermal
conductivity (blue).
Line style denotes the wave mode:
Alfv\'{e}n (solid curves), fast mode (dashed curves), and
slow mode (dotted curves).
All quantities are plotted as base-10 logarithms of the rates,
in units of s$^{-1}$, for the standard model with parameters listed
in Table \ref{table01}.
\label{fig18}}
\end{figure}
Figure \ref{fig18} illustrates how the components of the wave damping
rates vary with radial distance in the model of the fast solar wind
described in Section \ref{sec:global}.
For the fast and slow mode waves, the viscous and conductive damping
terms are of roughly comparable strength, but for the fast mode
the conductive damping wins out at large distances ($\beta \gg 1$).
The viscous term is most important for the Alfv\'{e}n mode, but
its magnitude remains small in comparison to the dominant terms for
fast and slow mode damping.
At the heights displayed here, Ohmic dissipation never appears to
be important in comparison to the other terms.
This situation is reversed, however, lower down in the chromosphere
\citep[see, e.g.,][]{Kh04}.
The curves in Figure \ref{fig18} are shown for the general case of
the transition to a collisionless plasma (i.e., with all rates
multiplied by $C$).
For $z \lesssim 0.1 \, R_{\odot}$ in the low corona, $C \approx 1$
and the general rates are identical to the unmodified \citet{B65}
rates.
For heights greater than $z \approx 1 \, R_{\odot}$, however, the
the rates multiplied by $C$ become about two orders of magnitude
smaller than the unmodified classical rates.

\section{C. Analytic Solutions to Advection-Diffusion Problems in
Limited Parameter Regimes}
\label{appen:ansol}

\subsection{C.1. Alfv\'{e}n Waves: Cascade and Source Terms}
\label{appen:ansol:egrow}

Equation (\ref{eq:difbperp}) is a reduced one-dimensional version
of the full advection-diffusion equation for Alfv\'{e}nic fluctuations.
The Appendix of \citet{CvB03} presented one method of solving this
equation in the low-wavenumber, strong turbulence ($\chi_{0} \ll 1$)
limit.
Here we derive a more general case for arbitrary $\chi_0$.
Ignoring both wave damping and mode coupling, and assuming
a steady state (i.e., $\partial b_{\perp}^{2} / \partial t = 0$),
Equation (\ref{eq:difbperp}) can be simplified to
\begin{equation}
  \frac{\partial \varepsilon}{\partial x} \, = \, \tilde{S}_{\rm A}
\end{equation}
where here we define $x = \ln k_{\perp}$ and we write the cascade rate as
\begin{eqnarray}
  \varepsilon & = & \frac{1}{\tau_{\rm A}} \left(
    \mu_{\perp} b_{\perp}^{2} - \alpha_{\perp}
    \frac{\partial b_{\perp}^{2}}{\partial x} \right) \label{eq:epsq1} \\
  & = & - \frac{\alpha_{\perp} k_{\perp}^{1+s}}{\tau_{\rm A}}
    \frac{\partial}{\partial k_{\perp}} \left( b_{\perp}^{2}
    k_{\perp}^{-s} \right) \label{eq:epsq2}
\end{eqnarray}
where $s = \mu_{\perp} / \alpha_{\perp}$.
Note that the ratio $s$ was called $\beta / \gamma$ by \citet{CvB03}
and \citet{LC09}.
The second form for $\varepsilon$ given in Equation (\ref{eq:epsq2})
helps to show that the power-law spectrum for $b_{\perp}$ in the
inertial range (i.e., where $\tilde{S}_{\rm A} = 0$ and
$\varepsilon$ is constant) should be independent of the value of $s$.
In the limiting cases of strong ($\chi_{0} \ll 1$) and weak
($\chi_{0} \gg 1$) turbulence, we use Equation (\ref{eq:tauAformal})
to find that $b_{\perp}$ is proportional to $k_{\perp}^{-1/3}$
and $k_{\perp}^{-1/2}$, respectively.

In regions of wavenumber space where the source term is nonzero,
$\varepsilon$ is not constant and the simple inertial-range scalings
do not apply.
If we assume that most of the fluctuation power is injected near
$x \approx x_{0} = \ln k_{0 \perp}$, then it makes sense to use a
compact Gaussian shape for the source term,
\begin{equation}
  \tilde{S}_{\rm A}(x) \, = \,
  \frac{\varepsilon_0}{\pi^{1/2} \sigma_0}
  \exp \left[ - \left( \frac{x - x_0}{\sigma_0}
  \right)^{2} \right]  \,\, ,
\end{equation}
where the dimensionless width of the Gaussian is specified by
$\sigma_{0} = 1$ in our models.
The constant $\varepsilon_0$ is varied arbitrarily to produce the
desired total fluctuation energy density $U_{\rm A}$.
\citet{CvB03} showed how the above form for the source function
integrates to a cascade rate
\begin{equation}
  \varepsilon (x) \, = \, \frac{\varepsilon_0}{2} 
  \left[ 1 + \mbox{erf} \left(
  \frac{x - x_0}{\sigma_0} \right) \right] \,\, .
\end{equation}
Finally, we define an auxiliary parameter
$q = b_{\perp}^{2} k_{\perp}^{-s}$ and integrate
Equation (\ref{eq:epsq2}) to obtain
\begin{equation}
  q^{3/2} \, = \, \frac{3}{2 \alpha_{\perp}} \int_{k_{\perp}}^{\infty}
  \frac{d\kappa \, (1 + \chi_{0}) \, \varepsilon(\kappa)}
  {\phi^{1/2} \kappa^{2 + (3s/2)}} \,\, .
  \label{eq:q32}
\end{equation}
In practice, we integrate this equation numerically and use the
definition of $q$ to obtain $b_{\perp}(k_{\perp})$.
In the energy containing range ($x \ll x_{0}$), we see that
$\varepsilon \rightarrow 0$, and thus $q$ is constant and
$b_{\perp}^{2} \propto k_{\perp}^s$.
The low-$k_{\perp}$ region of wavenumber space stands in contrast
to the inertial range because here the shape of the fluctuation
spectrum {\em does} depend on the value of $s$.

In the MHD strong-turbulence inertial range (where
$\varepsilon \approx \varepsilon_{0}$,
$\phi \approx 1$, and $\chi_{0} \ll 1$), we obtain the standard
solution $b_{\perp} \propto v_{\perp} \propto k_{\perp}^{-1/3}$.
However, when $k_{\perp} \rho_{p}$ increases past unity into the KAW
dispersive range, we see from Equation (\ref{eq:phidef}) that
for $\beta \gg 1$ there exists a sizable ``dispersion range''
in which $\phi \propto k_{\perp}^2$ and we obtain
\begin{equation}
  b_{\perp} \propto k_{\perp}^{-2/3}
  \,\, , \,\,\,\,
  v_{\perp} \propto k_{\perp}^{+1/3}
  \,\, .
\end{equation}
Converting these into the more commonly used one-dimensional spectra
(see Appendix \ref{appen:overview}), the MHD inertial range
has $e_{\rm A} \propto k_{\perp}^{-5/3}$, and the KAW inertial range
has $e_{\rm A} \propto k_{\perp}^{-7/3}$ for the magnetic fluctuations
and $e_{\rm A} \propto k_{\perp}^{-1/3}$ for the (electron) velocity
fluctuations.
These scalings have been described by, e.g.,
\citet{Bi96}, \citet{CvB03}, and \citet{Hw08}.
Note, however, that strong damping also begins to occur in the KAW
regime, so the above power laws may not be evident in the final
modeled spectra.

\subsection{C.2. Alfv\'{e}n Waves: Cascade and Dissipation Terms}
\label{appen:ansol:edecay}

We include the effects of high-$k_{\perp}$ dissipation in the
perpendicular cascade by assuming the damping acts only at
wavenumbers significantly above those where the source term is
dominant.
Thus, we assume that $\tilde{S}_{\rm A} = 0$ and we continue to
ignore mode coupling.
In this limiting case, the time-steady advection-diffusion
equation becomes
\begin{equation}
  k_{\perp} \frac{\partial \varepsilon}{\partial k_{\perp}}
  \, = \, - 2 \tilde{\gamma}_{\rm A} b_{\perp}^{2}  \,\, .
  \label{eq:pureadv}
\end{equation}
We note that \citet{Hw08} found it is important to include KAW
damping when solving for the steady-state wave power (and thus the
proton/electron energy partitioning) at large values of $k_{\perp}$.

In Appendix \ref{appen:ansol:egrow} we showed that the properties of
the inertial range spectra should be independent of the value of $s$.
In order to produce a closed-form solution, here we follow
\citet{Hw08} and assume that $s \rightarrow \infty$ (i.e., the
cascade proceeds purely by wavenumber advection).
Thus, by assuming that $\alpha_{\perp} = 0$, we can use
Equation (\ref{eq:epsq1}) to write the $b_{\perp}^{2}$ term on
the right-hand side of Equation (\ref{eq:pureadv}) as
\begin{equation}
  b_{\perp}^{2} \, \approx \,
  \frac{\varepsilon}{\mu_{\perp} k_{\perp} v_{\perp}}
\end{equation}
if we also assume $\chi_{0} \ll 1$ in the high-$k_{\perp}$ limit.
To retain the most generality in cases when $s$ is not infinitely
large, we can use Equation (\ref{eq:alcomp}) to replace $\mu_{\perp}$
by $\mu_{\perp \ast} = \mu_{\perp} + 2 \alpha_{\perp} / 3$.
This may not be completely accurate for the cases of weakest advection
(i.e., $s \approx 0$), but it is an improvement on ignoring
the influence of the $\alpha_{\perp}$ diffusion term altogether.

To simplify the modified transport equation, we take a further cue
from \citet{Hw08} and use the critical balance condition
$\omega \approx k_{\perp} v_{\perp}$ to rewrite the equation as
\begin{equation}
  \frac{k_{\perp}}{\varepsilon}
  \frac{\partial \varepsilon}{\partial k_{\perp}} \, = \,
  - \frac{2}{\mu_{\perp \ast}} \left( \frac{\tilde{\gamma}}{\omega}
  \right)_{\rm A}
\end{equation}
where the ratio $(\tilde{\gamma}/ \omega)_{\rm A}$ is the output
of the Vlasov-Maxwell dispersion analysis
discussed in Section \ref{sec:disp}.
For the KAW domain, this ratio is largely independent of
$k_{\parallel}$ and thus it can be treated as a function of
$k_{\perp}$ only.
When we solve this equation numerically, we start the integration
at a low enough value of $k_{\perp}$ that the damping is negligibly
small (i.e., at which $\varepsilon = \varepsilon_{0}$).
Thus, we integrate upwards in $k_{\perp}$ with
\begin{equation}
  \varepsilon \, = \, \varepsilon_{0} \, \exp \left[
  - \frac{2}{\mu_{\perp \ast}} \int \frac{dk_{\perp}}{k_{\perp}}
  \left( \frac{\tilde{\gamma}}{\omega} \right)_{\rm A} \right] \,\, .
  \label{eq:epsgamma}
\end{equation}
Finally, the damped solution for $\varepsilon (k_{\perp})$ is
used in Equation (\ref{eq:q32}) to obtain the damped power spectrum.

\subsection{C.3. Alfv\'{e}n Waves: Coupled Parallel and Perpendicular Transport}
\label{appen:ansol:gy}

The previous sections described the cascade as a function of
$k_{\perp}$ and ignored the behavior of the full power spectrum
$E_{\rm A} (k_{\parallel}, k_{\perp})$.
To first order, the strong predicted anisotropy of MHD turbulence
justifies this approach, but we are also concerned with the possible
leakage of power to high values of $k_{\parallel}$ and thus to high
frequencies.
Here we mainly follow the analysis of \citet{CvB03}, but we also
include the possible effects of weak turbulence when $\chi_{0} \gg 1$.
\citet{GS95} wrote the full power spectrum as a separable function
of two variables: $k_{\perp}$ and $\chi$, with
\begin{equation}
  E_{\rm A} (k_{\parallel}, k_{\perp}) \, = \,
  \frac{V_{\rm A} b_{\perp}(k_{\perp})}{k_{\perp}^3} g(\chi)
\end{equation}
and $\chi = k_{\parallel} V_{\rm A} / (k_{\perp} b_{\perp})$.
This definition allows the dimensionless function $g(\chi)$ to be
normalized to unity,
\begin{equation}
  \int_{-\infty}^{+\infty} d\chi \, g(\chi) \, = \, 1 \,\, ,
  \label{eq:gnorm}
\end{equation}
but in practice we usually calculate the normalization for $g(\chi)$
from the condition that the total power over all wavenumber space
integrates properly to $U_{\rm A}$.

It has been known for some time that the dominant contribution to
the integral in Equation (\ref{eq:gnorm}) should come from the region
where $| \chi | \lesssim 1$.
For values of $k_{\parallel}$ at which $| \chi | \lesssim 1$, the
\citet{GS95} solution for $b_{\perp} \propto k_{\perp}^{-1/3}$
gives a dominant perpendicular dependence for the intertial range
of $E_{\rm A} \propto k_{\perp}^{-10/3}$.
We also expect $g(\chi)$ to grow negligibly small for $|\chi| \gg 1$,
and thus for these large-$k_{\parallel}$ regions of wavenumber space
there should be very little Alfv\'{e}nic wave power.
\citet{Co02} found that numerical simulations of anisotropic MHD
turbulence were consistent with $g(\chi)$ being fit reasonably well
with either a simple exponential function ($g \sim e^{-\chi}$) or
a Castaing function (a convolution of multiple exponentials).
\citet{CvB03} derived an analytic solution to a simplified version
of Equation (\ref{eq:difeqA}), with
\begin{equation}
  g (\chi) \, = \, \frac{2\Gamma (n)}{3\Gamma (n - 0.5) \sqrt{\pi}}
  \left( 1 + \frac{4 \alpha_{\perp} \chi^{2}}{9 \alpha_{\parallel}}
  \right)^{-n}
  \label{eq:gkappa}
\end{equation}
and
\begin{equation}
  n \, = \, 1 + \frac{3 s}{4} \,\, .
\end{equation}
These expressions are appropriate for the MHD inertial range
where $v_{\perp} \propto k_{\perp}^{-1/3}$, but we use them for the
entire range of modeled wavenumbers.
The above form for $g(\chi)$ resembles a generalized Lorentzian,
or kappa distribution \citep[e.g.,][]{Va68,PL10} that is Gaussian for
small arguments and evolves to a power-law tail for large arguments.
We can simplify the argument of the power-law term by using
the values of the cascade parameters discussed in
Section \ref{sec:cascade:eqns}, with
$\alpha_{\perp} / \alpha_{\parallel} \approx 18.6 / (3s + 2)$.

An example choice for the dimensionless advection-diffusion ratio
($s = \mu_{\perp}/ \alpha_{\perp} = 2$) gives rise to a power-law
exponent $n = 5/2$, and the large-$k_{\parallel}$ behavior of the
Alfv\'{e}nic power spectrum is $E_{\rm A} \propto k_{\parallel}^{-5}$.
Smaller values of $s$ give shallower power-law slopes.
In fact, \citet{CvB03} and \citet{LC09} found that if $s$ could
be maintained at small values of order 0.1--0.3, there would be
sufficient high-$k_{\parallel}$ power to heat protons and heavy
ions in the corona via ion cyclotron resonance.
In the opposite limit of pure advection (i.e., $s \rightarrow \infty$
or $\alpha_{\perp} \rightarrow 0$, with $\alpha_{\parallel}$ and
$\mu_{\perp}$ remaining finite) Equation (\ref{eq:gkappa})
becomes a Gaussian,
\begin{equation}
  g(\chi) \, \propto \, \exp \left( -
  \frac{\mu_{\perp}}{3 \alpha_{\parallel}} \, \chi^{2} \right) \,\, .
  \label{eq:ggauss}
\end{equation}
\citet{Ch08b} also obtained a similar Gaussian solution
for the parallel spectrum under the assumption of pure advection.
Using the values of the cascade parameters discussed in
Section \ref{sec:cascade:eqns}, we can set
$\mu_{\perp} \approx 1.95$ in the limit of $s \rightarrow \infty$.
Thus, the ratio $\mu_{\perp} / \alpha_{\parallel} \approx 6.2$ and
we can write $g \sim e^{-2 \chi^{2}}$ in the pure-advection limit.

In this paper we modify the analysis described above in one
additional way.
Instead of using the usual critical balance parameter $\chi$
as the argument of $g(\chi)$, we instead use
\begin{equation}
  \chi_{\rm eff} \, = \, \frac{\chi}{\sqrt{1 + \chi_{0}^{2}}}
\end{equation}
where $\chi_0$ is defined in Equation (\ref{eq:chi0}).
In the strong turbulence regime ($\chi_{0} \ll 1$) this modification
makes no difference.
In the weak turbulence regime ($\chi_{0} \gg 1$) this has the effect
of extending the ``filled'' region of the spectrum (i.e.,
$g(\chi_{\rm eff}) \approx 1$) up through all wavenumbers with
$k_{\parallel} \lesssim k_{0 \parallel}$.

Finally, we must adjust the highest frequency part of the spectrum
to account self-consistently for the effects of ion cyclotron damping.
Because the cyclotron resonance at high $k_{\parallel}$ has a rapid
onset (see Figure \ref{fig11}(a)), we need only model its effects
over a limited range of wavenumber space.
We truncate the calculation of the spectrum at a maximum parallel
wavenumber
\begin{equation}
  \frac{k_{\parallel {\rm max}} V_{\rm A}}{\Omega_p}
  \, = \, \frac{0.72}{\beta^{0.43}}
  \label{eq:kparamax}
\end{equation}
at which $| \gamma_{\rm A} / \Omega_{p} | \approx 1$.
Above this wavenumber, we found that slowly-varying solutions to
the Vlasov-Maxwell dispersion relation cease to exist
\citep[see also][]{Sx92}.
Between $0.1 k_{\parallel {\rm max}}$ and $k_{\parallel {\rm max}}$,
we include the time-steady effect of resonant damping by assuming that
the local Alfv\'{e}nic wave power is produced solely by the nonlinear
coupling with the fast mode.
If only the coupling and damping are present, the time-steady
transport equation simplifies to
\begin{equation}
  \frac{\partial E_{\rm A}}{\partial t} \, \approx \,
  \frac{E_{\rm F} - E_{\rm A}}{\tau_{\rm AF}}
  - 2 \gamma_{\rm A} E_{\rm A} \, = \, 0 \,\, ,
  \label{eq:coupdamp}
\end{equation}
which can be solved analytically for $E_{\rm A}$.
However, we note that we already have a time-steady solution
for $E_{\rm A}$ in the presence of nonlinear coupling, but it
does not take into account the ion cyclotron damping.
Equation (\ref{eq:EAigamma}) gives that solution, which we now
call $E_{0 {\rm A}}$.
Thus, we insert it in place of $E_{\rm F}$ in
Equation (\ref{eq:coupdamp}) above, since that is the solution
toward which the coupling will drive the system in the absence
of damping.
We then use the analytic solution
\begin{equation}
  E_{\rm A} \, \approx \,
  \frac{E_{0 {\rm A}}}{1 + 2 \gamma_{\rm A} \tau_{\rm AF}}
  \label{eq:EAcycdamp}
\end{equation}
to account for ion cyclotron dissipation at high $k_{\parallel}$.
This solution gives rise to a significant reduction in the power
spectrum when the damping rate $\gamma_{\rm A}$ exceeds the 
rate at which power is supplied from the nonlinear coupling.

\subsection{C.4. Fast-Mode Waves: Cascade and Source Terms}
\label{appen:ansol:fast}

This section is conceptually similar to Appendix \ref{appen:ansol:egrow}
in that we ignore both damping and mode coupling, assume time-steady
conditions, and model the spectral transport of fast-mode waves as
a balance between diffusive cascade and the outer-scale source term.
In that limiting case, Equation (\ref{eq:difeqF}) becomes
\begin{equation}
  \frac{\partial \varepsilon}{\partial k} \, = \, k^{2} S_{\rm F}
  \label{eq:appBef1}
\end{equation}
where the cascade rate is defined here as
\begin{equation}
  \varepsilon \, = \,
  - \frac{4\pi \alpha_{\rm F} k^{8} \sin\theta}{V_{\rm A}}
  E_{\rm F} \frac{\partial E_{\rm F}}{\partial k}
  \,\, .
\end{equation}
When $S_{\rm F} = 0$, the cascade rate is constant along radial
rays of constant $\theta$, and thus $E_{\rm F} \propto k^{-7/2}$.
We follow \citet{Ch05} and others by assuming a Gaussian shape for
the fast-mode source term, but we also add a corresponding
$\sin\theta$ angle dependence, with
\begin{equation}
  S_{\rm F} (k) \, = \, S_{0} \,
  \exp \left[ - \left( \frac{k}{k_{0 {\rm F}}} \right)^{2} \right]
  \, \sin\theta \,\, .
\end{equation}
We eventually set $S_0$ at the level required to maintain
the spectrum at the known total energy density $U_{\rm F}$.
Our choice to constrain $k_{0 {\rm F}}$ to be equal to
$k_{0 \perp} = 1 / \lambda_{\perp}$ is discussed in
Section \ref{sec:cascade:eqns}.

Because both the left and right sides of Equation (\ref{eq:appBef1})
depend identically on $\sin\theta$, the resulting time-steady
solution for $E_{\rm F} (k)$ becomes independent of $\theta$.
This outcome was motivated by simulation results that show that
the fast-mode power spectrum is largely isotropic in wavenumber
space \citep{Co03,Sd09}.
It is also possible that additional isotropization of the fast-mode
spectrum can come from couplings with slow-mode waves \citep{Ch08a}
or from multi-scale ``wandering'' of the magnetic field that gives
rise to a continuously varying direction for $\theta = 0$ \citep{SK07}.
In future work, our method of artificially forcing isotropy via the
source term should be replaced with a more realistic description.

In any case, we cancel out both instances of $\sin\theta$ and integrate
Equation (\ref{eq:appBef1}) to obtain
\begin{equation}
  \varepsilon (k) \, = \, k_{0 {\rm F}}^{3} S_{0} \left(
  \frac{\sqrt{\pi}}{4} \mbox{erf} \, x - \frac{x e^{-x^{2}}}{2} \right)
  \label{eq:epsF}
\end{equation}
where here $x = k / k_{0 {\rm F}}$.
In the limit $x \gg 1$, the term in parentheses above approaches a
constant value of $\sqrt{\pi} / 4$.
In the limit $x \ll 1$, the term in parentheses is approximately
equal to $x^{3} / 3$.
Finally, we integrate the definition of $\varepsilon$ to get the
time-steady spectrum,
\begin{equation}
  \frac{E_{\rm F}^{2}}{2} \, = \, \frac{V_{\rm A}}{4\pi \alpha_{\rm F}}
  \int_{k}^{\infty} d\kappa \, \frac{\varepsilon(\kappa)}{\kappa^8}
\end{equation}
which we evaluate numerically using Equation (\ref{eq:epsF}) for
the cascade rate in the integrand.
In the energy containing range ($k \ll k_{0 {\rm F}}$), this yields
$E_{\rm F} \propto k^{-2}$, and thus $v_{k} \propto k^{+1/2}$.

\subsection{C.5. Fast-Mode Waves: Cascade and Dissipation Terms}
\label{appen:ansol:fdamp}

If we consider high values of $k$ above those affected by the
outer-scale source term, we can solve for the transition from the
inertial range to the dissipation range in the fast-mode power spectrum.
An analytic solution becomes possible if we rewrite the fast-mode
spectral transport time $\tau_{\rm F}$ as a function of $k$ and
$\theta$ only.
This can be done by using the time-steady inertial range scaling
for $v_k$, with
\begin{equation}
  v_{k}^{2} \, = \, v_{0}^{2} \left( \frac{k}{k_0} \right)^{-1/2}
  \,\, , \,\,\,\,
  \tau_{\rm F} \, = \frac{V_{\rm A}}{v_{0}^{2} \sin\theta
  \sqrt{k \, k_0}}
  \,\, .
  \label{eq:v0k}
\end{equation}
Note that Equation (8) of \citet{Sz07} gave the same result for the
fast-mode cascade timescale (but without the $\sin\theta$ term).
The normalization wavenumber $k_0$ is defined arbitrarily here; it
needs to be set well below the regime of strong damping, but well
above the outer-scale wavenumber $k_{0 {\rm F}}$ so that we can
justify ignoring the source term.

The above approximation gives
$D_{\rm F} \propto k^{5/2} \sin\theta$.
It is also straightforward to model the fast-mode damping rate as
being proportional to a constant power of the wavenumber, and thus
we assume $\gamma_{\rm F} = \gamma_{0} (k / k_{0})^{z}$.
Note that the normalizing constant $\gamma_0$ may depend on the
angle $\theta$ as well.
The time-steady version of Equation (\ref{eq:difeqF}) becomes
\begin{equation}
  \frac{\alpha_{\rm F} v_{0}^{2} k_{0}^{1/2} \sin\theta}
  {k^{2} V_{\rm A}}
  \frac{\partial}{\partial k} \left( k^{9/2}
  \frac{\partial E_{\rm F}}{\partial k} \right) \, = \,
  2 \gamma_{0} \left( \frac{k}{k_0} \right)^{z} E_{\rm F}
  \,\, .
\end{equation}
Defining the auxiliary variable
\begin{equation}
  x \, = \, \frac{2}{7} \left( \frac{k_0}{k} \right)^{7/2}
\end{equation}
helps to greatly simplify the differential equation.  Thus,
\begin{equation}
  \frac{\partial^{2} E_{\rm F}}{\partial x^2} \, = \,
  \frac{c_{\gamma} E_{\rm F}}{x^{(2z + 13)/7}}  \,\, ,
  \label{eq:d2Edx2}
\end{equation}
where
\begin{equation}
  c_{\gamma} \, = \, \frac{2 \gamma_{0} V_{\rm A}}
  {\alpha_{\rm F} v_{0}^{2} k_{0} \sin\theta}
  \left( \frac{2}{7} \right)^{(2z + 13)/7}
\end{equation}
is a constant that is essentially the ratio of the damping rate
to the cascade rate at the normalization wavenumber $k_0$.

For $z \geq 1$,
Equation (\ref{eq:d2Edx2}) is solved analytically with two
linearly independent terms proportional to the two types of modified
Bessel function ($I_n$ and $K_n$).
Knowing that the only physically realistic solution is one that
decreases monotonically with increasing $k$ (or with decreasing $x$),
we then use only one of those terms, which is given by
\begin{equation}
  E_{\rm F}(k) \, \propto \, k^{-7/4} \,
  K_{\zeta} \left[ 2\zeta \left( \frac{2}{7} \right)^{1/(2\zeta)}
  \sqrt{c_{\gamma}} \left( \frac{k}{k_0} \right)^{7/(4\zeta)} \right]
  \,\, .
\end{equation}
where $\zeta = 7 / (2z - 1)$.
Note that the transit-time damping rate of Equation (\ref{eq:gamttd})
gives $z = 1$ and $\zeta = 7$.
In the limiting case that the modified Bessel function of the second
kind has a small argument, we have $K_{\zeta}(x) \sim x^{-\zeta}$ and
thus $E_{\rm F} \propto k^{-7/2}$, independent of the value of $\zeta$.
This is the proper inertial-range solution in the case of either
low wavenumber ($k \ll k_0$) or weak damping ($c_{\gamma} \ll 1$).
The opposite case of a large argument gives exponentially steep
dissipation in the limit of large $k$ and/or large $\gamma_0$.
This kind of solution was also derived by \citet{HZ10}.

Another useful special case for the damping exponent is $z = 1/2$.
For this value of the exponent, Equation (\ref{eq:d2Edx2}) is solved
with two linearly independent power-law terms.
As above, we keep only the solution that does not diverge as
$x \rightarrow 0$ (i.e., as $k \rightarrow \infty$), and the
time-steady spectrum is given by
\begin{equation}
  E_{\rm F}(k) \, \propto \, k^{-7(1 + \sqrt{1 + 4 c_{\gamma}})/4}
  \,\, .
  \label{eq:EFzhalf}
\end{equation}
The weak-damping limit of $c_{\gamma} \ll 1$ gives the proper
inertial-range solution $E_{\rm F} \propto k^{-7/2}$,
but the presence of a nonzero value of $c_{\gamma}$ makes the
spectrum steeper.
This is one (possibly rare) case in which a physically motivated
source of damping gives rise to a power-law ``dissipation range.''

\end{document}